\documentclass[prd,aps,floats,floatfix,superscriptaddress,preprintnumbers,
showpacs,eqsecnum,nofootinbib,twocolumn]{revtex4}
\usepackage{amsfonts,amsmath,amssymb,latexsym,array,
theorem,mathrsfs,subfigure,bm,float,color,graphicx}
\usepackage{mathrsfs}

\definecolor{red}{rgb}{1,0,0} 
\definecolor{blue}{rgb}{0,0,1}
\definecolor{green}{rgb}{0,1,0}

\DeclareMathAlphabet{\mathpzc}{OT1}{pzc}{m}{it}

\newcommand{\ma}[1]{\mbox{$\mathcal{#1}$}}

\newcommand{\calhR}[1]{\raisebox{2ex}{\tiny ({\em h})}\hspace{-0.8em}{\ma R}}
\newcommand{\gsim}{\,\mbox{
\raisebox{-1.ex}{$\stackrel{\textstyle>}{\textstyle\sim}$}}\,}

\newcommand{\mpl}{m_{\mathrm{pl}}}

\begin{document}

%<<<<<<<<<<<<< TITLE >>>>>>>>>>>>>>>%
\title{
Black Holes and Thunderbolt Singularities 
with Lifshitz Scaling Terms
}

%<<<<<<<<<<<<< AUTHOR >>>>>>>>>>>>>>>%

\author{Yosuke {\sc Misonoh}}
\email{misonoh-at-aoni.waseda.jp}
\author{Kei-ichi {\sc Maeda}}
\email{maeda-at-waseda.jp}
\address{Department of Physics, Waseda University, 
Okubo 3-4-1, Shinjuku, Tokyo 169-8555, Japan}

%<<<<<<<<<<<<< ADDRESS >>>>>>>>>>>>>>>%

%<<<<<<<<<<<<< DATE >>>>>>>>>>>>>>>%
\date{\today}

%======================================%
%<<<<<<<<<<<<< ABSTRACT >>>>>>>>>>>>>>>%
%======================================%
%%%%%%%%%%%%%%%%%%%%%%%%%%%%%%%%%%%%%%%%%%%
%%%%%%%%%%%%%%%%%%%%%%%%%%%%%%%%%%%%%%%%%%%
%%%%%%%%%%%%%%%%%%%%%%%%%%%%%%%%%%%%%%%%%%%
%%%%%%%%%%%%%%%%%%%%%%%%%%%%%%%%%%%%%%%%%%%
\begin{abstract}
We study  a static, spherically symmetric and asymptotic flat spacetime, 
assuming the hypersurface orthogonal Einstein-aether theory with 
an ultraviolet modification motivated by the Ho\v{r}ava-Lifshitz theory, 
which is composed of the $z=2$ Lifshitz scaling terms such as 
scalar combinations of a three-Ricci curvature and 
the acceleration of the aether field.
For the case with the quartic term of 
the acceleration of the aether field, 
we obtain a two-parameter family of black hole solutions, 
which possess a regular universal horizon.
While, if three-Ricci curvature squared term 
is joined in ultraviolet modification,
 we find a solution with a thunderbolt singularity such that
 the universal horizon turns to be a spacelike singularity.

\end{abstract}
%%%%%%%%%%%%%%%%%%%%%%%%%%%%%%%%%%%%%%%%%%%
%%%%%%%%%%%%%%%%%%%%%%%%%%%%%%%%%%%%%%%%%%%
%%%%%%%%%%%%%%%%%%%%%%%%%%%%%%%%%%%%%%%%%%%
%%%%%%%%%%%%%%%%%%%%%%%%%%%%%%%%%%%%%%%%%%%

%<<<<<<<<<<<<< PACS NUMBERS >>>>>>>>>>>>>>>%
%\pacs{
%} 


\maketitle

%======================================%
%<<<<<<<<<<<< SECTION I  >>>>>>>>>>>>>>%
%======================================%
%%%%%%%%%%%%%%%%%%%%%%%%%%%%%%%%%%%%%%%%%%%
%%%%%%%%%%%%%%%%%%%%%%%%%%%%%%%%%%%%%%%%%%%
%%%%%%%%%%%%%%%%%%%%%%%%%%%%%%%%%%%%%%%%%%%
%%%%%%%%%%%%%%%%%%%%%%%%%%%%%%%%%%%%%%%%%%%
\section{Introduction}
%%%%%%%%%%%%%%%%%%%%%%%%%%%%%%%%%%%%%%%%%%%%%%%%%%%
%%%%%%%%%%%%%%%%%%%%%%%%%%%%%%%%%%%%%%%%%%%%%%%%%%%
%%%%%%%%%%%%%%%%%%%%%%%%%%%%%%%%%%%%%%%%%%%%%%%%%%%
Spacetime singularity is unavoidable in general 
relativity\cite{Hawking_Ellis}, which
 means the breakdown of our standard theory 
of gravity in ultraviolet region. 
 To establish the fundamental gravitational theory beyond general relativity
 is one of the most intriguing question of physics.
One may expect that this difficulty can be 
resolved by considering the quantum effect of gravity.
 However, unfortunately, the perturbative quantization approach of 
general relativity losses the renormalizability unlike 
the other fundamental interactions.
 In other words, there appears infinite numbers of divergent Feynman diagrams, 
and thus, the infinite counter terms are required to regularize the 
gravitational quantum effects. Hence one way to quantize gravity is 
non-perturbative approach such as the loop quantum gravity\cite{loop}
 or the dynamical triangulation\cite{triangulation}. 
 
Another way is to find a new renormalizable 
gravitational theory. String theory \cite{string} 
can be such a candidate, but 
it has not been completed. 
Recently, Ho\v{r}ava proposed a gravitational theory with Lifshitz 
scaling\cite{HL_original_paper}, which is anisotropic scaling between space 
and time, i.e., $t \to b^{-1}t\,,~ x^i \to b^{-z} x^i$.
 This scaling defines a scaling dimension $[t]=-z$ with $[x^i]=-1$, which is 
restored to the ordinary mass dimension when $z=1$.
 It is found that if we set $z$ to the number of the spatial dimension, the 
dimension of the gravitational constant becomes zero, which means the 
gravitational force acquires renormalizability at least at
a power-counting level.
 This gravitational theory is called Ho\v{r}ava-Lifshitz (HL) theory whose 
action includes higher spatial curvature terms up to cubic order as counter 
terms of renormalization.

 Although there is a need for further investigation to confirm whether HL 
theory is truly renormalizable or not\cite{renormalizability},
strong gravitational phenomena, such as  cosmological singularity
 avoidance \cite{HL_cosmological_sin_avoicance}
 and the black hole solution\cite{HL_BH_ref}
 have been studied by several authors.
 In particular, study regarding the spacetime structure is particularly
 intriguing frontier.
 Since the spacetime in HL theory losses local Lorentz symmetry due to 
Lifshitz scaling, the causal structure is drastically changed.
 If there is Lifshitz scaling with $z \neq 1$, the dispersion relation of the
 signal particle is modified as $\omega^2 \sim k^{2z}$, and then the sound 
 speed is given by $c \sim k^{z-1}$.
 In consequence, the sound speed almost diverges if the particle is in 
an extremely high energetic state. 
 One may consider it is impossible to define the casual horizon due to such 
an instantaneously propagating particle.
 However it is not the case.
 In the context of the Einstein-aether (\ae-) theory \cite{EA_original_paper} 
which has some equivalence to the infrared limit of HL 
theory\cite{khronon_theory}, there still exists a causal horizon for 
such an extreme energetic particle.
 Although \ae-theory itself is a toy model as a Lorentz violating 
gravitational theory, it is found that the HL theory is reduced to \ae-theory 
in infrared limit if the aether is restricted to be hypersurface orthogonal.
 In other words, the aether $u_\mu $ is constrained  
to a gradient of some scalar field $\varphi$ as
$u_\mu \propto \nabla_\mu \varphi$.

If we set $\varphi$ to time variable $t$, the HL action in ADM formalism 
is restored.
Therefore $\varphi$ is called a ``khronon”.
In this theory, the spacetime is expressed by a series of three 
dimensional spacelike hypersurface $\varphi=$ constant, denoted by 
$\Sigma_\varphi$. 
Particles with infinite propagating sound speed travel along to 
$\Sigma_\varphi$.
 Then, if there is a spacetime structure such that $\Sigma_\varphi$ 
parallels to a timelike Killing vector $\xi^\mu$, namely $u \cdot \xi =0$,
any particles  cannot escape from 
this surface at least in a spherically symmetric spacetime.
Therefore this surface is a static limit for such
energetic particles and it is called universal horizon\cite{EAHLBH,UH_BH}.
The properties  of the universal horizon have 
 been so far investigated on the following subjects : 
a static and spherically symmetric exact solution in \ae-theory with the 
universal horizon\cite{mec_UH,maximally_EA}, 
the existence of the universal horizon\cite{UH_LV,high_dim_UH,newlook},
a charged black hole solution \cite{charged_BH,C_BH}, 
thermodynamical aspects\cite{mec_UH,UH_tunneling,UH_Wald}, 
a ray trajectory in a black hole spacetime \cite{ray_tracing},
and a formation via gravitational collapse\cite{dynamical_UH,formation_UH}.

Then, one might ask a question whether the universal horizon exist even if 
the Lifshitz scaling terms such as higher
 spatial curvatures are present.
In this paper, we shall consider the backreaction to the black hole solution 
in \ae-theory by the Lifshitz scaling terms.
In other words, we investigate the black hole solution and properties of the
 universal horizon with the Lifshitz scaling terms, which is the 
ultraviolet modification of gravity.
As a first step to clarify the effect of gravitational Lifshitz scaling,
 we shall consider only the $z=2$ scaling terms.

This paper is organized as follows:
In \S \ref{action_equation}, the action we considered is shown.
We include the $z=2$ Lifshitz scaling terms such as 
 the quadratic spatial curvature to the action of \ae-theory with
 hypersurface orthogonal aether and give the basic equations. 
The propagating degree of freedoms in this theory, i.e., the graviton and the
 scalar-graviton are also discussed.
After giving the set up of our static, spherically 
symmetric system in \S III, 
we classify numerical solutions depending on the coupling constants 
of the Lifshitz scaling terms in \S \ref{num_sol}.
We then discuss the properties of a black hole solution and a thunderbolt 
singularity in \S V.
\S \ref{sum_paper} is devoted to conclusion of this paper.

%======================================%
%<<<<<<<<<<<< SECTION II  >>>>>>>>>>>>>>%
%======================================%
%%%%%%%%%%%%%%%%%%%%%%%%%%%%%%%%%%%%%%%%%%%
%%%%%%%%%%%%%%%%%%%%%%%%%%%%%%%%%%%%%%%%%%%
%%%%%%%%%%%%%%%%%%%%%%%%%%%%%%%%%%%%%%%%%%%
%%%%%%%%%%%%%%%%%%%%%%%%%%%%%%%%%%%%%%%%%%%
\section{\ae-theory with Lifshitz scaling}
\label{action_equation}
%%%%%%%%%%%%%%%%%%%%%%%%%%%%%%%%%%%%%%%%%%%%%%%%%%%
%%%%%%%%%%%%%%%%%%%%%%%%%%%%%%%%%%%%%%%%%%%%%%%%%%%
%%%%%%%%%%%%%%%%%%%%%%%%%%%%%%%%%%%%%%%%%%%%%%%%%%%
%%%%%%%%%%%%%%%%%%%%%%%%%%%%%%%%%%%%%%%%%%%%%%%%%%%
%%%%%%%%%%%%%%%%%%%%%%%%%%%%%%%%%%%%%%%%%%%%%%%%%%%
\subsection{non-projectable HL gravity v.s. the \ae-theory 
with Lifshitz scaling}
%%%%%%%%%%%%%%%%%%%%%%%%%%%%%%%%%%%%%%%%%%%%%%%%%%%
%%%%%%%%%%%%%%%%%%%%%%%%%%%%%%%%%%%%%%%%%%%%%%%%%%%
In order to see the behavior of the black hole solution with the backreaction 
from Lifshitz scaling for $z \neq 1$,  
we shall consider non-projectable HL gravity theory. 
In \cite{NPHL_action}, its most general action 
is given as
\begin{eqnarray}
I_{\rm HL}=\int dtd^3x N\sqrt{g_3} \left({\cal L}_K+{\cal L}_P\right)
\end{eqnarray}
with
\begin{eqnarray}
{\cal L}_K&:=& \alpha\left(\mathcal{K}_{ij}\mathcal{K}^{ij}-\lambda 
\mathcal{K}^2
\right)
\nonumber \\
{\cal L}_P&:=& -\left(
{\cal V}_{z=1}+
{\cal V}_{z=2}+ 
{\cal V}_{z=3}\right) 
\,,
\end{eqnarray}
where $N$, $g_{3,ij}$ and $\mathcal{K}_{ij}$ are a lapse function,
 3-metric, and an extrinsic curvature, respectively,
and the potentials are defined by 
\begin{eqnarray}
{\cal V}_{z=1}&:=&\gamma_0 \mathcal{R}+\gamma_1 \Phi_i\Phi^i
\nonumber \\
{\cal V}_{z=2}&:=&\gamma_3 (\Phi_i\Phi^i)^2+\cdots
+\gamma_6 (\Phi_i\Phi^i)\mathcal{R}+\cdots
+\gamma_{10} \mathcal{R}^2
\nonumber \\
{\cal V}_{z=3}&:=&\gamma_{11}(\Phi_i\Phi^i)^3
%+\gamma_{12}(\Phi_i\Phi^i)^2\mathcal{R}
+\cdots+\gamma_{36}\mathcal{R}^{ij}\mathcal{D}_i\mathcal{D}_j\mathcal{R}\,.
\end{eqnarray}
 $\alpha$, $\lambda$ and $\gamma_n$ ($n=0, 1, 3 \cdots, 36$) are 
the coupling constants.
The potential terms include not only the higher-order terms of the
 spatial curvatures $\mathcal{R}_{ij}$, $\mathcal{R}:=\mathcal{R}^i_{\,i}$ 
but also the non-linear terms of the gradient of a lapse function
$\Phi_i:=\mathcal{D}_i\ln N$.
${\cal V}_{z=2}$ and ${\cal V}_{z=3}$ consist 
of 8 and 26 independent terms, respectively \cite{NPHL_action}.

 The IR limit of non-projectable 
HL gravity theory 
is   equivalent to the \ae-theory 
with a hypersurface orthogonality condition\cite{khronon_theory},
from which a spacetime is foliated by three dimensional 
spacelike hypersurface $\Sigma_\varphi$. 
Then the aether field $u_\mu$ is described 
by a gradient of a khronon field $\varphi$ as
\begin{eqnarray}
u_\mu := {\nabla_\mu \varphi \over \sqrt{- (\nabla^\alpha \varphi)
(\nabla_\alpha \varphi)}}\,. 
\label{u_const_HO}
\end{eqnarray}
Since the lapse function $N$ in the HL gravity theory 
is related to this khronon field $\varphi$ in the \ae-theory as
$N=\exp[\varphi]$, the aether field 
corresponds to the gradient of 
a lapse function $\Phi^i$ in 
the low energy IR limit. 

Hence we will study the  \ae-theory with 
additional Lifshitz scaling terms in order to discuss 
black hole solutions in the non-projectable HL gravity.
Although the singular behavior on the Killing horizon can be avoided by 
adopting Painleve-Gullstrand coordinate, it must be singular at 
the universal horizon where the aether becomes normal to 
the timelike Killing vector.
On the universal horizon, the khronon field 
$\varphi$ diverges, in other words, there is 
no continuous time coordinate beyond this horizon in the
(3+1)-decomposition.
We then reformulate the theory in covariant manner 
rather than the ADM (3+1)-decomposition.
It is of great use to avoid coordinate singularities.

 In this paper,
 as a first step, we shall restrict the ultraviolet modification terms only 
to simple scalar terms with $z=2$ scaling such as $\mathcal{R}^2$
 instead of considering all possible terms.
The reason why only  scalar terms are included is that those terms give the
 $k^4$ dependence in the dispersion relation of the scalar-graviton.
Then one expects that the property of the horizon for the scalar-graviton which
 is generally singular in \ae-theory will be drastically altered.
It would be appropriate terms
 to see the backreaction effect by the Lifshitz scaling.

%%%%%%%%%%%%%%%%%%%%%%%%%%%%%%%%%%%%%%%%%%%%%%%%%%%
%%%%%%%%%%%%%%%%%%%%%%%%%%%%%%%%%%%%%%%%%%%%%%%%%%%
\subsection{The action and disformal transformation}
%%%%%%%%%%%%%%%%%%%%%%%%%%%%%%%%%%%%%%%%%%%%%%%%%%%
%%%%%%%%%%%%%%%%%%%%%%%%%%%%%%%%%%%%%%%%%%%%%%%%%%%
We consider the Einstein-aether gravity theory with 
$z=2$ Lifshitz scaling terms, which action is given by  
\begin{eqnarray}
I &=& {1 \over 16 \pi G} \int d^4 x \sqrt{-g} \left[
 \mathcal{L}_{(\mathrm{IR})} +\mathcal{L}_{(\mathrm{UV})} \right]\, 
\notag \\
\mathcal{L}_{(\mathrm{IR})} &=& R -M^{\mu \nu}_{~~\alpha \beta} 
\left( \nabla_{\mu} u^{\alpha} \right) \left(\nabla_{\nu} u^{\beta}\right) \,,
 \notag \\
\mathcal{L}_{(\mathrm{UV})} &=& -\mpl^{-2} \left(\beta_1 \dot{u}^4 
+\beta_2 \dot{u}^2 \mathcal{R} + g_2 \mathcal{R}^2 
 \right) \,. \label{full_action}
\end{eqnarray}
where $R$ is a four dimensional Ricci scalar curvature,
 $G$ is a gravitational 
constant
\footnote{
Note that the Newton gravitational constant $G_{\rm N}$ is different 
from the gravitational constant $G$ appeared in the action.
Taking the weak field limit, we find their relation (\ref{G_N}).
},
$\mpl:=1/\sqrt{G}$ is a Planck mass, which may corresponds to
 a typical Lorentz violating scale, 
and the aether field 
$u^\mu$ is a dynamical unit timelike vector field.
$M^{\mu \nu}_{~\alpha \beta}$ is defined by
\begin{eqnarray}
&&M^{\mu \nu}_{~~\alpha \beta} :=   c_{13} \delta^{\mu}_{~\beta}  
\delta^{\nu}_{~\alpha} + c_2 \delta^{\mu}_{~\alpha}  \delta^{\nu}_{~\beta} 
-c_{14} u^{\mu} u^{\nu} g_{\alpha \beta} \notag \,,
\\
\end{eqnarray}
where $c_{13}:=c_1+c_3, c_{14}:=c_1+c_4 $ with $\{c_i\}$ $(i=1$-$4)$ being 
the coupling constants in the \ae-theory.

The \ae-theory given by ${\cal L}_{\rm (IR)}$ 
with a hypersurface orthogonality condition
 is equivalent to the IR limit of non-projectable 
HL gravity theory
\cite{khronon_theory}.
The relation between both coupling constants is given by 
\begin{eqnarray}
&& \alpha={1-c_{13} \over 16 \pi G}\,,~~\lambda={1-c_2 \over 1-c_{13}}\,, 
\notag \\
&& {\gamma_0 \over \alpha}=-{1 \over 1-c_{13}}\,,~~{\gamma_1 \over \alpha}
=-{c_{14} \over 1-c_{13}}\,.
\end{eqnarray}

A covariantized three dimensional Ricci scalar 
curvature $\mathcal{R}$ is  given by
\begin{eqnarray}
\mathcal{R} &=& R - (\nabla_\alpha u_\beta)(\nabla^\beta u^\alpha) +
 (\nabla_\alpha u^\alpha)^2 \notag \\
&&~~~ + 2\nabla_\alpha [\dot{u}^\alpha - u^\alpha (\nabla_\beta u^\beta) ] \,, 
\label{3-scalar_curvature}\end{eqnarray}
where $\dot{u}^\mu := u^\alpha \nabla_\alpha u^\mu$ is an acceleration of 
the aether corresponding to $\Phi^i$, and $\dot u^2:=\dot u_\mu \dot u^\mu$. 

 $\mathcal{L}_{(\mathrm{UV})}$ is introduced as 
an ultraviolet modification motivated by HL theory with $z=2$, which is composed of the scalar combination of $\mathcal{R}$ and $\dot{u}^2$.
The coupling constants in the $z=2$ Lifshitz scaling 
are rewritten as
\begin{eqnarray}
\beta_1:=16\pi \gamma_{3}\,,~~\beta_2:=16\pi \gamma_{6}\,,~~
g_2:=16\pi \gamma_{10}\,.
\end{eqnarray}
We ignore the other coupling constants, i.e.,
$\gamma_{4}=\gamma_{5}=\gamma_{6}=\gamma_{7}=\gamma_{8}=\gamma_{9}=0$.

 Performing quadratic order perturbation of the action (\ref{full_action}) 
around Minkowski spacetime,  we find two types of the propagating degree of 
freedom.
One is a usual helicity-2 polarization which corresponds to the graviton. 
The other is helicity-0 polarization what we shall refer to as 
a scalar-graviton.
The dispersion relations are given by
 \begin{eqnarray}
\omega_{G}^2 &=& {1 \over 1-c_{13}}k^2  \,, \label{dis_hel_2} \\
\omega_{S}^2 &=& \, {(c_{13}+c_2)(2-c_{14}) \over c_{14}(1-c_{13})
(2+c_{13}+3c_{2}) }k^2 \notag \\ &&~ + {8(c_{13}+c_2) g_2 \over 2+c_{13}
+3c_{2}} \left( {k^2 \over \mpl } \right)^2 \,. \label{dis_hel_0}
 \end{eqnarray}
 Note that the infrared portions which is proportional to $k^2$ correspond to
 \ae-theory's one\cite{ae_wave}.

 As is the case of the \ae-theory\cite{ae_res}, we  find an invariance in 
the above model under the following disformal transformation;
 \begin{eqnarray}
\hat{g}_{\mu \nu} = g_{\mu \nu} +(1-\sigma)u_\mu u_\nu \,,~
\hat{u}^\mu =\sigma^{-1/2}u^\mu \,,
\label{disformal}
 \end{eqnarray}
where $\sigma>0$.
 This transformation can be simplified by introducing 
 three metric on the spacetime  hypersurface $\Sigma_\varphi$,
 which is defined by 
\begin{eqnarray}
\gamma_{\mu \nu}:= g_{\mu \nu} + u_\mu u_\nu
\,.
\end{eqnarray} 
The disformal transformation (\ref{disformal}) is rewritten as
 \begin{eqnarray}
\hat{\gamma}_{\mu \nu} = \gamma_{\mu \nu}\,,~
\hat{u}^\mu =\sigma^{-1/2}u^\mu \,.
\label{field_red}
 \end{eqnarray}
 This transformation (\ref{field_red}) 
means a rescaling of timelike separation between 
two spacelike hypersurfaces with fixing three-dimensional space. 
 Under the transformation,  the action is invariant if  
each coupling constant changes as 
 \begin{eqnarray}
&&\hat{c}_{13}-1 = \sigma(c_{13}-1)\,,~\hat{c}_{13}+\hat{c}_2
=\sigma(c_{13}+c_2)\,,~
\notag \\
&& \hat{c}_{14}=c_{14}\,,~\hat{g}_{2}=g_2\,,~
\hat{\beta}_{1}=
%\sigma^2 
\beta_1\,,~
\hat{\beta}_{2}=
%\sigma^4 
\beta_2\,. ~~~~~
\label{cc_trans_law}
 \end{eqnarray}
 Remarkably, the coefficients proportional to $k^2$ in (\ref{dis_hel_2}) 
and (\ref{dis_hel_0}) are changed to $\sigma^{-1}$ times after the 
transformation, whereas, $k^4$ terms are invariant.
  This means the propagating speeds of each gravitons in infrared limit are
scaled as $\sigma^{-1/2}$ but that of the graviton 
in ultraviolet limit is unchanged.
  This property holds even if the all possible higher curvature terms 
motivated by HL theory are considered 
(see Appendix.{\ref{sec_field_redefinition}}).

%%%%%%%%%%%%%%%%%%%%%%%%%%%%%%%%%%%%%%%%%%%%%%%%%%%
%%%%%%%%%%%%%%%%%%%%%%%%%%%%%%%%%%%%%%%%%%%%%%%%%%%
\subsection{The basic equations}
%%%%%%%%%%%%%%%%%%%%%%%%%%%%%%%%%%%%%%%%%%%%%%%%%%%
%%%%%%%%%%%%%%%%%%%%%%%%%%%%%%%%%%%%%%%%%%%%%%%%%%%
To derive the basic equations, we shall start from taking the variation 
of action (\ref{full_action}) with respect to $g^{\mu \nu}$ and $u^\mu$,
i.e., 
\begin{eqnarray}
\delta I=  {1 \over 16 \pi G} \int d^4 x \sqrt{-g} \left[ E_{\mu \nu} 
\cdot \delta g^{\mu \nu} +2\AE_{\mu}\cdot \delta u^\mu  \right]\,, 
\end{eqnarray}
where,
\begin{eqnarray}
E_{\mu \nu} &:=& E^{(\mathrm{IR})}_{\mu \nu} - 
\mpl^{-2 }[ g_2 E^{(g_2)}_{\mu \nu}  + \beta_1 E^{(\beta_1)}_{\mu \nu}  
+ \beta_2 E^{(\beta_2)}_{\mu \nu}] \,, \notag \\ 
\label{variation_begin} \\
\AE_{\mu} &:=& \AE^{\mathrm{(IR)}}_\mu - \mpl^{-2 }[g_2  \AE^{(g_2)}_\mu 
+ \beta_1 \AE^{(\beta_1)}_\mu
  + \beta_2 \AE^{(\beta_2)}_\mu] \,, \notag \\
\end{eqnarray}
The infrared portions, $E^{\mathrm{(IR)}}_{\mu \nu}$
and $\AE^{\mathrm{(IR)}}_{\mu}$, are defined by
\footnote{
The round and square brackets in the tensoral index are
 a symmetrization and anti-symmetrization symbols, respectively,
 i.e., $A_{(\mu \nu)} := {1 \over 2} (A_{\mu \nu}
 + A_{\nu \mu})$ and $A_{[\mu \nu]} := {1 \over 2}(A_{\mu \nu} - A_{\nu \mu})$. 
}
\begin{eqnarray}
E^{(\mathrm{IR})}_{\mu \nu}  &:=& G_{\mu \nu}  -c_{14} \dot{u}_{\mu} 
\dot{u}_{\nu} + {1 \over 2}J^{\alpha}_{~\beta} \left( \nabla_{\alpha} 
u^{\beta} \right) g_{\mu \nu} \notag \\
&& + \nabla_{\alpha} \left[ J_{(\mu}^{~~\alpha} u_{\nu)} - J_{(\mu \nu)}
 u^{\alpha} - J^{\alpha}_{~(\mu} u_{\nu)} \right]\,, \notag \\ \\
\AE^{(\mathrm{IR})}_{\mu}  &:=& \nabla_{\alpha} J^{\alpha}_{~~\mu} 
+ c_{14} \dot{u}_{\alpha} \left( \nabla_{\mu} u^{\alpha} \right)  \,, 
\end{eqnarray}
where $G_{\mu \nu} := R_{\mu \nu} -{1 \over 2}R g_{\mu \nu}$ is Einstein 
tensor and
\begin{eqnarray}
J^{\mu}_{~~\nu} &:=& M^{\mu \alpha}_{~~\nu \beta} \left( \nabla_{\alpha} 
u^{\beta}\right) \notag \\
&=& c_{13} \nabla_\nu u^\mu + c_2 (\nabla_\alpha u^\alpha) 
\delta^{\mu}_{~\nu} -c_{14} u^\mu \dot{u}_{\nu} \,. 
\end{eqnarray}
The ultraviolet portions,
 $E^{(g_2)}_{\mu \nu}$, $E^{(\beta_1)}_{\mu \nu}$, $E^{(\beta_2)}_{\mu \nu}$,
 $\AE^{(g_2)}_\mu$, $\AE^{(\beta_1)}_\mu$ and $\AE^{(\beta_2)}_\mu$ are
 obtained as 
 \newpage
\begin{widetext}
\begin{eqnarray}
{1 \over 2}E^{(g_2)}_{\mu \nu} &=& \mathcal{R} R_{\mu \nu} 
-{1 \over 4}\mathcal{R}^2 g_{\mu \nu} + (\nabla^2 \mathcal{R})g_{\mu \nu}
 + \nabla_\alpha(a^\alpha \mathcal{R})g_{\mu \nu} 
-\nabla_{(\mu} \nabla_{\nu )}  \mathcal{R} \notag \\
&& +\nabla_\alpha \left[ u_\mu u_\nu (\nabla^\alpha \mathcal{R}) 
-u^\alpha (\nabla_{(\mu} u_{\nu)})\mathcal{R} -u_{(\mu}(\nabla_{\nu)} u^\alpha)
\mathcal{R} +(\nabla^\alpha u_{(\mu})u_{\nu)}\mathcal{R} -2u^\alpha u_{(\mu}
 (\nabla_{\nu)} \mathcal{R} ) \right] \,, \\
\notag \\
E^{(\beta_1)}_{\mu \nu} &:=& -{1 \over 2}\dot u^4 g_{\mu \nu} 
-2\dot u^2 \dot u_\mu \dot u_\nu +\nabla_\alpha 
\left[ 4 \dot u^2 u^\alpha u_{(\mu} \dot u_{\nu)} -2 \dot u^2 
\dot u^\alpha u_\mu u_\nu \right] \,, \\
E^{(\beta_2)}_{\mu \nu} &:=& -{1 \over 2}\dot u^2 \mathcal{R} g_{\mu \nu}
 -\mathcal{R} \dot u_\mu \dot u_\nu +\dot u^2 \mathcal{R}_{\mu \nu} 
-\nabla_{(\mu} \nabla_{\nu)} \dot u^2 +\nabla_\alpha \Big[ 2 \mathcal{R}
 u^\alpha u_{(\mu} \dot u_{\nu)} -\mathcal{R} \dot u^\alpha u_\mu u_\nu 
+\nabla^\alpha (\dot u^2 \gamma_{\mu \nu}) \notag \\ \
&& -\dot u^2 u^\alpha (\nabla \cdot u)g_{\mu \nu} +\nabla_{\beta} (\dot u^2 u^\alpha u^\beta)g_{\mu \nu} +\dot u^2 u^\alpha \nabla_{(\mu} u_{\nu)} 
+\dot u^2 u_{(\mu} \nabla_{\nu)} u_\alpha 
-\dot u^2 (\nabla^\alpha u_{(\mu}) u_{\nu)} 
-2 \nabla_{(\mu} (u_{\nu)} \dot u^2 u^\alpha ) \Big] \,, \notag \\ \\
{1 \over 2} \AE^{(g_2)}_\mu &:=& (\nabla_\alpha \nabla_\mu u^\alpha 
 -\nabla_\mu \nabla_\alpha u^\alpha)\mathcal{R} +(\nabla_\alpha u^\alpha)
(\nabla_\mu \mathcal{R}) - (\nabla_\mu u^\alpha)(\nabla_\alpha \mathcal{R})
 \,, \\
\AE^{(\beta_1)}_\mu &:=& 
2\dot u^2 \dot u_\alpha (\nabla_\mu u^\alpha) -2 \nabla_\alpha
 (\dot u^2 \dot u_\mu u^\alpha)\,, \\
\AE^{(\beta_2)}_\mu &:=& \mathcal{R} \dot u_\alpha (\nabla_\mu u^\alpha)
 -\nabla_\beta (\dot u_\mu u^\beta \mathcal{R}) 
+\dot u^2 (\nabla_\alpha \nabla_\mu u^\alpha) 
-\nabla_\alpha (\dot u^2 \nabla_\mu u^\alpha) 
+\nabla_\alpha \nabla_\mu (\dot u^2 u^\alpha) 
+ \nabla_\mu (\dot u^2 \nabla \cdot u) \notag \\
&& -\dot u^2 \nabla_\mu (\nabla \cdot u) 
-\nabla_\mu \nabla_\alpha (\dot u^2 u^\alpha) 
 \,.
\label{variation_end}
\end{eqnarray}
\end{widetext}
Note that $E_{\mu \nu}=0$ and $\AE_{\mu}=0$ are not the basic equations,
 because the constraint of the aether field has not been taken into account.
To find the basic equations, we usually 
have to introduce a Lagrange multiplier.
Instead expressing the aether by $u^\mu = U^\mu / 
\sqrt{-U_\alpha U^\alpha}$
 where $U^\mu$ is an arbitrary timelike vector field,
we find the basic equations from a variation of $U^\mu$.
Since a variation of $u^\mu$ is given by
\begin{eqnarray}
\delta u^\mu &=& -{1 \over 2}u^\mu u_\alpha u_\beta (\delta g^{\alpha \beta}) 
\nonumber \\
&&
+ {(\delta^\mu_{~\alpha} + u^\mu u_\alpha) \over \sqrt{-U^\gamma U_\gamma}}
(\delta U^\alpha) \,,
\end{eqnarray}
we find  the basic equations as
\begin{eqnarray}
&&E_{\mu \nu} -(\AE_\alpha u^\alpha)u_\mu u_\nu =0\,, \label{ae_eq_1} \\
&&(g_{\mu \alpha} +u_\mu u_\alpha)\AE^\alpha =0\,, \label{ae_eq_2}\\
&&u_\alpha u^\alpha =-1 \label{ae_eq_3}\,.
\end{eqnarray}
Here we rewrite the basic equations in terms of the aether field $u^\mu$ 
with the normalization condition (\ref{ae_eq_3}).

If the aether field is hypersurface orthogonal 
as we have assumed here, we can take a variation with respect to  
the khronon field $\varphi$ instead of  $U^\mu$.
Since the aether field $u^\mu$ is given by Eq. (\ref{u_const_HO}), 
the variation of $\varphi$ is found by using the relation: 
 \begin{eqnarray}
\delta u^\mu &=& \left[u_{(\alpha} \delta^\mu_{~\beta)}+{1 \over 2}u^\mu 
u_\alpha u_\beta \right] (\delta g^{\alpha \beta}) 
\nonumber 
\\
&&
+{(g^{\mu \alpha} +u^\mu u^\alpha)  \over \sqrt{-g^{\alpha \beta}
 (\nabla_\alpha \varphi) (\nabla_\beta \varphi)}}\nabla_\alpha (\delta \varphi)
\,.
\end{eqnarray}
The resultant basic equations are 
\begin{eqnarray}
E_{\mu \nu} +u_\mu u_\nu (\AE_\alpha u^\alpha) +2\AE_{(\mu}u_{\nu)}=0\,, 
 \label{kh_eq_1} 
\end{eqnarray}
and
\begin{eqnarray}
\nabla_\mu \left[ {(g^{\mu \nu} +u^\mu u^\nu) \AE_\nu \over 
\sqrt{-(\nabla^\alpha \varphi) (\nabla_\alpha \varphi)}} \right]=0\,,
\label{kh_eq_2} 
\end{eqnarray}
with the definition (\ref{u_const_HO}). 

In the case of the hypersurface orthogonal aether field, 
although the basic equations are given by Eqs. (\ref{kh_eq_1})
and (\ref{kh_eq_2}) with (\ref{u_const_HO}), 
these equations contain higher-derivative terms of the khronon field $\varphi$.
If spacetime is  static and spherically symmetric, however, 
we may find simpler equations, which are the original basic equations
 (\ref{ae_eq_1})-(\ref{ae_eq_3}).
It is because  the hypersurface orthogonality of the aether is automatically
 satisfied for spherically symmetric spacetime, and then the original 
basic equations are reduced to the basic equations with 
hypersurface orthogonality
\footnote{
The equality of these set of equations holds if the spacetime is regular 
everywhere. This can be proven by considering volume integral of
 (\ref{kh_eq_2}) and using Gauss's theorem\cite{ae-HL_equality}.
 }. 
Although those equations are equivalent,
Eqs. (\ref{ae_eq_1})-(\ref{ae_eq_3}) are written in terms of the aether 
field $u^\mu$,
then those are the second-order differential equations of $u^\mu$.
For this reason, we shall adopt the (\ref{ae_eq_1})-(\ref{ae_eq_3}) as 
the basic equations in the rest of this paper.

%======================================%
%<<<<<<<<<<<< SECTION III  >>>>>>>>>>>>>>%
%======================================%
%%%%%%%%%%%%%%%%%%%%%%%%%%%%%%%%%%%%%%%%%%%
%%%%%%%%%%%%%%%%%%%%%%%%%%%%%%%%%%%%%%%%%%%
%%%%%%%%%%%%%%%%%%%%%%%%%%%%%%%%%%%%%%%%%%%
%%%%%%%%%%%%%%%%%%%%%%%%%%%%%%%%%%%%%%%%%%%
\section{spherically symmetric ``black hole": Set Up}
\label{set_up}
%%%%%%%%%%%%%%%%%%%%%%%%%%%%%%%%%%%%%%%%%%%%%%%%%%%
%%%%%%%%%%%%%%%%%%%%%%%%%%%%%%%%%%%%%%%%%%%%%%%%%%%
%%%%%%%%%%%%%%%%%%%%%%%%%%%%%%%%%%%%%%%%%%%%%%%%%%%
We discuss a static and spherically symmetric spacetime 
with asymptotically flatness.
In order to avoid a coordinate singularity at horizon, we adopt 
the following metric ansatz like the Eddington-Finkelstein type: 
\begin{eqnarray}
ds^2 = -T(r)dv^2 + 2B(r) dvdr + r^2 d\Omega^2 \,, 
 \label{EF_ansatz}
\end{eqnarray}
where $v$ is an ingoing null coordinate and $B\geq 0$.
The aether field in this coordinate system is assumed to be 
\begin{eqnarray}
u^\mu = \Big( a(r),b(r),0,0 \Big)
\,,
 \label{ae_ansatz}
\end{eqnarray}
where the function $b(r)$ is fixed by the normalization condition
(\ref{ae_eq_3}) as 
\begin{eqnarray}
b(r)= {a(r)^2 T(r)-1 \over 2 a(r)B(r)} \,.
\end{eqnarray}
In this spacetime, there exists a timelike Killing vector 
 $\xi^\mu :=(1,0,0,0)$ associated with 
the time translational invariance.

Since the basic equations (\ref{ae_eq_1})-(\ref{ae_eq_3}) in this ansatz 
take quite complicated form, we omit to show it explicitly.
Instead, the structure of the basic equation is illustrated.
Substituting (\ref{EF_ansatz}) into the basic equation 
(\ref{ae_eq_1})-(\ref{ae_eq_2}), we find there are five non-trivial and 
independent set of equations : $(v,v)$, $(v,r)$, $(r,r)$ and 
$(\theta, \theta)$ components of (\ref{ae_eq_1}) and $s^\mu$ projection of 
(\ref{ae_eq_2}), where $s^\mu$ is a ``radial" spacelike unit vector 
perpendicular to $u^\mu$.
From the discussion in \cite{initial_const}, we find following two 
 constraint equations :  
\begin{eqnarray}
C^v =0\,,~~
C^r =0 \label{initial_constraint} \,,
\end{eqnarray}
where $C^\mu$ is defined by
\begin{eqnarray}
C^\mu := E^{r \mu} -2 \left( \AE_\alpha u^\alpha \right) u^r u^\mu 
-u^r \AE^\mu \,.
\end{eqnarray}
These equations include one fewer $r$ derivatives than the rest portion 
of the basic equations, 
and they are automatically preserved by solving the other equations with 
respect to $r$-evolution if (\ref{initial_constraint}) are satisfied on 
an ``initial" constant $r$-surface.
The rest of the equations, namely $(v,v)$ and $(\theta, \theta)$ components 
of (\ref{ae_eq_1}) and $s^\mu$ component of (\ref{ae_eq_2}) give the evolution 
equations with respect to $T(r)$, $B(r)$ and $a(r)$. 

%%%%%%%%%%%%%%%%%%%%%%%%%%%%%%%%%%%%%%%%%%%%%%%%%%%
%%%%%%%%%%%%%%%%%%%%%%%%%%%%%%%%%%%%%%%%%%%%%%%%%%%
\subsection{Asymptotic behaviour}
\label{BH_mass}
%%%%%%%%%%%%%%%%%%%%%%%%%%%%%%%%%%%%%%%%%%%%%%%%%%%
%%%%%%%%%%%%%%%%%%%%%%%%%%%%%%%%%%%%%%%%%%%%%%%%%%%
Since we assume an asymptotic flatness, 
the asymptotic values of the variables are given by 
\begin{eqnarray}
T(r) \rightarrow 1\,, ~ B(r) \rightarrow 1 \,,~
 a(r) \rightarrow 1 \,,~b(r) \rightarrow 0\,.
\end{eqnarray}
In order to investigate the asymptotic behavior of the solution,
 we perform an asymptotic expansion around Minkowski spacetime,
that is, the functions $T(r)$, $B(r)$ and $a(r)$ are expanded as a 
series of $1/r$ as 
\begin{eqnarray}
T(r) &=&1+{T_1 \over r}+{T_2 \over r^2}+{T_3 \over r^3}+{T_4 \over r^4}
+\cdots \,, \notag \\
B(r) &=&1+{B_1 \over r}+{B_2 \over r^2}+{B_3 \over r^3}+{B_4 \over r^4}
+\cdots \,, \notag \\
a(r)&=&1+{a_1 \over r}+{a_2 \over r^2}+{a_3 \over r^3}+{a_4 \over r^4}
+\cdots \,.
\end{eqnarray}
Substituting these series into the basic equations,
and solving them order by order, we find the expansion coefficients as
\begin{widetext}
\begin{eqnarray*}
&&
T_1={\rm arbitrary}\,,~~T_2=0\,,~~
T_3={c_{14} T_1^3 \over 48}\,,~~
\nonumber \\
&&
T_4=
{ \{ (4c_{14}+19)c_{14} -54 c_{13}  \}T_1^4 +192\mpl^{-2}(c_{14}g_2 -\beta_2)T_1^2 +48(c_{14}-2c_{13})(4a_2-3T_1^2)a_2 \over
 192(2-c_{14})}\,,~~~~~~~~~~~~~~~~~ 
\notag \\[.5em]
&&
B_1=0\,,~~
B_2= {c_{14} T_1^2 \over 16} \,,~~
B_3= - {c_{14}T_1^3 \over 12}\,,~~
\notag \\
&&
B_4={ 3c_{14} ( c_{14}^2 +14c_{14} -36c_{13}+4 ) T_1^4  +256 \mpl^{-2}(2c_{14}-1)(c_{14}g_2 -\beta_2)T_1^2 +192(1-c_{13})c_{14}a_2(4a_2 -3T_1^2 ) \over 512 (c_{14}-2)}\,,~~~~ 
\end{eqnarray*}
\begin{eqnarray*}
&&
a_1=-{T_1 \over 2}\,,~~a_2 ={\rm arbitrary}\,,~~a_3=
 -\left( {c_{14}-6 \over 96}T_1^3 +T_1 a_2 \right)
\,,~~~~~~~~~~~~~
 \notag \\
&&a_4={1 \over 1920 (c_{14}-2)c_{123} } \Bigg[ [5c_2 \{ 5c_{14} (2c_{14}-1)+24 \} +18 c_{14} (c_{14}-2) +c_{13}\{  32c_{14}^2 +11c_{14} +30(4-9c_2) \} -270c_{13}^2 ]T_1^4 
~~~~~~~~~~~~~~\notag \\
&&~~~~~~
+48[[ (2-c_{14}) c_{14} + 10c_2 (c_{14}-5) + c_{13} \{ 30c_{123} +c_{14} (c_{14} +8 ) -50 \} ] a_2  +20 c_{123}  \mpl^{-2} (c_{14} g_2 -\beta_2 ) ] T_1^2 \Bigg]  \notag \\
&&~~~~~~
+ {\left(1+c_{13}-c_{14} \over 2-c_{14} \right) a_2^2 }
 \,.
\label{BC}
\end{eqnarray*}
\end{widetext}
The asymptotic behavior of the function $b(r)$ is given by 
\begin{eqnarray}
b(r) &=&{b_2 \over r^2}+{b_4 \over r^4}
+\cdots \,, 
\end{eqnarray}
with 
\begin{eqnarray}
b_2&=& a_2 - {3 T_1^2\over 8} \,,~~
\notag \\
b_4&=&{c_{14} ( 3c_{123} +2c_2 +2 )(3 T_1^2 -8a_2)T_1^2 \over 640}\,.
\end{eqnarray}
The important point is every order of these functions, at least up to the 
eighth order, is described only by two arbitrary coefficients,
 $T_1$ and $a_2$ as in the case of the \ae-theory\cite{EABH}. 
Additionally, the effect of the $g_2$ and $\beta_2$ terms, 
that is, the contribution from the fourth spatial derivative terms 
first appears  in the fourth order coefficients $T_4, B_4$ and $a_4$. 
$\beta_1$ appears after the fifth order
coefficients, which we 
have not shown here because they are so lengthy.

The free parameter $T_1$ is accosted with a black holes mass.
From the discussion in \cite{noether_charge_ae, Hamiltonian_HL, GW_HL}, 
the black hole mass $M$ as a Noether charge with respect to time translational
 symmetry is given by 
\begin{eqnarray}
M=-{T_1 \over 2G} \left(1- {c_{14} \over 2} \right) = -{T_1 \over 2G_{\rm N}} \,,
\label{ae_BH_mass}
\end{eqnarray}
 where 
\begin{eqnarray}
G_{\rm N} := G\left( 1- {c_{14} \over 2} \right)^{-1}
\label{G_N}
\end{eqnarray}
 is the observed Newton constant.
\footnote{
We shall refer $M_{\rm pl}:=1/\sqrt{G_{\rm N}}$ 
as a {\it observed Planck mass} which is related to 
the observed Newton constant $G_{\rm N}$.
Note that the Planck mass $\mpl := 1/\sqrt{G}$ appeared 
in (\ref{full_action}) is rather than related to the Lorentz violating scale.
}

The parameter $a_2$ can be fixed from the analyticity 
of a black hole horizon for the scalar-graviton in 
the infrared limit, but it becomes a free parameter 
when we include the $z=2$ Lifshitz scaling terms
as will be discussed later.
This free parameter $a_2$
may characterize the distribution of 
 an aether cloud around a black hole. 
 
%%%%%%%%%%%%%%%%%%%%%%%%%%%%%%%%%%%%%%%%%%%%%%%%%%%
%%%%%%%%%%%%%%%%%%%%%%%%%%%%%%%%%%%%%%%%%%%%%%%%%%%
\subsection{Black hole horizons}
\label{BH_horizon}
%%%%%%%%%%%%%%%%%%%%%%%%%%%%%%%%%%%%%%%%%%%%%%%%%%%
%%%%%%%%%%%%%%%%%%%%%%%%%%%%%%%%%%%%%%%%%%%%%%%%%%%
 In the \ae-theory or the HL gravity theory, 
the metric horizon, which is the $r$-constant null surface of 
(\ref{EF_ansatz}),
is not generally an event horizon.
 As shown in Eqs. (\ref{dis_hel_2}) and (\ref{dis_hel_0}), 
the sound speeds of the graviton and the scalar-graviton depend 
on the coupling constants. As a result, without tuning of the couplings,
they generally differ from unity.
 The metric horizon only means a static limit for a propagating mode 
with the sound speed being unity.
 For this reason, we first 
reconsider the horizons of the aether black hole. 

%%%%%%%%%%%%%%%%%%%%%%%%%%%%%%%%%%%%%%%%%%%%%%%%%%%
\subsubsection{horizons in the infrared limit}
%%%%%%%%%%%%%%%%%%%%%%%%%%%%%%%%%%%%%%%%%%%%%%%%%%%
Firstly, we shall consider the low-energy infrared limit of the graviton 
and scalar-graviton.
Since the relevant parts in (\ref{dis_hel_2}) and (\ref{dis_hel_0}) are 
$k^2$ terms, the sound speed of the graviton $c_{\rm G}$ and that of 
the scalar-graviton $c_{\rm S}$ in the infrared limit are given by
\begin{eqnarray}
c_{\rm G}^2 \sim {1 \over 1-c_{13}}\,,~ c_{\rm S}^2 
\sim {(c_{13}+c_2)(2-c_{14}) 
\over c_{14}(1-c_{13})(2+c_{13}+3c_{2}) }\,. \label{speed_graviton_IR}
\end{eqnarray}
Note that under the transformation (\ref{field_red}), each sound speed 
is changed as
\begin{eqnarray}
\hat{c}_{\rm G}^2 = \sigma^{-1} c_{\rm G}^2 \,,~~
{\rm and}~~~\hat{c}_{\rm S}^2 = \sigma^{-1} c_{\rm S}^2\,.
\end{eqnarray}
Therefore, if we set $\sigma=c_{\rm G}^2$ or $c_{\rm S}^2$, we find a 
frame in which either 
the sound speed of the graviton or that of the scalar-graviton is unity.
Thus, we can adjust the horizon for the graviton or that of the scalar-graviton 
to the metric horizon by an appropriate disformal transformation.
Explicitly, by performing the follwing disformal transformations;
\begin{eqnarray}
g^{[g]}_{\mu \nu} &:=& g_{\mu \nu} +(1-c_{\rm G}^2) u_\mu u_\nu\,,~ 
 \label{eff_met_g}
 \\
g^{[s]}_{\mu \nu} &:=& g_{\mu \nu} +(1-c_{\rm S}^2) u_\mu u_\nu\,. 
\label{eff_met_s}
\end{eqnarray}
 the graviton horizon or the scalar-graviton horizon are located 
on the $r$-constant null surfaces of the effective metrics 
(\ref{eff_met_g})
 and (\ref{eff_met_s}), respectively.
In the Eddington-Finkelstein ansatz (\ref{EF_ansatz}), the 
$r$-constant null surfaces 
of the graviton and the scalar-graviton are given by 
\begin{eqnarray}
T_{\rm G}(r)=0\,,~~T_{\rm S}(r)=0\,,
\end{eqnarray}
respectively, where
\begin{eqnarray}
T_{\rm G}:=-g^{[g]}_{v v}&=&T-(1-c_{\rm G}^2) 
\left[{1+a^2 T \over 2 a}\right]^2
\,, ~~~~
\\
T_{\rm S}:=-g^{[s]}_{v v}&=&T-(1-c_{\rm S}^2) 
\left[{1+a^2 T \over 2 a}\right]^2
\,,
\end{eqnarray}
 respectively.
In Appendix.\ref{trans_law_EF}, we present the transformation of 
Eddington-Finkelstein type metric (\ref{EF_ansatz}) 
under the disformal transformation 
(\ref{field_red}).

%%%%%%%%%%%%%%%%%%%%%%%%%%%%%%%%%%%%%%%%%%%%%%%%%%%
\subsubsection{horizons in the ultraviolet region}
%%%%%%%%%%%%%%%%%%%%%%%%%%%%%%%%%%%%%%%%%%%%%%%%%%%
In turn, we shall focus on the propagation of the graviton and
 the scalar-graviton in the high energy limit. 
Although the sound speed of the graviton is the same as that in the 
infrared limit,
the sound speed of the scalar-graviton in the high energy limit 
turns to be  
\begin{eqnarray}
c_{\rm S}^2 \sim  {8(c_{13}+c_2) g_2 \over 2+c_{13}+3c_{2}}
 \left( {k \over \mpl } \right)^2 \,, 
\label{mod_sclar_grav_speed}
\end{eqnarray}
which depends on the three momentum $k$.
Thus, the sound speed can increase to infinitely high 
in an ultimately excited state.
In this situation, the $r$-constant 
null surface given by Eq. (\ref{eff_met_s}) 
is no longer an event horizon.
An event horizon of a black hole must be the surface whose outside region 
is causally disconnected from the inside
for any propagation modes even  
with an infinite sound speed.
Otherwise, an inside singularity  is exposed. 

The above case can be resolved to consider the special aether configuration.
The ultimately excited scalar-graviton should propagate along the three 
dimensional spacelike hypersurface. 
In other words, any future directed signal must not propagate against 
the direction which $\varphi$ decreases.
Thus, such an excitation mode must be trapped inside a surface where the 
hypersurface $\Sigma_\varphi$ is parallel to the timelike Killing vector 
$\xi^\mu$, namely, $u \cdot \xi$ vanishes.
This is the concept of the universal horizon which is 
regarded as a real black hole horizon
 in Lorentz violating spacetime\cite{EAHLBH}.

%======================================%
%<<<<<<<<<<<< SECTION IV >>>>>>>>>>>>>>%
%======================================%
%%%%%%%%%%%%%%%%%%%%%%%%%%%%%%%%%%%%%%%%%%%%%%%%%%%
%%%%%%%%%%%%%%%%%%%%%%%%%%%%%%%%%%%%%%%%%%%%%%%%%%%
%%%%%%%%%%%%%%%%%%%%%%%%%%%%%%%%%%%%%%%%%%%%%%%%%%%
%%%%%%%%%%%%%%%%%%%%%%%%%%%%%%%%%%%%%%%%%%%%%%%%%%%
\section{spherically symmetric ``black hole": Solutions}
\label{num_sol}
%%%%%%%%%%%%%%%%%%%%%%%%%%%%%%%%%%%%%%%%%%%%%%%%%%%
%%%%%%%%%%%%%%%%%%%%%%%%%%%%%%%%%%%%%%%%%%%%%%%%%%%
%%%%%%%%%%%%%%%%%%%%%%%%%%%%%%%%%%%%%%%%%%%%%%%%%%%
%%%%%%%%%%%%%%%%%%%%%%%%%%%%%%%%%%%%%%%%%%%%%%%%%%%
 To find a black hole solution with the $z=2$ Lifshitz scaling terms,
we shall solve the basic equations numerically. 
 Our strategy is as follows. (i) To impose the boundary conditions near 
the asymptotically flat region by applying (\ref{BC}), that is, to give 
``initial" values of the variables $T(r), B(r),$ and $a(r)$ 
and their derivatives at infinity.
(ii) To integrate from an appropriate distant spatial point 
toward the center of a spherical object.

%%%%%%%%%%%%%%%%%%%%%%%%%%%%%%%%%%%%%%%%%%%%%%%%%%%
\subsection{black hole solution in the infrared limit {\rm :} \\
the case of $g_2=\beta_1=\beta_2=0$}
\begin{figure}[h]
\begin{center}
\includegraphics[width=70mm]{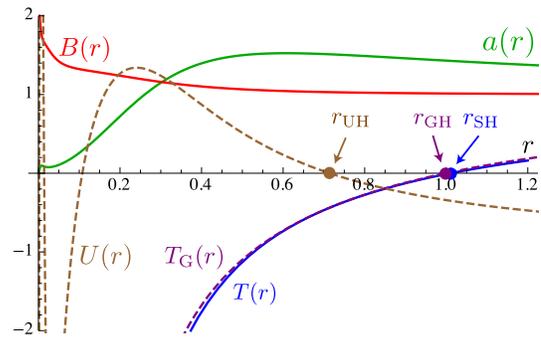}
\end{center}
\caption{Black hole in the
\ae-theory\cite{EABH}. 
The coupling constants are set to 
$c_{13}=0.100, c_{2}=-6.135 \times 10^{-4},
 c_{14}=0.100$. 
We tune the parameter as $a_2 = 1.112 \times 10^{-3}$, 
which gives a regular scalar-graviton horizon. 
The graviton sound speed and the scalar-graviton's one are 
$c_{\rm G}^2 =1.111$ and $c_{\rm S}^2 =1.000$. 
The black hole mass is normalized as $G_{\rm N} M=0.5$.\\
The blue, red and green curves indicate the functions $T(r), B(r)$ 
and $a(r)$,
 respectively. The scalar-graviton horizon is given by  $r_{\mathrm{SH}} = 1.010$.
The dashed purple curve indicates $T_{\rm G}$, from which we find the 
graviton horizon as $r_{\mathrm{GH}} = 0.998$.
The dashed brown curve indicates $U(r):=u \cdot \xi$,
which zero point gives the position of the universal horizons.
The outermost universal horizon radius is $r_{\rm UH} = 0.720$.\\
}
\label{sol_g2zero}
\end{figure}

%%%%%%%%%%%%%%%%%%%%%%%%%%%%%%%%%%%%%%%%%%%%%%%%%%%
First we show the result for the case of the \ae-theory, i.e., 
 $g_2=\beta_1=\beta_2=0$.
It gives a black hole solution in the low-energy infrared limit.
The numerical black hole solution is shown in 
FIG.\ref{sol_g2zero}, which was already found in  \cite{EABH},
for the coupling constants $c_{13}=0.100, c_{2}=-6.135 \times 10^{-4},
 c_{14}=0.100$.
In this solution, the black hole mass is chosen as $G_{\rm N} M=0.5$.
Note that our unit is fixed by setting $T_1=-1$, so that the normalization 
length is $r_M :=2G_{\rm N} M=1$.

Since, the  scalar-graviton sound speed  is set to unity, 
the scalar-graviton horizon  coincides with the metric horizon : $T(r_{\mathrm{SH}})=0$.
We find $r_{\mathrm{SH}}=1.010=2.02 G_{\rm N}M$, which is a little larger than 
the Schwarzschild radius.

The graviton horizon locates inside that of the scalar-graviton, 
i.e., $r_{\mathrm{GH}}= 0.998<r_{\mathrm{SH}}$. 
It is because the graviton sound speed 
 is faster than the scalar-graviton's one.
The most outer universal horizon is formed inside these two horizons, i.e.,
$r_{\rm UH} = 0.720<r_{\mathrm{GH}}, r_{\mathrm{SH}}$.
Additionally, more than one inner universal horizons is formed due to
the rapid oscillatory behavior of the function $U(r)$ near the central singularity.
It means, this solution has a causally disconnected region for 
low energy particles even if the 
particle with $z \neq 1$ Lifshitz scaling is taken into account
  \footnote{
 Note that an instantaneous propagating mode appears 
 when the interaction between khronon and matter field is taken into account 
 even if the higher spatial derivative terms in action are 
absent\cite{UH_BH, g_b_healthy}.
  In our discussion, however,
 we focus only on the gravitational part of the theory 
 whose action is given by (\ref{full_action})
 without $\mathcal{L}_{(\rm{UV})}$. 
 }.
In this sense, this solution is regarded as a black hole 
in the low-energy infrared limit.

The important point of this solution is that 
a physical singularity  generally appears
 on the scalar-graviton horizon (= the metric horizon  in the present case),
if we do not tune the parameter $a_2$.
In order to regularize the scalar-graviton horizon, 
we must choose an appropriate value for the parameter $a_2$ as a boundary condition, 
which is $a_2 = 1.112 \times 10^{-3}$.
Then, the function $B_{[\ae]}$, which is the ``coefficient" of $1/T_{\rm S}$ 
in the evolution equation of $B(r)$, must vanish
 on the scalar-graviton horizon $r=r_{\mathrm{SH}}$.
We present the detailed analysis of the regularity on the horizons in Appendix \ref{horizon_regularity}.
As a result, for a regular $\ae$-black hole solution, there remains
 only one free parameter $T_1$, which is associated with the 
black hole mass, just as is the case of the Schwarzschild solution in general 
relativity.

\newpage
%%%%%%%%%%%%%%%%%%%%%%%%%%%%%%%%%%%%%%%%%%%%%%%%%%%
\subsection{black hole solutions with Lifshitz scaling 
{\rm :} \\ the case of $\beta_1 \neq 0$ and $g_2=\beta_2=0$}
\label{a4_BH}
%%%%%%%%%%%%%%%%%%%%%%%%%%%%%%%%%%%%%%%%%%%%%%%%%%%
When the higher-order aether correction $\dot{u}^4$ 
is taken into account, i.e. $\beta_1 \neq 0$  
the solution turns to depend on  
$a_2$ as well as $T_1$ unlike \ae-black holes.
Therefore one may consider the ultraviolet correction $\dot{u}^4$ 
does cure the singular behavior
 on the scalar-graviton horizon appeared for the infrared-limit theory.
In fact, if we assume only $\dot{u}^4$ term ($g_2=\beta_2=0$), we find that 
there is no singular behavior on any horizon in general.
The detailed discussion is developed in Appendix. \ref{horizon_regularity}.

\begin{figure}[h]
\begin{center}
\includegraphics[width=70mm]{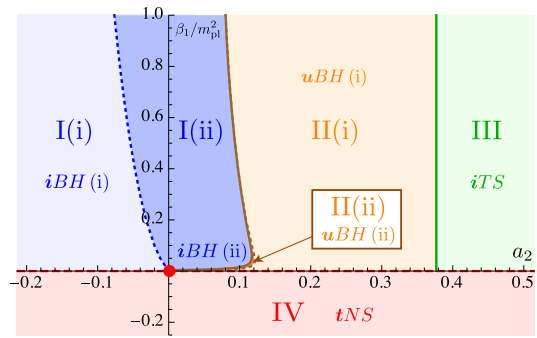} \\ (a) 
The phase diagram of the solutions \\~\\
\includegraphics[width=70mm]{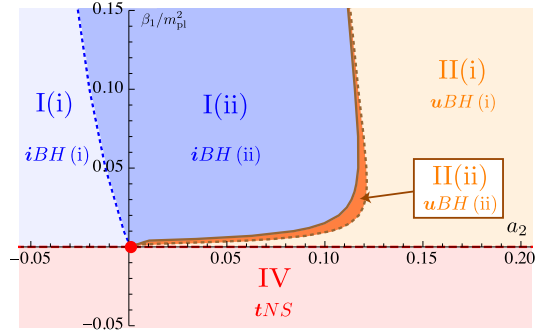} \\ (b) The enlarged phase 
diagram near the \ae-black hole
\end{center}
\caption{
The phase diagram of the solution in the ($ a_2, \beta_1 $) 
parameter plane, where
The coupling constants are chosen as the same as Fig.\ref{sol_g2zero}. 
The red dashed line which is $\beta_1=0$ indicates the case of
 \ae-theory.
The \ae-black hole with a regular scalar-graviton horizon is shown by
the red circle.
The genuine black hole solutions which possess the universal horizon are 
discovered in the region II(i) and II(ii).
}
\label{sol_a4_region}
\end{figure}

In this subsection, we consider the case of  $\beta_1 \neq 0$ with 
$g_2=\beta_2=0$.  
There are five types of ``black hole" solutions, which are classified
 in two dimensional parameter space, namely, in $(a_2,\beta_1)$ plane.
Note that we use the unit of $T_1=-1$.
We give a classification of these solutions 
in TABLE \ref{sol_a4_table}, which phase diagram of these solutions 
is shown in Fig. \ref{sol_a4_region}. 

\begin{table}[h]
\begin{center}
\begin{tabular}{lc|c|c|c|cccc}
\hline \hline
& {\it solution} & {\it region} & {\it horizons} & {\it singularity} &\it{BH}\\
\hline \hline
&$\bm{iBH}$\,(i)& I(i)    & $r_{\mathrm{GH}}<r_{\mathrm{SH}}$ \& no $r_{\rm UH}$& $r=0$ & $\Delta$ \\
&$\bm{iBH}$\,(ii)& I(ii)      &$r_{\mathrm{GH}}<r_{\mathrm{SH}}$ \& no $r_{\rm UH}$ 
& $0<r<r_{\mathrm{GH}}$ & $\Delta$ \\
&$\bm{uBH}$\,(i)& II(i) &$r_{\rm UH}<r_{\mathrm{GH}}<r_{\mathrm{SH}}$& $r=0$ & $\bigcirc$  \\
&$\bm{uBH}$\,(ii)& II(ii) &$r_{\rm UH}<r_{\mathrm{GH}}<r_{\mathrm{SH}}$& $0<r<r_{\rm UH}$ & $\bigcirc$ \\
&$\bm{iTS}$ & III     &no horizons& $r=r_{\mathrm{SH}}$ & $\times$\\
&$\bm{tNS}$ & IV      &no horizons& $r=0$   & $\times$ \\
\hline \hline
\end{tabular}
\caption{The classification of the solution for the case of $\beta_1 \neq 0$ 
with $\beta_2=g_2=0$.
The region I-IV correspond to the areas shown in Fig. \ref{sol_a4_region}.
$r_{\mathrm{SH}}, r_{\mathrm{GH}}$ and $r_{\rm UH}$ are the positions of the scalar-graviton horizon, 
the graviton horizon, 
and the universal horizon, respectively.
$\Delta$ means that the solution describes a black hole for 
gravitons and scalar-gravitons, but may becomes naked  
for high-energetic Lifshitz scaling test particles with $z>1$.
}
\label{sol_a4_table}
\end{center}
\end{table}

\begin{figure}[h]
\begin{center}
\includegraphics[width=70mm]{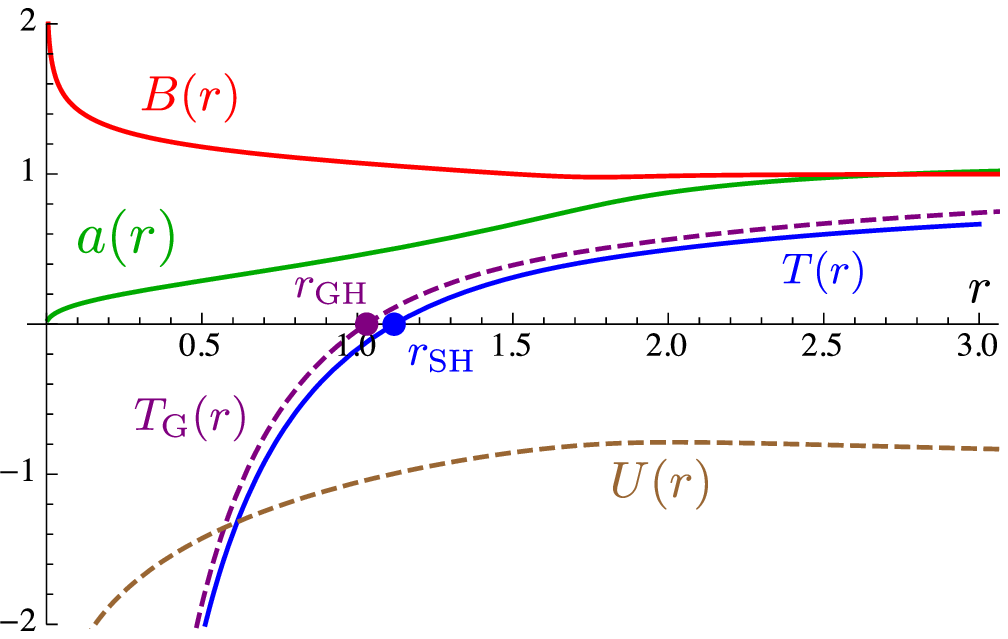} \\ (a) 
The evolution of the each components of the metric and aether.\\~\\
\includegraphics[height=50mm]{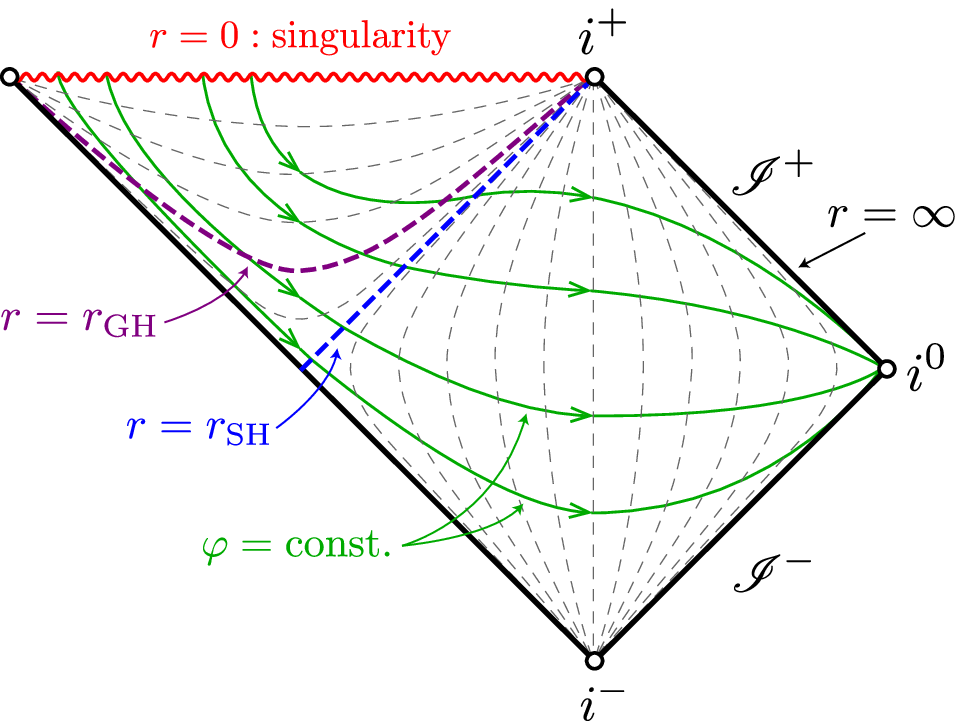} \\ (b) 
The Carter-Penrose diagram 
\end{center}
\caption{
The typical example of $\bm{iBH}$\,(i) (region I(i)). 
In the top figure (a), We choose
$(a_2, \beta_1/\mpl^2)=(-1.000,1.000)$.
The remaining coupling constants and the boundary conditions are set to
 the same values as those of Fig. \ref{sol_g2zero}.
The blue, red and green curves indicate the functions $T(r), B(r)$ and 
$a(r)$, respectively.
The dashed purple and dashed brown curves indicate  $T_{\rm G}$
and $U(r):=u \cdot \xi$.
The graviton horizon is found at $r_{\mathrm{GH}} =1.030$, while 
the scalar-graviton horizon exists at $r_{\mathrm{SH}} = 1.120$.
Since $U(r)$ does not vanish, 
there is no universal horizon, namely, no causal boundary.
In the bottom figure (b), 
the dashed curves indicates $r=$constant surface, 
especially, the blue, purple and brown curves represent the scalar-graviton 
horizons, the graviton horizons and the universal horizons, respectively.
The spacetime singularities are represented by the waving curves.
The future (past) null infinity, the future (past) timelike infinity and 
the spacelike infinity are indicated by $\mathscr{I}^+$ $(\mathscr{I}^-)$,
 $i^+$ $(i^-)$ and $i^0$, respectively. 
The ultimately excited particle with $z > 1$ Lifshitz scaling propagates 
along the $\varphi=$constant surface, which is indicated by the solid green
 curves. 
}
\label{sol_I_a4_1}
\end{figure}

We explain each solution in due order:\\[.5em]
(1) $\bm{iBH}$\,(i)
[{\it an infrared black hole with a central singularity}]:
If $\beta_1$ and $a_2$ are set to be in the light blue colored region 
in Fig. \ref{sol_a4_region} (region I(i)), we find a kind of 
black hole which possesses the graviton horizons. 
The typical numerical example is shown in Fig. \ref{sol_I_a4_1}(a).
There are graviton and scalar-graviton horizons. 
In the present case, the dispersion relations of graviton and scalar-graviton
 are given by $\omega^2 \sim k^2$,
which means these horizons coincide with \ae-theory's ones. 
Hence the gravitons and scalar-gravitons cannot escape 
from the inside of these horizons.
For low-energy particles, they also play a role of horizon too.
In this sense, one may regard this solution  as a type of black hole,
which we call an {\it infrared black hole} ($\bm{iBH}$).

The spacetime and aether field are regular except at the center.
However, since there exists no universal horizon,
 this solution has no causally disconnected region. 
The universal horizon turns to be genuine causal boundary due to 
the Lifshitz scaling with $z \neq 1$.
Therefore the singularity at the center is exposed if 
non-gravitational propagating modes with the $z>1$ Lifshitz scaling
 are taken into account.
Thus we conclude that this solution does not describes 
a true black hole in the strict sense but a type of naked singularity
even if the graviton horizons exist.

To clarify this situation, we shall depict the spacetime structure.
Note that the Carter-Penrose diagram itself does not describe the causal structure of the solution
due to the lack of Lorentz invariance.
However, since it would be useful to understand the spacetime structure, 
we will show it for this solution.
In Fig. \ref{sol_I_a4_1}(b), we illustrate the Carter-Penrose diagram for this solution,
in which null rays propagate on $\pm 45 \circ$ direction.
The metric (and scalar-graviton) horizon, which is one of the horizons 
in the Carter-Penrose diagram, 
is a horizon for the $z=1$ Lorentz invariant particles or for the low-energetic infrared particles. 
For the $z>1$ Lifshitz scaling high-energetic particles, however, it is no longer horizon, 
but the spacelike universal horizon will take its place.
\\ \\
(2) $\bm{iBH}$\,(ii)
[{\it an infrared black hole with a singular spherical shell} ]:
 This solution can be found in the deep blue colored region in 
Fig.\ref{sol_a4_region} (region I(ii)).
Although graviton and scalar-graviton horizons are formed, a singularity 
appears at $r=r_{\rm ss}>0$.
As mentioned in (1), this singularity is not causally disconnected 
from infinity due to the absence of the universal horizon.
Therefore, although this solution 
behaves as a black hole for gravitons and scalar-gravitons 
as well as Lorentz invariant $z=1$ particles, 
it turns to be a naked singularity 
for high-energetic particles with the $z>1$ Lifshitz scaling. 
In this sense, we also classify this solution as $\bm{iBH}$.

The difference from the case (1) is that 
the singularity shapes a spherical shell rather than a spacetime point with 
infinitesimal volume in $\bm{iBH}$\,(i).
In this paper, we shall refer it as a {\it singular shell}.

\begin{figure}[h]
\begin{center}
\includegraphics[width=70mm]{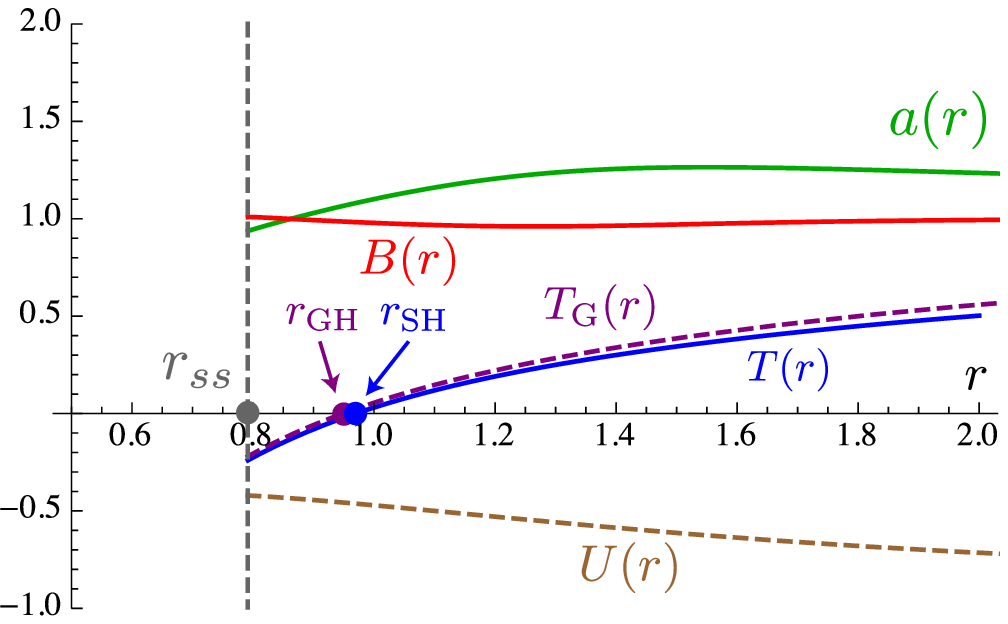} \\ (a) The evolution of the each components of the metric and aether.\\~\\
\includegraphics[height=50mm]{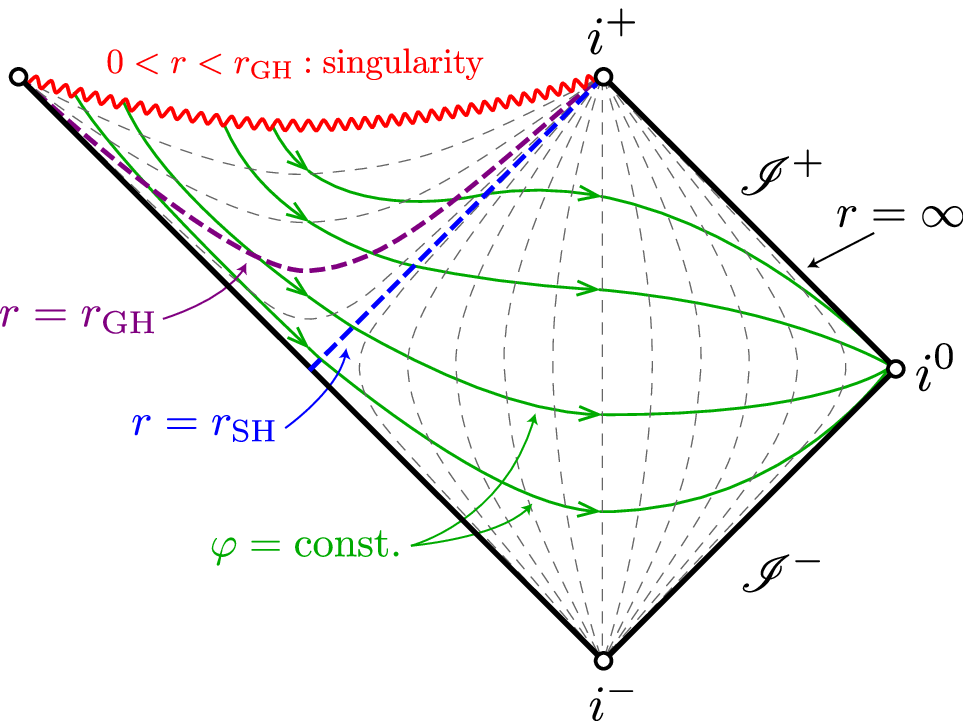} \\ (b) The Carter-Penrose diagram 
 \end{center}
\caption{
The typical example of $\bm{iBH}$\,(ii) (region I(ii)), 
In the top figure(a), We choose, 
$(a_2, \beta_1/\mpl^2)=(0,1.000)$.
The remaining coupling constants and the boundary conditions are set to 
the same values as those of Fig. \ref{sol_g2zero}.
The blue, red and green curves indicate the functions $T(r), B(r)$ and 
$a(r)$, respectively.
The dashed purple and dashed brown curves indicate the $T_{\rm G}$
and $U(r)$.
The graviton horizon is found at $r_{\mathrm{GH}} =0.949$, while 
the metric horizon exists at $r_{\mathrm{SH}} = 0.970$.
The singularity appears at $r_{\mathrm{ss}}=0.792$, which gives the radius 
of the singular shell.
In the bottom figure(b), the conformal structure is depicted.
The meaning of the curves and symbols in this figure are same as those of
FIG. \ref{sol_I_a4_1}(b).
}
\label{sol_I_a4_2}
\end{figure}

To see the cause of this singularity, 
we shall focus on the structure of the evolution equation.
We find that the evolution equation of $B(r)$
which is a linear-order differential equation with respect to $r$ (see Appendix \ref{horizon_regularity}) 
turns to be singular at $r=r_{\rm ss}$.
More specifically, the divergence of $B'(r)$ results 
in this type of singularity. 
The typical numerical example and the Carter-Penrose diagram are shown in Fig. \ref{sol_I_a4_2}(a) 
and (b), respectively.
\\ \\
(3) $\bm{uBH}$\,(i)
[{\it an ultimate black hole  with a central singularity} ]:
The black hole solution which possesses the graviton, scalar-graviton and 
universal horizon and no singularity except the center is found in the light
 orange colored region in Fig. \ref{sol_a4_region} (region II(i)).
Since the central singularity is hidden by the universal horizon, this type 
of the solution is a real black hole.
Any particles with the $z>1$ Lifshitz scaling as well as
the Lorentz invariant $z=1$ particles cannot escape from 
the inside of the universal horizon. 
We then call it an ultimate black hole ($\bm{uBH}$).

\begin{figure}[h]
\begin{center}
\includegraphics[width=70mm]{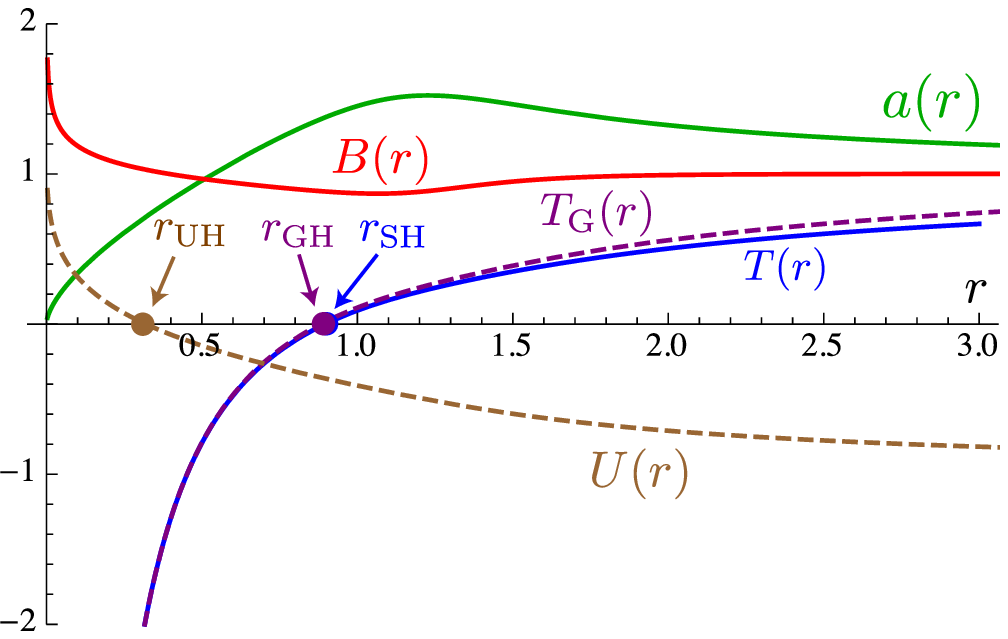} \\ 
(a) The evolution of the each components of the metric and aether.\\~\\
\includegraphics[height=50mm]{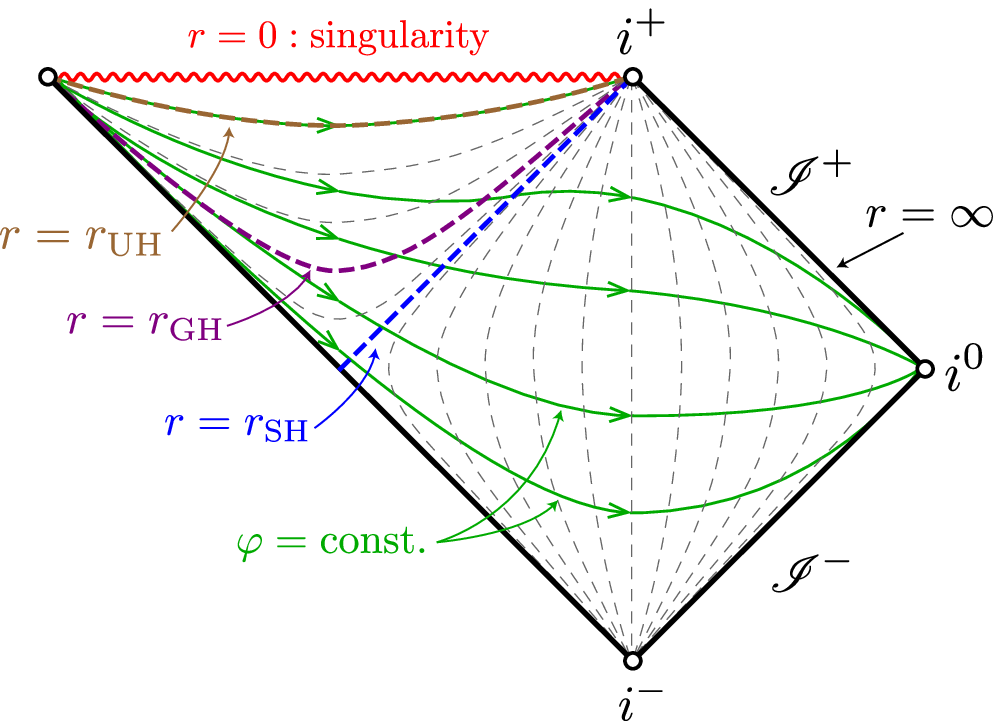} \\ 
(b) The Carter-Penrose diagram 
\end{center}
\caption{
The typical example of $\bm{uBH}$\,(i) (region II(i)).
In the top figure(a), 
we choose $(a_2, \beta_1/\mpl^2)=(0.200,1.000)$.
The remaining coupling constants and the boundary conditions are set 
to the same values as those of Fig. \ref{sol_g2zero}.
The blue, red and green curves indicate the functions $T(r), B(r)$ and 
$a(r)$, respectively.
The dashed purple and dashed brown curves indicate $T_{\rm G}$
and $U(r)$.
The graviton and the scalar-graviton horizons are found at $r_{\mathrm{GH}} =0.882$ 
and $r_{\mathrm{SH}}=0.896$, while 
the universal horizon exists at $r_{\mathrm{UH}} = 0.320$.
Since there are a universal horizon and a central singularity, 
this solution is referred to a black hole solution.
In the bottom figure(b), the conformal structure is depicted.
The meaning of the curves and symbols in this figure are same as those of
FIG. \ref{sol_I_a4_1}(b).
}
\label{sol_I_a4_3b}
\end{figure}

We shall show the typical example in Fig. \ref{sol_I_a4_3b}(a).
This solution is much similar to the \ae-black hole illustrated 
in Fig. \ref{sol_g2zero} 
except the oscillatory behavior of $U(r)$ near the central singularity.
Moreover, we also show the Carter-Penrose diagram of this solution in FIG. \ref{sol_I_a4_3b}(b).
\\ \\
(4) $\bm{uBH}$\,(ii) 
[{\it an ultimate black hole with a singular spherical shell} ]:
When $a_2$ and $\beta_1$ are in deep orange colored region in 
Fig.\ref{sol_a4_region} (region II(ii)), we find the solution 
with graviton, scalar-graviton and universal horizon, however, there are
a singular shell inside the universal horizon.
Since the singular shell is covered by the universal horizon, any information
 from the singularity never be leaked into the outside.
Thus we can regard this type of the solution as a real black hole.
Then we also classify this solution as an ultimate black hole ($\bm{uBH}$).

\begin{figure}[h]
\begin{center}
\includegraphics[width=70mm]{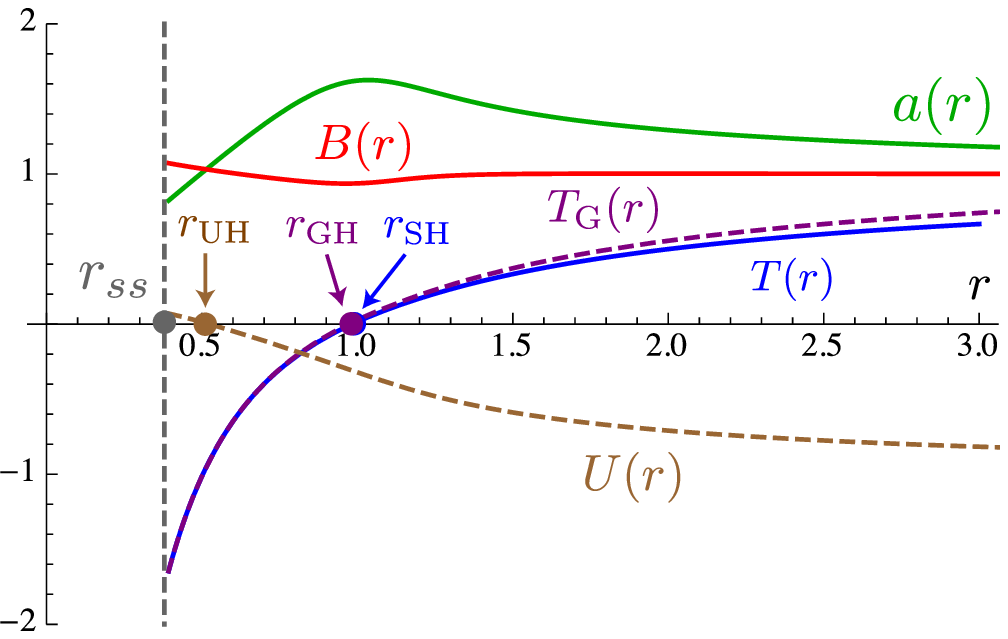} \\ 
(a) The evolution of the each components of the metric and aether.\\~\\
\includegraphics[height=50mm]{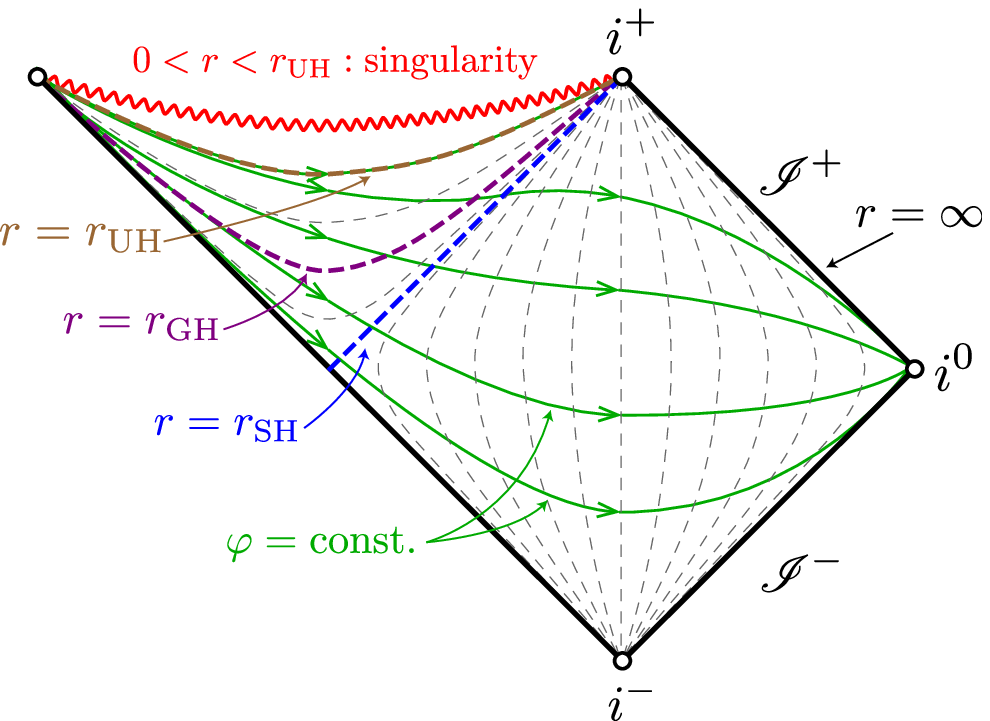} \\ 
(b) The Carter-Penrose diagram 
\end{center}
\caption{
The typical example of $\bm{uBH}$\,(ii) (region II(ii)). 
In the top figure(a),
we choose
$(a_2,\beta_1/\mpl^2)=(0.117,1.000)$.
The remaining coupling constants and the boundary conditions are set to 
the same values as those of Fig. \ref{sol_g2zero}.
The blue, red and green curves indicate the functions $T(r), B(r)$ and $a(r)$, 
respectively.
The dashed purple and dashed brown curves indicate $T_{\rm G}$ and $U(r)$.
The graviton and scalar-graviton horizons are found at $r_{\mathrm{GH}} =0.972$ and 
$r_{\mathrm{SH}}=0.983$, while the universal horizon exists at $r_{\mathrm{UH}} = 0.524$.
The singularity appears at $r_{\rm ss}=0.392$, which gives the radius of the
 singular shell.
In the bottom figure(b), the conformal structure is depicted.
The meaning of the curves and symbols in this figure are same as those of
FIG. \ref{sol_I_a4_1}(b).
}
\label{sol_I_a4_3a}
\end{figure}

\begin{figure}[h!]
\begin{center}
\includegraphics[width=70mm]{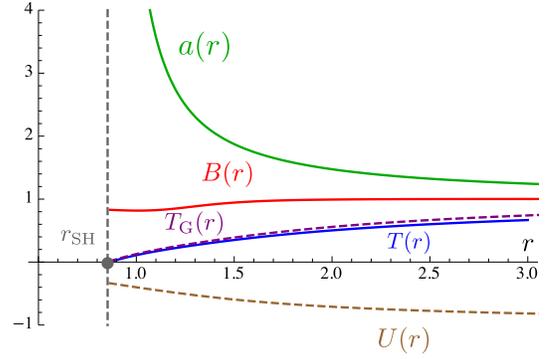} \\ 
(a) The evolution of each components of the metric and  aether.\\~\\
\includegraphics[width=70mm]{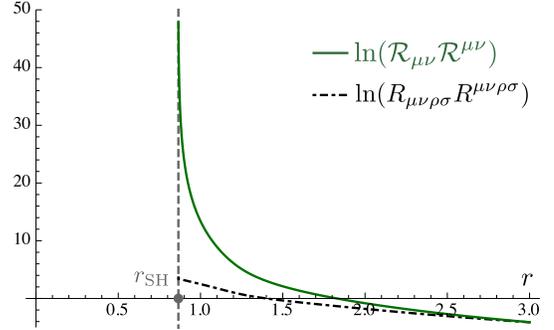} \\ 
(b) The corresponding evolutions of the curvatures.
 \\~\\
\includegraphics[height=50mm]{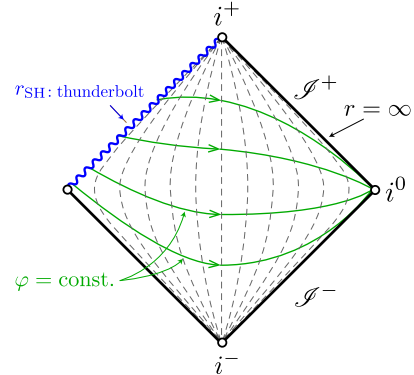} \\ 
(c) The Carter-Penrose diagram 
\end{center}
\caption{
The typical example of $\bm{iTS}$ (region III).
In the top figure(a) and the middle figure(b), 
we choose $(a_2, \beta_1/\mpl^2)=(0.500,1.000)$.
The remaining coupling constants and the boundary conditions are set to 
the same values as those of Fig. \ref{sol_g2zero}.
In the top figure (a), the blue, red and green curves indicate the functions 
$T(r), B(r)$ and $a(r)$, respectively.
The dashed purple and dashed brown curves indicate the $T_{\rm G}$ and $U(r)$.
Near the spherical shell with the radius $r_{\rm SH}=0.865$, 
the quadratic scalar of three-dimensional Ricci tensor 
$\mathcal{R}_{\mu \nu}\mathcal{R}^{\mu \nu}$ diverges rather than 
four-dimensional Kretchmann invariant 
$\mathcal{R}_{\mu \nu}\mathcal{R}^{\mu \nu}$,
which are denoted by the solid dark green 
curve and dot-dashed black curve in the middle figure(b), respectively. 
In the bottom figure(c), the conformal structure is depicted.
The meaning of the curves and symbols in this figure are same as those of
FIG. \ref{sol_I_a4_1}(b).
}
\label{sol_IV_a4}
\end{figure}

The typical example and the Carter-Penrose diagram of this solution are shown in Fig. \ref{sol_I_a4_3a}(a)
and (b), respectively.
\\ \\
(5) $\bm{iTS}$
[{\it an infrared thunderbold singularity}]:
For the solutions
 in the light green colored region in Fig. \ref{sol_a4_region} 
(region III), 
a singularity always appears at the null  Killing horizon of 
the scalar-graviton. 
On the singular shell, the function $a(r)$, namely, the $v$ component of the 
aether field diverges if we set $a_2$ to be larger value than a critical one.
Furthermore, it is found that the critical value of $a_2$ 
($\sim 0.377$) which induces the 
$a(r)$ divergence seems to be universal under any choice of $\beta_1$
as shown in Fig. \ref{sol_a4_region}).
We shall present one typical example and the corresponding Carter-Penrose diagram in Fig. \ref{sol_IV_a4}.
Near the spherical shell with the radius $r_{\rm{SH}}=0.865$, 
the quadratic scalar of three-dimensional Ricci tensor 
$\mathcal{R}_{\mu \nu}\mathcal{R}^{\mu \nu}$ diverges.
Hence it is a physical singularity.

One may wonder why such a singular behavior is occurred on the scalar-graviton
 horizon.
More specifically, from Appendix \ref{horizon_regularity}, we have known that 
the basic equations possess no dependence on the negative power of $T_{\rm S}$ 
unlike the \ae-case.
Therefore, the scalar-graviton horizon where $T_{\rm S}=0$ should be regular 
in general.
To see in detail, we shall expand the evolution equation of $a(r)$ around 
$r_{\rm SH}$. 
From our numerical analysis in Fig. \ref{sol_IV_a4}, we find 
$a^{-1}(r_{\mathrm{SH}}+\epsilon) \sim 
\delta $ and $a'(r_{\mathrm{SH}}+\epsilon) 
\sim \delta^{-2}$, where $\epsilon\ll 1$ and $\delta\ll 1$.
In Fig. \ref{sol_IV_a4}, we have check the divergence numerically 
at least for $\epsilon\gsim 1.000 \times 10^{-3}$ and $\delta 
\gsim  4.132 \times 10^{-3}$.
Near the singularity, where $T_{\rm S}(r_{\rm SH}+\epsilon)
:= \Delta_\epsilon \ll 1$, 
the dominant term in the evolution equation of $a(r)$ is given by
\begin{eqnarray}
a''(r) \sim {T_{\rm S}' (r_{\rm SH})^3  \over B(r_{\rm SH})^2 
\Delta_\epsilon}a^{5} \,.
\end{eqnarray}
Thus, we find that $a(r)$ shows singular behavior when $\Delta \to 0$,
 where the limit gives the scalar-graviton horizon.
Since this singularity is due to the aether field rather than the spacetime 
metric.

This singularity seems to be closely analogous to the 
{\it thunderbolt} singularity.
 The thunderbolt singularity 
is proposed in the context of quantum black hole evaporation.
 In \cite{Hawking_Stewart}, the thunderbolt singularity 
is firstly invented in 
semi-classical analysis of $(1+1)$-dimensional dilaton-coupled gravity 
with scalar field \cite{CGHS} which is renormalizable theory of 
quantum gravity.
 Furthermore, this type of singularity is also discovered in 
$(1+1)$-dimensional quantum field theory via complete quantized analysis
\cite{NS_TB}.

In such a situation, a null singularity appears on the event horizon. 
Although a causally disconnected region is not formed because of
 the existence of a singularity, this singularity itself is not 
detected by any outside observer. As a result, 
it is not a naked singularity. They call it a thunderbolt singularity.

In our case, the singularity of the aether field 
appears on a Killing horizon is null.
Hence it behaves similar to a thunderbolt singularity for 
Lorentz invariant $z=1$ particles. 
We call it 
an infrared thunderbolt singularity ($\bm{iTS}$).
The thunderbolt composes of the singular aether field. 
\\ \\
(6) $\bm{tNS}$\,
[{\it a timelike naked singularity without horizon}]:\\
The singular shell without any horizon is found 
in the light red colored region in Fig. \ref{sol_a4_region} (region IV), 
namely, $\beta_1<0$.
Since the singularity is timelike and 
there is no horizon, this type of the solution 
 completely exposes its singularity.
Note that that singularity is originated from the divergence of the 
evolution equation $B(r)$ as is the case in 
$\bm{iBH}$\,(ii) and $\bm{uBH}$\,(ii).

\begin{figure}[h]
\begin{center}
\includegraphics[width=70mm]{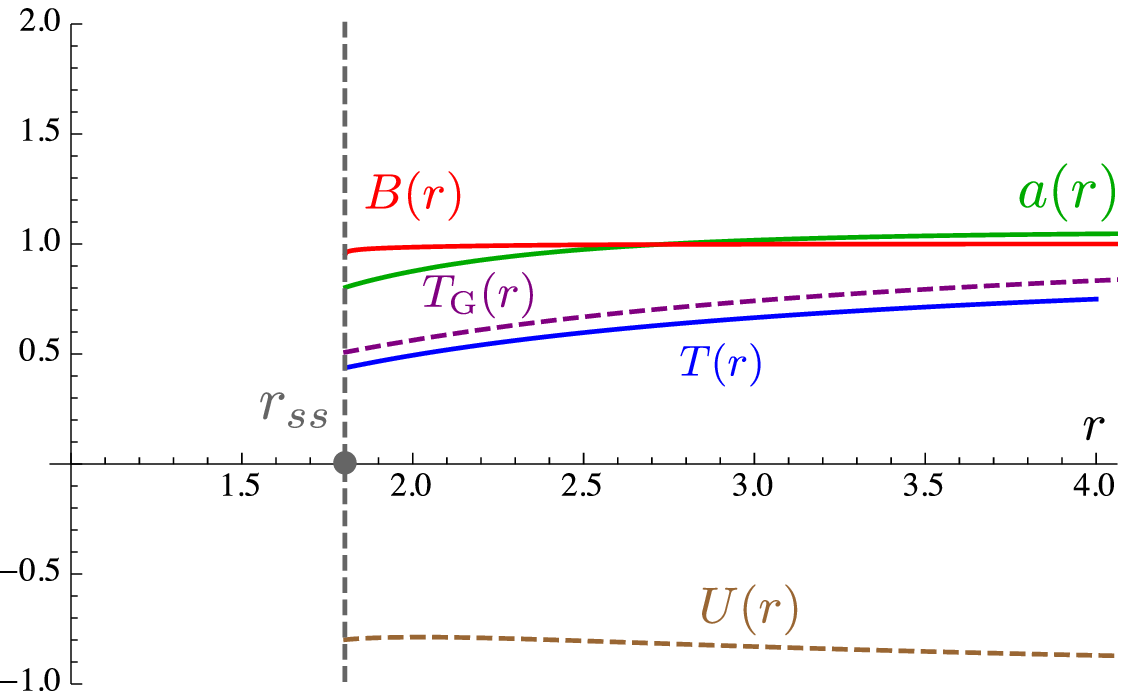} \\ (a) The evolution of each components of the metric and aether. \\~\\
\includegraphics[height=50mm]{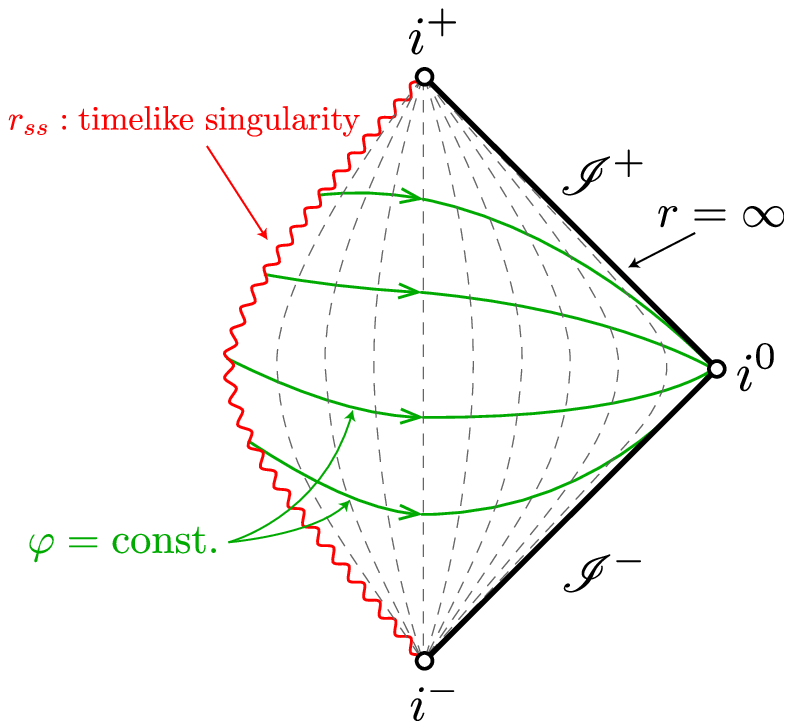} \\ (b) The Carter-Penrose diagram
\end{center}
\caption{
The typical example of $\bm{tNS}$ (region IV). 
In the top figure (a), we choose $(a_2, \beta_1/\mpl^2)=(-1.000,-1.000)$.
The remaining coupling constants and the boundary conditions are set to 
the same values as those of Fig. \ref{sol_g2zero}.
The blue, red and green curves indicate the functions 
$T(r), B(r)$ and $a(r)$, respectively.
The dashed purple and dashed brown curves indicate the $T_{\rm G}$ and $U(r)$.
Note that this solution shows singular behavior at $r_{\rm ss}=1.804$.
In the bottom figure (b), the conformal structure is depicted.
The meaning of the curves and symbols in this figure are same as those of
FIG. \ref{sol_I_a4_1}(b).
}
\label{sol_V_a4}
\end{figure}

We shall show the typical example in Fig. \ref{sol_V_a4} .

\newpage
%%%%%%%%%%%%%%%%%%%%%%%%%%%%%%%%%%%%%%%%%%%%%%%%%%%
\subsection{Ultimate thunderbolt singularity ($\bm{uTS}$)\\ 
{\rm :} the case of $g_2 \neq 0$ and $\beta_1=\beta_2 = 0$}
%%%%%%%%%%%%%%%%%%%%%%%%%%%%%%%%%%%%%%%%%%%%%%%%%%%
Next, we consider only
 the spatial higher curvature correction term $\mathcal{R}^2$ 
to take into account 
the back reaction effect of the Lifshitz scaling in the high energy limit.
That is the case of $g_2 \neq 0$ and $\beta_1 = \beta_2 = 0$.
From the discussion in section \ref{BH_horizon}, the graviton horizon 
does not change and its position is  
still at $r_{\mathrm{GH}}$ where $T_{\rm G}(r_{\mathrm{GH}})=0$. 
On the other hand, the scalar-graviton horizon is shifted 
from the Killing horizon to the universal
 horizon because of  $k^4$ term 
in the sound speed (\ref{mod_sclar_grav_speed}).

\begin{figure}[h!]
\begin{center}
\includegraphics[width=70mm]{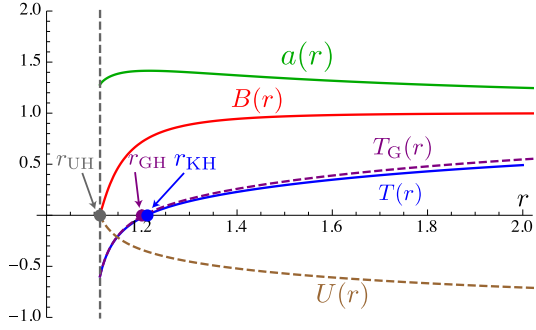} \\ 
(a) The evolution of each components of metric and aether.\\~\\
\includegraphics[width=70mm]{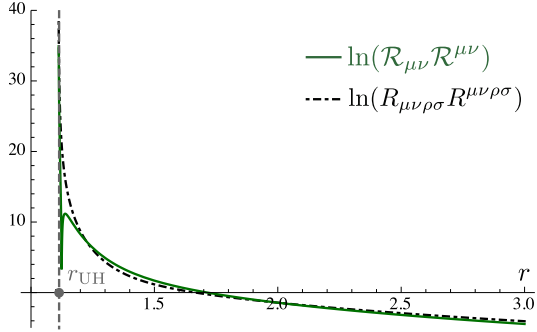}  \\ 
(b) The corresponding evolution of the curvatures.\\~\\
\includegraphics[height=50mm]{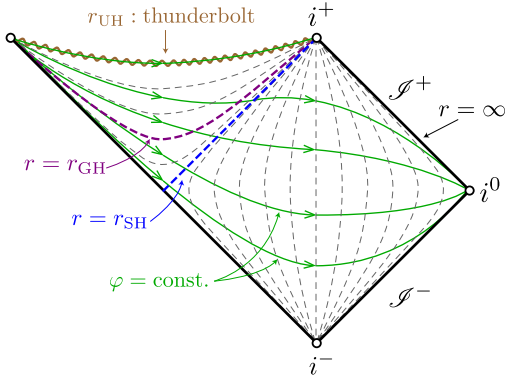} \\ 
(c) The Carter-Penrose diagram
\end{center}
\caption{The spherically symmetric solution with higher spatial curvature 
correction. 
The setting of the coupling constants and the boundary conditions 
are the same as those of Fig. \ref{sol_g2zero} except $g_2/\mpl^2 =-1.000$.
In the top figure (a), 
the blue, red and green curves indicate the functions $T(r), B(r)$ and 
$a(r)$, respectively.
The dashed purple and dashed brown curves indicate the $T_{\rm G}$
and $U(r)$.
The graviton horizon is found at $r_{\mathrm{GH}} =1.248$, while 
the metric horizon exists at $r_{\rm KH} = 1.250$.
The calculation has been broken down at $r_{\mathrm{UH}} = 1.157$ 
where $U(r)$ approaches to zero.
In the middle figure (b), the dotted gray and dotted black curves indicate 
$\ln (R_{\mu \nu \rho \sigma} 
R^{\mu \nu \rho \sigma})$ and $\ln (\mathcal{R}_{\mu \nu} 
\mathcal{R}^{\mu \nu})$, respectively. 
In the bottom figure (c), the conformal structure is depicted.
The meaning of the curves and symbols in this figure are same as those of
FIG. \ref{sol_I_a4_1}(b).
}
\label{sol_g2_1}
\end{figure}

We first assume $g_2<0$. 
In Fig. \ref{sol_g2_1}(a), one numerical solution is shown for $g_2=-1$.
The setting of the coupling constants and the boundary conditions are 
the same as those of Fig. \ref{sol_g2zero} except $g_2 =-1$.
Notably, we find that 
a singular behavior on the metric horizon found 
 for untuned arbitrary value of $a_2$ in the \ae-gravity theory vanishes.
The metric horizon turns to be regular for any value of $a_2$.
Instead, a singular behavior is found inside the metric horizon.
The $(v,r)$ component of the metric, $B(r)$  vanishes there. 
The aether field $u^\mu$ aligns perpendicular to the timelike Killing vector
 $\xi^\mu$ near the singular point, which corresponds to 
the universal horizon with $u\cdot\xi=0$.

We calculate the four-dimensional Kretchmann invariant $R_{\mu \nu \rho
 \sigma} R^{\mu \nu \rho \sigma}$ and the quadratic scalar 
of the three-dimensional  Ricci tensor 
$\mathcal{R}_{\mu \nu} \mathcal{R}^{\mu \nu}$, which are shown in Fig. 
\ref{sol_g2_1}(b).
Near the singular point, those two scalar functions diverge. 
Thus, this singular point is a physical singularity rather than a coordinate 
singularity. 
So we find that the universal horizon becomes singular.
Additionally, the structure of this spacetime is depicted in FIG. \ref{sol_g2_1}(c).

\begin{figure}[hbt]
\begin{center}
\includegraphics[width=75mm]{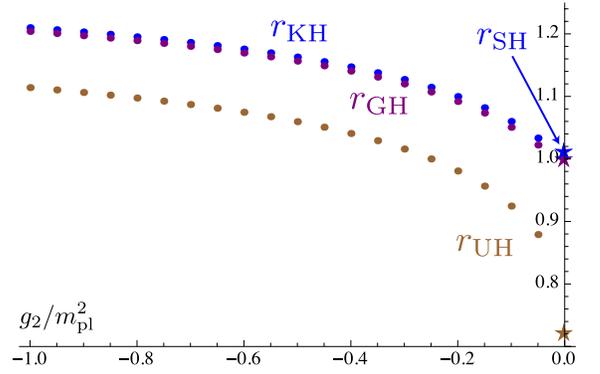} 
\end{center}
\caption{The relation between several horizon radii and $g_2$. 
The coupling constants and the boundary conditions are the same as  Figs.  
\ref{sol_g2zero} and \ref{sol_g2_1} except $g_{2}$.
The purple, blue and brown star marks indicate the graviton horizon radius 
$r_{\mathrm{GH}}$, the scalar-graviton horizon radius $r_{\mathrm{SH}}$ and the singular universal horizon 
radius $r_{\rm UH}$ in the \ae-gravity theory, respectively. 
The purple, blue and brown round circles
 indicate the graviton horizon radius $r_{\mathrm{GH}}$,
 the metric horizon radius $r_{\rm KH}$, and 
 the radius of the singular universal horizon 
$r_{\rm UH}$.
Numerically the position of $r_{\rm UH}$ is
 evaluated at the point of $U(r)= -0.025$, because 
$U(r) =0$ is singular.
}
\label{Hradius_g2}
\end{figure}

What causes this type of singularity ? 
In order to clarify this question, 
we first show the relation between $g_2$ 
and the radii of the graviton horizon $r_{\mathrm{GH}}$, the metric horizon $r_{\rm KH}$  
and the singular universal horizon 
 $r_{\rm UH}$ in Fig. \ref{Hradius_g2}. 
We also give 
the detailed data of the singular universal horizon radius $r_{\rm UH}$ near $g_2 =0$ 
in Table \ref{ru_g2_table}.
\begin{table}[hbt]
\begin{center}
\begin{tabular}{ccccc}
\hline \hline
& $g_{2}/\mpl^{2}$ & $r_{\rm UH}$ & $g_{2}/\mpl^{2}$ & 
$r_{\rm UH}$ \\
\hline \hline
& 0                                & 0.7200  &$-5.00 \times 10^{-4}$
& 0.7944 \\
&$-1.00 \times 10^{-4}$& 0.7783  &$-6.00 \times 10^{-4}$& 0.7962  \\
&$-2.00 \times 10^{-4}$& 0.7854  &$-7.00 \times 10^{-4}$& 0.7977  \\
&$-3.00 \times 10^{-4}$& 0.7893  &$-8.00 \times 10^{-4}$& 0.7990  \\
&$-4.00 \times 10^{-4}$& 0.7922  &$-9.00 \times 10^{-4}$& 0.8002  \\
\hline \hline
\end{tabular}
\caption{The detailed value of $r_{\mathrm{UH}}$ for near $g_2 =0$. 
The coupling constants and the boundary conditions are same as Fig. 
\ref{Hradius_g2} except $g_2$.
The values of $r_{\mathrm{UH}}$ are evaluated at the point where 
$U(r) = -0.025$.
}
\label{ru_g2_table}
\end{center}
\end{table}
From these results, we find that the universal horizon radii
change smoothly from $g_2 =0$ case to $g_2 \neq 0$ cases.

One may wonder whether it can be regular if we tune the 
free parameter $a_2$ just as the scalar-graviton horizon 
in the \ae-gravity theory.
To see this, we shall perform the expansion of the basic equations around 
the universal horizon.
Focusing on the coefficients of the most highest $r$ derivative terms, 
$T'''(r)$, $B'''(r)$ and $a'''(r)$ terms in the $(\theta, \theta)$ component 
of the Einstein equation, we find all of them have been vanished at 
the universal  horizon (see Appendix.\ref{horizon_regularity}).
This result does not depend on the value of $a_2$.
This fact means that the singularity on the universal horizon 
cannot be remedied by tuning the free parameter $a_2$ unlike the \ae-case.
While there is no singular behavior on the scalar-graviton horizon 
$r_{\mathrm{SH}}$ in the infrared limit where $T_{\rm S}$ vanishes.

Hence it is not quite unnatural to consider that  the singular behavior 
which appears on the scalar-graviton radius with $T_{\rm S}=0$
in the infrared limit is shifted to the universal
horizon when we include the higher curvature term from the Lifshitz scaling.
Namely, the dispersion relation of the scalar-graviton with nonzero $g_2$ 
gives the infinite sound speed of the scalar-graviton.  
The similar situation is 
found for the exact solution with $c_{14}=0$ in the
\ae-gravity theory \cite{mec_UH}, for which 
 the sound speed
of  the scalar-graviton 
 (\ref{speed_graviton_IR})
becomes infinite and 
the scalar-graviton horizon coincides with the universal horizon. 
In this case, however, 
this singularity can be removed by choosing an appropriate value of $a_2$.

We shall turn our attention to the physical interpretation of this solution.
 Recall that the universal horizon is defined by surface where 
$u \cdot \xi =0$, i.e., a static limit for the ultimately excited dispersive 
particle with $z \neq 1$ Lifshitz scaling.
 In other words, only the particle which possesses infinite energy can stay 
on the surface, and any future-directed signal on the surface cannot 
goes outward even if the particle is spacelike with infinite energy.
 Namely, the information on the universal horizon must not be leaked out to 
outside of the horizon.
  Although this solution cannot be regarded as a black hole solution 
whose spacetime singularity is isolated by an event horizon, the cosmic 
censorship hypothesis is not violated on this account.

This singularity is very similar to the 
 thunderbolt singularity if we replace a null event horizon 
with a spacelike universal horizon.
The universal horizon is a real horizon for the 
$z>1$ Lifshitz scaling particles. 
So the singular universal horizon cannot be detected by  any 
outside observers. Although there is no causally disconnected 
region, it is not a naked singularity. 
We then call it a ultimate thunderbolt singularity ($\bm{uTS}$). 
We may speculate that the appearance of the 
thunderbolt singularity on the universal horizon  
via $\mathcal{R}^2$ term indicates
 quantum gravitational loop correction in the  
Lorentz violating system as in quantization of 
the Lorentz invariant system with a thunderbolt singularity.

\begin{figure}[h]
\begin{center}
\includegraphics[width=70mm]{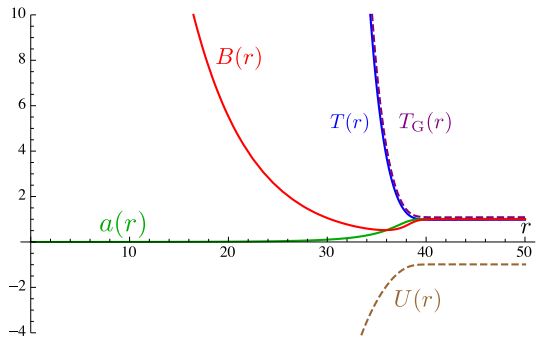} 
\end{center}
\caption{
The typical example of the solution with positive $g_2$. We set
 $a_2=1.123 \times 10^{-3}$ and $g_2/\mpl^2=1.000$.
The remaining coupling constants and the boundary conditions are set to 
the same values as those of Fig. \ref{sol_g2_1}.
The blue, red and green curves indicate the functions $T(r), B(r)$ and $a(r)$,
 respectively.
The dashed purple and dashed brown curves indicate the $T_{\rm G}$ and $U(r)$.
The function $a(r)$ drops to zero, while $T(r)$ and $B(r)$ diverges for smaller
radius $r$.
}
\label{sol_g2_minus}
\end{figure}

Besides, one may wonder about solutions for the positive value of $g_2$.
In fact, this case is less interesting.
Namely, all black hole horizons which exist in the case of $g_2=0$  
completely disappear. 
The property of the solution is quite unphysical, i.e., the functions $T(r)$
 and $B(r)$ show  positive divergence at smaller radius
 $r \sim \mathcal{O}(10G_{\rm N} M)$,
 while the function $a(r)$ drops to zero without forming any horizon.
In other words, there does not exist any type of black holes discussed before
(See Fig. \ref{sol_g2_minus}).

%%%%%%%%%%%%%%%%%%%%%%%%%%%%%%%%%%%%%%%%%%%%%%%%%%%
\subsection{Solutions with $z=2$ Lifshitz scaling terms\\ 
{\rm :} the case of $g_2 \neq 0$ and $\beta_1,\beta_2 \neq 0$}
%%%%%%%%%%%%%%%%%%%%%%%%%%%%%%%%%%%%%%%%%%%%%%%%%%%
%%%%%%%%%%%%%%%%%%%%%%%%%%%%%%%%%%%%%%%%%%%%%%%%%%%
If we have only  $\dot u^4$ term, 
we find $\bm{uBH}$ for an appropriate value of 
$a_2$. The universal horizon is not singular. 
On the other hand, when we have only $\mathcal{R}^2$ term, 
the universal horizon becomes singular, giving a thunderbolt singularity
 $\bm{uTS}$. 
One may wonder what happens 
if both $z=2$ Lifshitz scaling terms, $\dot u^4$ and $\mathcal{R}^2$,
exist ($g_2\neq 0$ and $\beta_1\neq 0$).
Since $\mathcal{R}^2$ and $\dot{u}^2 \mathcal{R}$ give the highest derivative 
terms in the equation of motion, namely, $T''(r)$, $B''(r)$ and $a'''(r)$,  
the $\bm{uTS}$ spacetime with $|\beta_1| \ll |g_2| $ is not so different from
the original one.
On the other hand, if $|\beta_1|$ is not so small compared with
 $|g_2|$, the spacetime tends to generate
a singularity caused by the aether field.
More specifically, a singular spherical shell appears before forming the
 thunderbolt singularity on the universal horizon.
Thus, we can only find $\bm{iBH}$\,(ii) and $\bm{tNS}$ spacetime in this 
situation.

Finally, we mention the case of $\beta_2 \neq 0$.
Although we have shown that there must not exist a regular 
universal horizon in this case (see Appendix \ref{horizon_regularity}), 
any horizons cannot be found as far as our numerical analysis.
In other words, the spacetime produces a singular spherical shell 
before forming any horizon, namely, $\bm{tNS}$ solution.

%======================================%
%<<<<<<<<<<<< SECTION V >>>>>>>>>>>>>>%
%======================================%
%%%%%%%%%%%%%%%%%%%%%%%%%%%%%%%%%%%%%%%%%%%
%%%%%%%%%%%%%%%%%%%%%%%%%%%%%%%%%%%%%%%%%%%
\section{Properties of Solutions} 
\label{properties}
%%%%%%%%%%%%%%%%%%%%%%%%%%%%%%%%%%%%%%%%%%%
We shall discuss the properties of obtained 
black hole  and thunderbolt singularity solutions
from several view points.

%%%%%%%%%%%%%%%%%%%%%%%%%%%%%%%%%%%%%%%%%%%%%%%%%%%
\subsection{Distribution of the aether field} 
\label{Aether Distribution}
%%%%%%%%%%%%%%%%%%%%%%%%%%%%%%%%%%%%%%%%%%%%%%%%%%%
In our solutions, there are two free parameters, $M$ and $a_2$.
The mass is a conserved quantity, which characterizes 
the solution. Althogh the different value of $a_2$ gives 
the different solution, $a_2$ may not correspond to 
any conserved  quantity. In order to understand the 
physical meaning of $a_2$, we consider the energy 
density distribution of the aether field.

We define the effective energy density and pressures 
of the aether field by
\begin{eqnarray}
\rho_{\rm \ae}&:=&T^{\rm [\ae]}_{\mu\nu}u^\mu u^\nu
\,,
\nonumber \\
P^{[r]}_{\rm \ae}&:=&T^{\rm [\ae]}_{\mu\nu}s^\mu s^\nu
\,,
\nonumber \\
P^{[\perp]}_{\rm \ae}&:=&{1 \over 2}
 T^{\rm [\ae]}_{\mu\nu}(g^{\mu\nu}+u^\mu u^\nu
-s^\mu s^\nu)
\,.
\end{eqnarray}
When we calculate $\rho_{\rm \ae}, P^{[r]}_{\rm \ae}, P^{[\perp]}_{\rm \ae}$
 we use the Einstein equations: 
\begin{eqnarray}
8\pi G T^{\rm [\ae]}_{\mu\nu}=G_{\mu\nu}
\,.
\end{eqnarray}
It makes easy to evaluate $T^{\rm [\ae]}_{\mu\nu}$ once we obtain
 the solutions.

\begin{figure}[h]
\begin{center}
\includegraphics[width=70mm]{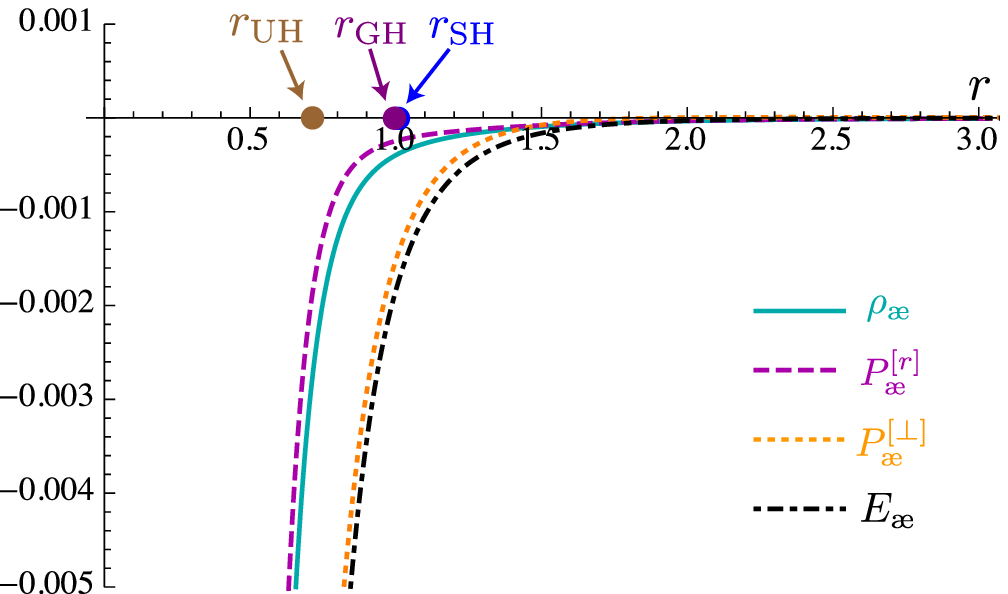}
\end{center}
\caption{
The distributions of the aether cloud for the \ae-black hole.
The solid cyan, dashed magenta, dotted yellow and dot-dashed black curve
 indicate 
$\rho_{\rm \ae}$, $P^{[r]}_{\rm \ae}$, $P^{[\perp]}_{\rm \ae}$ and 
$E_{\rm \ae}$, respectively.
The coupling constants and the boundary conditions are set to the 
same values as those of 
FIG. \ref{sol_g2zero}.
}
\label{ep_plot_ae}
\end{figure}

First we show $\rho_{\rm \ae}$, $P^{[r]}_{\rm \ae}$, 
and $P^{[\perp]}_{\rm \ae}$ for 
 the \ae-black hole in Fig. \ref{ep_plot_ae} as a reference.
In the \ae-black hole, 
all aether quantities are always negative. 
 Additionally, the following quantity $E_{\rm \ae}$ is introduced 
in order to examine the strong energy condition,
\begin{eqnarray}
E_{\rm \ae} &:=& \left( T^{\rm [\ae]}_{\mu \nu} -{1 \over 2}T^{\rm [\ae]} 
g_{\mu \nu} \right) u^\mu u^\nu \notag \\
&=& {1 \over 2} \left( 3\rho_{\rm \ae} - P^{[r]}_{\rm \ae} 
-2P^{[\perp]}_{\rm \ae} \right)  \,.
\end{eqnarray}
It is also negative definite, which means
that the strong energy condition is broken.

\begin{figure}[h!]
\begin{center}
\includegraphics[width=70mm]{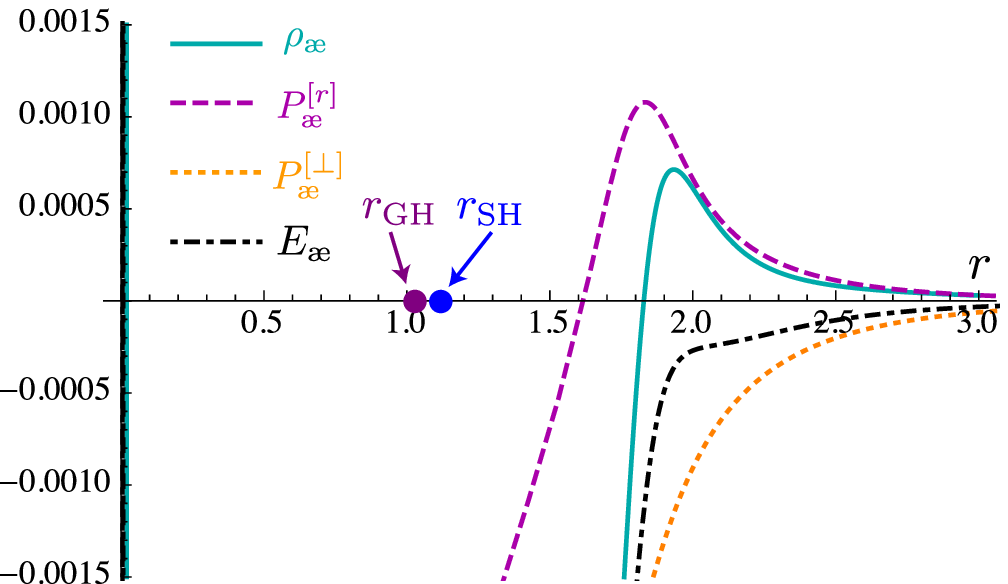} \\
(a) $\bm{iBH}$ (i)  \\~\\ 
\includegraphics[width=70mm]{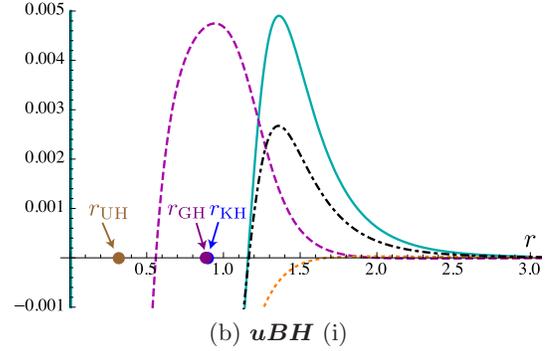} \\ 
(b) $\bm{uBH}$ (i) \\~\\
\includegraphics[width=70mm]{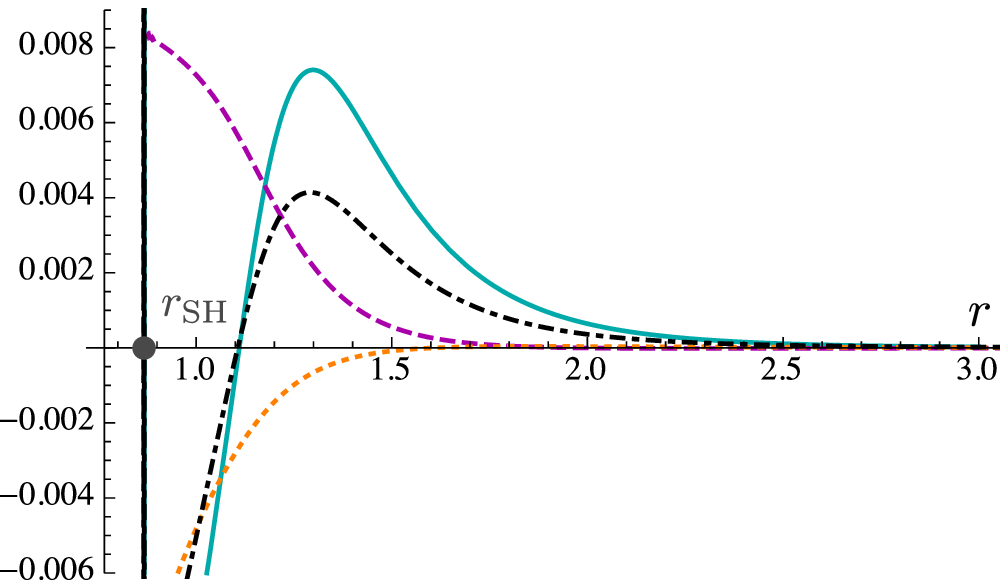} \\ 
(c) $\bm{iTS}$
\end{center}
\caption{
The distributions of the aether cloud for the solutions which are found in 
$\beta_1 \neq 0$, $\beta_2=g_2=0$.
We illustrate the typical example of ${\bm{iBH}}$ (i) (the top figure (a)), 
the ${\bm{uBH}}$ (i) (the middle figure (b))
the ${\bm{iTS}}$ (the bottom figure (c)) 
instead of showing all kind of the solutions.
The solid cyan, dashed magenta, dotted yellow and dot-dashed black curve 
indicate 
$\rho_{\rm \ae}$, $P^{[r]}_{\rm \ae}$, $P^{[\perp]}_{\rm \ae}$ and 
$E_{\rm \ae}$, respectively.
The coupling constants and the boundary conditions for (a), (b) and (c)
are set to the same values as those of 
FIG. \ref{sol_I_a4_1},
\ref{sol_I_a4_3b},
and \ref{sol_IV_a4}, respectively.
}
\label{ep_plot}
\end{figure}

\begin{figure}[h]
\begin{center}
\includegraphics[width=70mm]{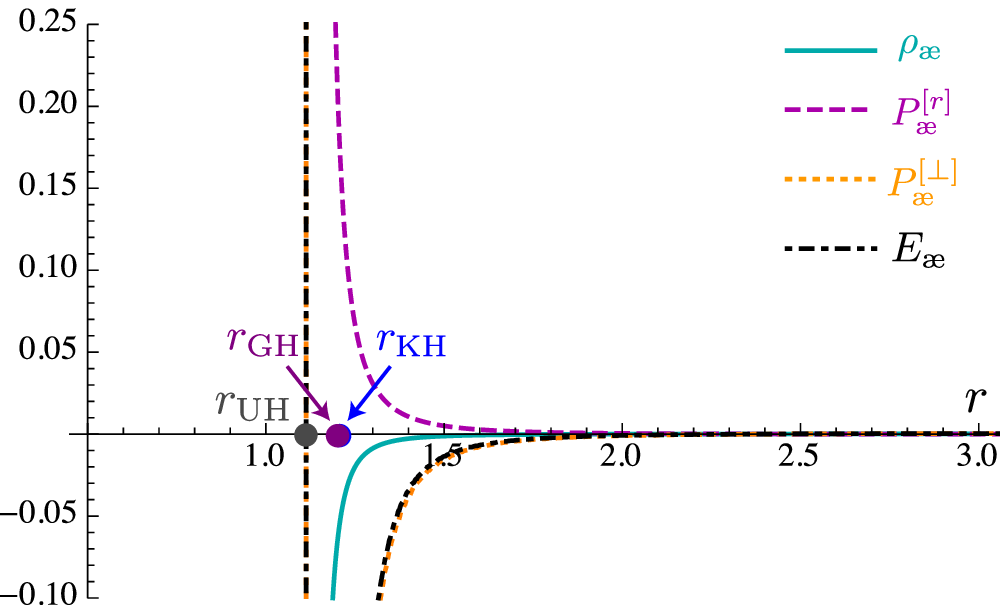}
\end{center}
\caption{
The distributions of the aether cloud for the $\bm{uTS}$ spacetime.
The solid cyan, dashed magenta, dotted yellow and dot-dashed black curve 
indicate 
$\rho_{\rm \ae}$, $P^{[r]}_{\rm \ae}$, $P^{[\perp]}_{\rm \ae}$ and 
$E_{\rm \ae}$, respectively.
The coupling constants and the boundary conditions are set to the 
same values as those of 
FIG. \ref{sol_g2_1}.
}
\label{ep_plot_uTS}
\end{figure}

We then  show 
$\rho_{\rm \ae}$, $P^{[r]}_{\rm \ae}$, $P^{[\perp]}_{\rm \ae}$ and 
$E_{\rm \ae}$ for each solution in Figs. \ref{ep_plot} and 
\ref{ep_plot_uTS}.

When we add the $z=2$ Lifshitz scaling terms, the distribution 
of the aether field is drastically changed. 
For the case of 
$\beta_1 \neq 0$, $\beta_2=g_2=0$ 
(for $\bm{iBH}$, $\bm{uBH}$ and $\bm{iTS}$ spacetime), 
the localized aether cloud is formed, i.e.,
 $\rho_{\rm \ae}$ and $P^{[r]}_{\rm \ae}$ are localized 
near the graviton and the scalar-graviton horizons.
We may speculate that the positivity of the aether 
density and radial pressure 
relax a singular behavior at the scalar-graviton horizon, 
which exists in \ae-theory with general value of $a_2$.
Note that the strong energy condition is satisfied for some finite 
radial region.
Furthermore,
 referring FIG. \ref{sol_a4_region} and the aether distribution 
FIG. \ref{ep_plot} (a), (b) and (c), 
for the larger value of $a_2$, the more dense aether cloud forms. 
As a result, 
the $\bm{iTS}$ spacetime is emerged in the large $a_2$ region 
 due to the gravitational collapse of the aether cloud.

For the case of 
$g_2 \neq 0$, $\beta_1=\beta_2=0$ 
(for $\bm{uTS}$ spacetime), 
the distribution is different. 
Referring FIG. \ref{ep_plot_uTS}, 
although the radial pressure is positive, the energy density 
becomes negative. The strong energy condition is broken in 
the whole spacetime.
Note that the radial pressure diverges where the shell singularity appears.

%%%%%%%%%%%%%%%%%%%%%%%%%%%%%%%%%%%%%%%%%%%%%%%%%%%
\subsection{Preferable Black Holes} 
\label{thermo_BH}
%%%%%%%%%%%%%%%%%%%%%%%%%%%%%%%%%%%%%%%%%%%%%%%%%%%
%%%%%%%%%%%%%%%%%%%%%%%%%%%%%%%%%%%%%%%%%%%%%%%%%%%
Although black hole thermodynamics in \ae-gravity theory has been 
discussed in the last decade, the complete understanding have not yet 
achieved.
Hence, in this section, 
we only discuss which solution is more preferable 
from the view point of the thermodynamical stability

In the previous section, 
we find two-parameter ``black hole" solutions:
One free parameter is a black hole mass $M$, which is 
used to normalize the variables, and the other free parameter 
is $a_2$.
However we have only one Noether charge 
with respect to time translational symmetry, which is the black hole mass $M$
given by (\ref{ae_BH_mass}).
As we showed in the previous subsection, 
the parameter $a_2$ describes the distribution of 
the aether field, but does not provide a conserved quantity.
It just describes a cloud of the aether field around a black hole 
or a thunderbolt singularity. $a_2$ describes 
a different configuration of the aether cloud.

Hence, fixing a black hole mass $M$ and changing $a_2$, we may find 
 most preferable configuration of the aether field, which 
gives a stable solution.
To find such a solution, we adopt the view point of 
thermodynamical stability, i.e., we assume that
the maximum entropy determines the stable configuration. 

However, the definition of the black hole entropy is unclear 
due to the unavailability of Wald's Noether charge method on the black hole
 horizons in \ae-theory and its extension.
In addition, 
according to \cite{Wald_entropy}, the black hole entropy is modified by the
 higher curvature terms.
Namely, it is given by the integration of functional derivative of the action
 with respect to four dimensional Riemann tensor denoted by
$E^{\mu \nu \rho \sigma}$ over the bifurcation surface $\mathcal{B}$.
In our case, 
\begin{eqnarray}
E^{\mu \nu \rho \sigma} = (1 -\beta_2 \dot{u}^2 -2g_2 \mathcal{R})
 g^{\mu \rho}g^{\nu \sigma} \,.
\end{eqnarray}
Unfortunately, this modification factor $(1  -\beta_2 \dot{u}^2 
 -2g_2 \mathcal{R})$ diverges near
the universal horizon due to the singular behavior of the three 
curvature $\mathcal{R}$.

Hence here we consider only the case without the 
higher-curvature correction terms ($g_2=\beta_2=0$). 
Then we simply assume the black hole entropy is given by the area of the 
horizon $\mathcal{A}(r_H)$, 
where $r_H$ is one of  horizon radii
\footnote{
According to \cite{Wald_entropy}, the black hole entropy is modified by the higher curvature.
Namely, it is given by the integration of functional derivative of the action with respect to four dimensional Riemann tensor over the bifurcation surface $\mathcal{B}$.
When $\dot{u}^4$ term is considered, the entropy should be same as \ae-theory's one.
It is because $\dot{u}^4$ never produces any additional term by the functional derivative.
}.
If we find more appropriate definition of the black hole entropy, 
our result would be changed.

Here we adopt the universal horizon to evaluate the black hole entropy:
\begin{eqnarray}
\mathcal{S}_{\rm uBH}:=
{\mathcal{A}(r_{\rm UH})\over 4G_{\rm N}}={\pi r^2_{\rm UH}\over G_{\rm N}}
\,.
\end{eqnarray}

\begin{figure}[h!]
\begin{center}
\includegraphics[width=75mm]{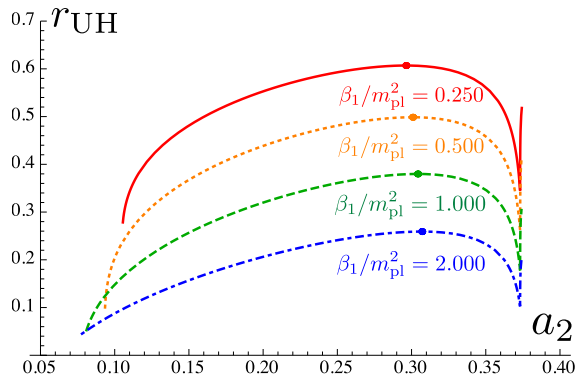} 
\end{center}
\caption{ The radii of the universal horizons $r_{\mathrm{UH}}$ 
with respect to $a_2$ for $\bm{uBH}$ spacetime. 
The solid red, dotted orange, dashed green and dot-dashed blue curve indicate
for $\beta_1/\mpl^2=0.250, 0.500, 1.000$ and $2.000$ case, respectively.
The remaining coupling constants and the boundary conditions are set to 
the same values as those of Fig. \ref{sol_g2zero}.
The points which give the largest value of $r_{\mathrm{UH}}$ for each $\beta_1$ are indicated by the circle plot on the each curves. 
The turning point where the curve has the sharp edge is given by
 $a_{2\mathrm{(turn)}}=0.373$ for every $\beta_1$.
}
\label{th_a4}
\end{figure}

In Fig. \ref{th_a4}, we show the radii of the universal horizon $r_{\mathrm{UH}}$  
with respect to $a_2$ for $\bm{uBH}$ spacetime.
The property of the universal horizon is summarized as follows : 
Each universal horizon radius is a convex upward function
 with respect to
 $a_2$ except the right edge. 
The maximum value of
 black hole entropy for each value of $\beta_1$ is given 
at the top of the convex, which is denoted by $r_{\mathrm{UH}\mathrm{(max)}}$. 
  At the end of the convex function which is called a 
"turning point" denoted by
 $a_{2\mathrm{(turn)}}$, 
the function sharply bounce and turns to increase until 
the regular universal horizon disappears.
Beyond this point, we find $\bm{iTS}$ solution.
  Remarkably, the value of $a_2$ at turning point and the right edge of 
these plot are
invariant with respect to $\beta_1$, which are given by $a_{2\mathrm{(turn)}}
 = 0.373$ and
 $a_{2(\mathrm{end})}=0.377$, respectively.
  Note that $a_{2(\mathrm{end})}$ corresponds to the border between 
$\bm{uBH}$ and $\bm{iTS}$ spacetimes.
 
 We give the detailed data of the thermal quantities with the maximum entropy
and at the turning point 
for each value of $\beta_1$ in TABLE \ref{a4_table}.  
\begin{table}[hbt]
\begin{center}
\begin{tabular}{cc|cc|cc}
\hline \hline
& $\beta_1/\mpl^2$ & $a_{2\mathrm{(max)}}$ & $r_{\mathrm{UH (max)}}$ &
  $a_{2\mathrm{(turn)}}$ & $r_{\mathrm{UH (turn)}}$ \\
\hline \hline
&$0.250  $ &$0.296$ & $0.607$& $0.373$ & $0.345$ \\
&$0.500  $ &$0.301$ & $0.499$& $0.373$ & $0.262$ \\
&$1.000  $ &$0.304$ & $0.380$& $0.373$ & $0.178$ \\
&$2.000  $ &$0.307$ & $0.259$& $0.373$ & $0.103$ \\
\hline \hline
\end{tabular}
\caption{
 the detailed data of the universal horizon radii at the maximum point and
 at the turning point
 for each value of $\beta_1$.
}
\label{a4_table}
\end{center}
\end{table}

When we take the maximum value of the universal horizon radii 
with respect to $a_2$, we may find the most preferable 
black hole solution for a given mass $M$.
It is because such a maximum point may give a stable solution  
from the view point of the black hole thermodynamics.

\begin{figure}[h!]
\begin{center}
\includegraphics[width=75mm]{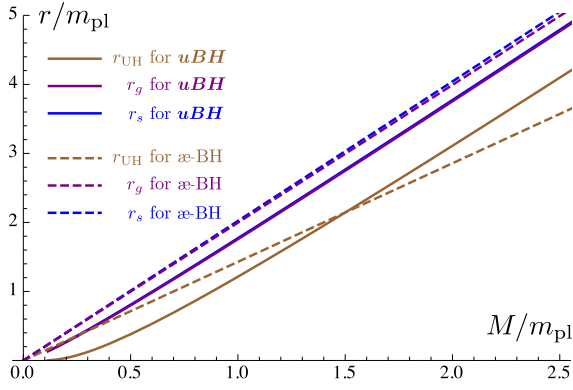}  
\end{center}
\caption{
The mass dependence of the horizon radii for the most preferable black hole solutions with $\beta_1 / \mpl^2 =1$ are indicated by solid curves.
$a_2$ is fixed so that the maximum universal horizon radii is obtained. 
The remaining coupling constants and the boundary conditions are set to 
the same values as those of Fig. \ref{sol_g2zero}.
The brown, purple and blue curves indicate 
the radii of the universal horizon, the graviton horizon and 
the scalar-graviton horizon, respectively. 
The radius of the $\bm{uBH}$ universal horizon turn to be greater than 
that of \ae-black hole in $M>M_{\mathrm{crit}}=1.497 \mpl$ region.
}
\label{M-r_relation}
\end{figure}

Then we find a series of these most preferable black hole 
solutions in terms of $M$.
In Fig. \ref{M-r_relation}, we plot the horizon radii of 
such a black hole v.s. 
the gravitational mass $M$ for $\beta_1/\mpl^2=1$.
Additionally, that of the \ae-black hole, namely, 
$\beta_1 = \beta_2 = g_2 =0$ case, 
are also shown as reference. 
For all  horizons (the graviton, scalar-graviton and universal 
horizons) of the \ae-black hole,
their radii seem to increase linearly in terms of the mass $M$, 
which is the same  as that of Schwarzschild solution.

Whereas, for our $\bm{uBH}$, the universal horizon radius increases 
with a higher power-law function of $M$ than the linear one
in the small mass region, and it   approaches a linear function 
for large values of $M$.
The $\bm{uBH}$ universal horizon radius 
is smaller than \ae-black holes's one for $M>M_{\mathrm{crit}}=1.497\mpl$.
We shall refer $M_{\mathrm{crit}}$ as a {\it critical mass}.
Note that 
the universal horizon radius seems to vanish for small value of $M$, 
but it is not clear whether it vanishes at a finite mass $M_{\rm min}(>0)$,
or at $M=0$.

One may wonder whether our ${\bm uBH}$ solution will 
recover the \ae-black hole or not 
when $\beta_1$ approaches to zero.
To see this, we give the detailed data of $r_{\rm{UH (max)}}$ 
and corresponding $a_{2\mathrm{(max)}}$ near $\beta_1 =0$ in Table \ref{beta_1-rUH_relation}. 
\begin{table}[hbt]
\begin{center}
\begin{tabular}{cccc}
\hline \hline
& $\beta_1/\mpl^2$ & $a_{2\mathrm{(max)}}$ & $r_{\mathrm{UH (max)}}$ \\
\hline \hline
&$0         $ &$1.112 \times 10^{-3}$ & $0.720$ \\
&$0.001  $ &$0.229$                       & $1.001$ \\
&$0.005  $ &$0.248$                       & $0.949$ \\
&$0.010  $ &$0.259$                       & $0.916$ \\
&$0.050  $ &$0.281$                       & $0.799$ \\
&$0.100  $ &$0.288$                       & $0.727$ \\
\hline \hline
\end{tabular}
\caption{
The detailed values of $r_{\rm{UH}}$ and corresponding $a_2$ of 
the most preferable $\bm{uBH}$ solutions for $\beta_1\ll 1$.
The case of $\beta_1=0$ corresponds to the \ae-black hole.
The coupling constants and the boundary conditions are set to the 
same values as those of 
FIG. \ref{sol_g2zero}.
}
\label{beta_1-rUH_relation}
\end{center}
\end{table}
From this table, we find that the value of $a_{2_\mathrm{(max)}}$ 
gradually decreases  as $\beta_1\rightarrow 0$.
Whereas, the maximum value of the universal horizon radius increases 
in such a limit.
Thus, we can conclude that the $\bm{uBH}$ solution cannot be 
smoothly connected to the \ae-black hole.

%%%%%%%%%%%%%%%%%%%%%%%%%%%%%%%%%%%%%%%%%%%%%%%%%%%
\subsection{Smarr's formula and Black Hole Temperature} 
\label{temperature_BH}
%%%%%%%%%%%%%%%%%%%%%%%%%%%%%%%%%%%%%%%%%%%%%%%%%%%
%%%%%%%%%%%%%%%%%%%%%%%%%%%%%%%%%%%%%%%%%%%%%%%%%%%
Concerning the black hole first law in \ae-theory, 
it is found that the aether field prevents from establishing the black hole 
mass-entropy relation 
on the Killing horizon\cite{noether_charge_ae} 
via Noether charge method\cite{NC_is_entropy, Wald_entropy}. 
Notwithstanding, the Smarr's formula in \ae-theory has been proposed only 
in static and spherically symmetric configuration\cite{mec_UH}, which is 
established by applying Gauss's law to the aether field equation.
It is found that the aether portion of the basic equation can
 be reduced 
to Maxwell-like form in static and spherically symmetric spacetime :
  \begin{eqnarray}
 \nabla_\alpha \mathcal{F}^{\alpha \mu}=0\,, 
 \label{Maxwell-like_eq}
 \end{eqnarray} 
 where, $\mathcal{F}_{\mu \nu}$ is given by
 \begin{eqnarray}
  \mathcal{F}_{\mu \nu}&:=& 2q u_{[\mu} s_{\nu]}\,, \\
   q&:=& \left[ {c_{14} \over 2} -c_{13} \right] (\dot{u} \cdot s) 
(u \cdot \xi) +(1-c_{13}) \kappa \notag \\
 &&+ {c_{123} \over 2} (\nabla \cdot u) (s \cdot \xi) \,, 
\label{Maxwell_def}
 \end{eqnarray}
  $s^\mu$ is a spacelike unit vector perpendicular to $u^\mu$ 
  and a surface gravity $\kappa$ which is given by 
  $\xi^\alpha \nabla_\alpha \xi^\mu = \kappa \xi^\mu$ 
  is equivalent to the following form due to a spacetime symmetry : 
 \begin{eqnarray}
 \kappa = \sqrt{-{1 \over 2}(\nabla_\alpha \xi_\beta)
( \nabla^\alpha \xi^\beta)}\,.
 \end{eqnarray}
 Note that the structure of (\ref{Maxwell-like_eq}) is similar 
to Maxwell's equation, 
 thus, we can perform the flux integration. 
  According to Gauss's law, the integration of $\mathcal{F}_{\mu \nu}$ over
  $\mathcal{B}_r$  which is a two sphere at radius $r$ must produce same value 
for any $r$.
 Considering the integration over spatial infinity $\mathcal{B}_\infty$ and 
$\mathcal{B}_r$, then we obtain 
\begin{eqnarray}
 M = q(r) {\mathcal{A}(r) \over 4\pi G_{\rm N}} \,,
 \label{Smarr_eq0}
\end{eqnarray}
where  $\mathcal{A}(r):= 4\pi r^2$ is a surface  area of $\mathcal{B}_r$.
Note that this relation is held for any $r$.
When we evaluate the r.h.s. of Eq. (\ref{Smarr_eq0}),
we find the Smarr's formula:
\begin{eqnarray}
M = \mathcal{T}^{(\ae)}_{\rm uBH}\mathcal{S}_{\rm uBH}
\,,
\label{Smarr_eq}
\end{eqnarray}
where 
$\mathcal{T}^{(\ae)}_{\rm uBH}=q(r_{\rm UH})/\pi$ is a black hole temperature.

 Turning to our attention to the case of including higher curvature and aether
 effects,
 although the Maxwell-like aether equation is not found,
 we may obtain the relation between the black hole mass and the entropy 
(or the area of the universal horizon) by 
evaluating the deviation from the \ae-black hole's one.
 Note that our thermodynamical analysis is limited to the black hole 
solution only with $\beta_1$ term. 
 If we consider the surface gravity or black hole temperature on the singular
 horizon, these quantities must diverge.
 Thus, we shall not discuss the solutions without a regular universal horizon.
 
We shall presume the mass-entropy relation as follows:
\begin{eqnarray}
 M &=& 
\mathcal{T}^{(z=2)}_{\rm uBH}  \mathcal{S}_{\rm uBH}
\label{mod_Smarr_rel}
\end{eqnarray}
where
$ \mathcal{T}^{(z=2)}_{\rm uBH}:=\mathcal{T}^{(\ae)}_{\rm uBH} + 
\delta \mathcal{T}^{(z=2)}_{\rm uBH}$. 
$\delta \mathcal{T}^{(z=2)}_{\rm uBH}$ is introduced 
as the correction by the $z=2$ Lifshitz scaling term  
from the \ae-black hole temperature.

\begin{figure}[h!]
\begin{center}
\includegraphics[width=85mm]{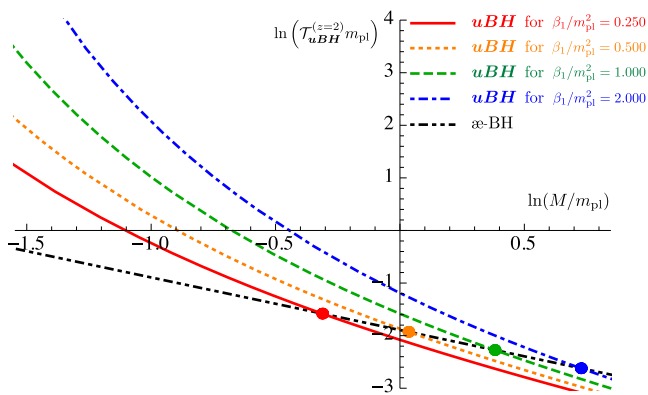}  
\end{center}
\caption{
The mass-temperature on the universal horizon 
for the most preferable black hole solutions. 
The solid red, dotted orange, dashed green and dot-dashed blue curves
indicate the temperatures $ \mathcal{T}^{(z=2)}_{\rm uBH}$
for $\beta_{1}/\mpl^2 = 0.250, 0.500, 1.000$ and $2.00$, respectively. 
The values of $a_2$ for each case are fixed 
so that the maximum universal horizon radii is obtained. 
The remaining coupling constants and the boundary conditions are set to 
the same values as those of Fig. \ref{sol_g2zero}.
The \ae-case is also shown as a reference, which is indicated by 
the two-dot chain black line. 
The critical mass for each $\beta_1$ are indicated by the colored dot.
}
\label{M-T_relation}
\end{figure}

In Fig. \ref{M-T_relation}, we  show the mass-temperature relation 
of the most preferable black hole solutions defined in the previous
subsection.
We also show  \ae-case\cite{mec_UH} as a reference.
From this plot, we find that the \ae-black hole 
temperature seems 
to be inversely proportional to the black hole mass $M$, which is  the 
same as the Schwarzschild black hole in GR.
Whereas, that of our $\bm{uBH}$ solution is clearly far from the
 inverse $M$ law.
More specifically, the black hole temperature of $\bm{uBH}$ turns to be lower 
than that of \ae-case for the range of $M>M_{\mathrm{crit}}$,
where $M_{\mathrm{crit}}$ is the critical mass defined in the previous 
subsection.
In TABLE. \ref{Mcrit_List}, we show the values of $M_{\mathrm{crit}}$ and 
corresponding $r_{\mathrm{UH}}$ in terms of $\beta_1$.
\begin{table}[hbt]
\begin{center}
\begin{tabular}{cccccc}
\hline \hline
& $\beta_1/\mpl^2$ & $M_{\mathrm{crit}}/\mpl$ &$r_{\mathrm{UH}}/\mpl$ \\
\hline \hline
& $1.000 $ & $1.497$ & $2.134$ \\
& $0.900 $ & $1.420$ & $2.029$ \\
& $0.800 $ & $1.339$ & $1.913$ \\
& $0.700 $ & $1.253$ & $1.789$ \\
& $0.600 $ & $1.160$ & $1.656$ \\
& $0.500 $ & $1.059$ & $1.512$ \\
& $0.400 $ & $0.947$ & $1.352$ \\
& $0.300 $ & $0.820$ & $1.171$ \\
& $0.200 $ & $0.670$ & $0.956$ \\
& $0.100 $ & $0.473$ & $0.676$ \\
\hline \hline
\end{tabular}
\caption{
 The critical masses and corresponding universal horizon 
radii in terms of $\beta_1$.
}
\label{Mcrit_List}
\end{center}
\end{table}

From FIG. \ref{M-T_relation}, 
the temperatures of the $\bm{uBH}$ solutions seem to 
obey the inverse $M$ law in large $M$ region, 
while it may decrease exponentially in small $M$ region.
To examine this behavior, we shall assume that the mass-temperature relation 
is given by the following functional form : 
\begin{eqnarray}
\mathcal{T}^{(z=2)}_{\rm uBH} &\approx& \mu \left( e^{\nu /(M-M_{\rm min})} -1 \right) \notag \\
&\sim& 
\begin{cases}
 \mu e^{\nu /(M-M_{\rm min})}  &~{\rm for}~ M-M_{\rm min} \ll \nu \\
 \mu \nu /(M-M_{\rm min}) &~{\rm for}~ M-M_{\rm min} \gg \nu 
\end{cases} \,, \notag \\
\label{approx_T}
\end{eqnarray} 
where  $\mu$, $\nu$ and $M_{\rm min}$ are some fitting parameters.
Since the universal horizon radius is related to the black hole mass as
 (\ref{mod_Smarr_rel}),
the universal horizon should disappear when $M=M_{\rm min}$,
where $\mathcal{T}^{(z=2)}_{\rm uBH}$ diverges.

The curves in FIG. \ref{M-T_relation} 
are approximately reproduced
by the above functional form (\ref{approx_T})
if we choose the fitting parameters given in TABLE. \ref{M-T_fitting}.
\begin{table}[hbt]
\begin{center}
\begin{tabular}{cccccc}
\hline \hline
& $\beta_1/\mpl^2$ & $\mu ~ \mpl$ &$\nu/\mpl$ & $M_{\rm min}/\mpl$ & $\mathcal{T}^{(z=2)}_{\rm{uBH} }(M\gg\mpl)$   \\
\hline \hline
& $0.250 $ & $0.136$ & $0.649$ & $6.561 \times 10^{-7}$ & $0.088/M$ \\
& $0.500 $ & $0.096$ & $0.919$ & $9.261 \times 10^{-7}$ & $0.088/M$ \\
& $1.000 $ & $0.068$ & $1.299$ & $9.468 \times 10^{-7}$ & $0.088/M$ \\
& $2.000 $ & $0.048$ & $1.837$ & $9.998 \times 10^{-7}$ & $0.088/M$ \\
\hline
& $0        $ & N/A   & N/A     & N/A    & $\mathcal{T}^{(\ae)}_{\rm uBH}=0.151/M$ \\
\hline \hline
\end{tabular}
\caption{
 The fitting parameters in $\mathcal{T}^{(z=2)}_{\rm uBH}$ 
 and the asymptotic behaviors in large $M$ region for each value of $\beta_1$.
 The coupling constants and the boundary conditions are 
 set to the same values as those of FIG. \ref{M-T_relation}.
 We also show the temperature of the \ae-black hole $\mathcal{T}^{(\ae)}_{\rm uBH}$ 
 as a reference in the bottom line.
}
\label{M-T_fitting}
\end{center}
\end{table}
In our numerical result,
it is found that $M_{\rm min}$ is extremely small, 
or possibly vanishes.
Moreover, it is notable that the values of $\mu \nu$ seem to be universal 
for $\beta_1>0$,
namely, all asymptotic behaviors in large $M$ region,  
which is denoted by $\mathcal{T}^{(z=2)}_{\rm{uBH} }(M\gg \mpl)$, 
are the same, but differs from that of the \ae-black hole.

%======================================%
%<<<<<<<<<<<< SECTION V >>>>>>>>>>>>>>%
%======================================%
%%%%%%%%%%%%%%%%%%%%%%%%%%%%%%%%%%%%%%%%%%%
%%%%%%%%%%%%%%%%%%%%%%%%%%%%%%%%%%%%%%%%%%%
%%%%%%%%%%%%%%%%%%%%%%%%%%%%%%%%%%%%%%%%%%%
%%%%%%%%%%%%%%%%%%%%%%%%%%%%%%%%%%%%%%%%%%%
\section{conclusion and discussion}
\label{sum_paper} 
%%%%%%%%%%%%%%%%%%%%%%%%%%%%%%%%%%%%%%%%%%%%%%%%%%%
%%%%%%%%%%%%%%%%%%%%%%%%%%%%%%%%%%%%%%%%%%%%%%%%%%%
%%%%%%%%%%%%%%%%%%%%%%%%%%%%%%%%%%%%%%%%%%%%%%%%%%

Without Lorentz symmetry, the Killing horizon is no longer the event horizon in
the static and spherically symmetric spacetime 
due to the presence of the superluminal propagating modes.
However, in the context of the \ae-theory or 
infrared limit of the non-projectable HL gravity,  
a black hole solution can be still constructed 
by considering the universal horizon 
which is a static limit for an instantaneously propagating particle.
In this paper, we have studied the backreaction to the black hole solution 
in \ae-theory
by the Lifshitz scaling terms.
We have analyzed the ultraviolet modification of the \ae-black holes
including the simple scalar terms with $z=2$ Lifshitz scaling,
specifically, the quadratic term of spacial curvature 
associated with the hypersurface orthogonal 
aether field ($\mathcal{R}^2$), 
the quartic term of  the aether acceleration ($\dot{u}^4$), and
the product of the spacial curvature and the quadratic term of the 
aether acceleration ($\dot{u}^2 \mathcal{R}$).

Only for the case with 
$\dot{u}^4$ term ($\beta_1 \neq 0$ and $\beta_2=g_2=0$), 
we have succeeded in finding a black hole solution 
with regular universal horizon,
which is referred as $\bm{uBH}$.
In contrast to the \ae-case, the black hole solutions are
 obtained without tuning the boundary parameter of the aether field $a_2$.
In other words, $\bm{uBH}$ solutions exist in a finite range in 
the $(\beta_1, a_2)$ parameter plane (see FIG. \ref{sol_a4_region}).
If we select the parameter $a_2$ beyond this range, 
a singular shell appears before forming a universal horizon.
While, considering $\dot{u}^2 \mathcal{R}$ and/or
 $\mathcal{R}^2$ terms ($\beta_2 \neq 0$ and/or $g_2 \neq 0$), 
any black hole solution with regular universal horizon cannot be constructed
due to the divergence of the basic equations on the universal horizon 
(see Appendix.\ref{horizon_regularity}).
However, including only $\mathcal{R}^2$ term with negative $g_2$, 
we have found the solution with a thunderbolt singularity ($\bm{uTS}$),
whose universal horizon turns is still singular but 
the singularity is not observed any outside observers.
Although this solution cannot be regarded as a black hole, 
the cosmic censorship hypothesis is not violated.
Since the thunderbolt singularity had been discovered 
in the context of the quantum gravity in lower dimension, 
the emergence of this solution is not so strange
since the $z=2$ Lifshitz scaling terms are 
regarded as quantum gravitational corrections. 

We then have studied  several properties of our solutions including 
spacetime structures with their Carter-Penrose diagrams.
To investigate the physical meaning of $a_2$, 
we have shown the effective energy momentum tensor of 
the aether field $T^{[\ae]}_{\mu \nu}$,
i.e., the effective energy density and pressure.
From these results, the parameter $a_2$ seems closely related to 
the distribution of the spherical aether cloud.
Namely, large value of $a_2$ configures a dense aether cloud 
and eventually induces a collapse of the aether field, which results in
a formation of a "timelike" singular shell ($\bm{tNS}$).
Moreover, we may speculate that regularity on 
the (infrared limit of) scalar graviton horizon is 
recovered  due to the localization of the aether field
 near the horizon, 
which does not appeared in \ae-case.

Finally, we have explored the thermodynamical 
aspect of the $\bm{uBH}$ solution.
Since the maximum universal horizon radius is 
obtained by choosing an appropriate value of $a_2$
for a given mass $M$,
the parameter $a_2$ may be fixed so that the area of the black hole, 
which may be regarded as the black hole entropy, 
becomes maximum. 
This solution may provide the most 
preferable black hole because it can be  thermodynamical stable.
Additionally, the Smarr's formula and the black hole temperature 
are also examined.
When we speculate the mass-temperature relation as (\ref{mod_Smarr_rel}),
it is found that the temperature does not obey the inverse $M$ law at least
 for small $M$ unlike \ae-case.
It increases exponentially as $M$ approaches to $M_{\rm  min}$, for which 
the horizon area vanishes.

Although there are several coupling constants in the action we consider (\ref{full_action}), 
we have manipulated only ultraviolet coupling constants, i.e., $\beta_1$, $\beta_2$ and $g_2$
from $z=2$ Lifshitz scaling terms. 
The coupling constant $c_{13}$, $c_{2}$ and $c_{14}$ are all fixed so that the \ae-black hole
is restored in infrared limit.
However, we shall emphasize that the existence of the other types of black hole solution
should not  be excluded in this theory with the different values of the coupling constants $c_1$-$c_4$.
For example, non-black hole solution  in \ae-theory
(a static and spherically symmetric solution
with Killing horizon but without a universal horizon found in \cite{EABH}) 
may turn to form a causal boundary due to the Lifshitz scaling terms.

Turning our attention to more energetic region, 
it is obvious that the $z=3$ Lifshitz scaling terms which are required by power-counting 
renormalizability of gravity turn to be dominant rather than $z=2$ terms. 
Namely, our analysis in this paper should correspond to the intermediate region
between the energy scale described in \ae-theory and the such a ultimately high energy scale.
If we consider more ultraviolet modification terms in action, 
we may obtain the additional intriguing spacetime such as a 
singularity-free solution
which is discovered in the context the early universe in HL theory.

%======================================%
%<<<<<<<<<<<< acknowledgement >>>>>>>>>>>>>>%
%======================================%
%%%%%%%%%%%%%%%%%%%%%%%%%%%%%%%%%%%%%%%%%%%
%%%%%%%%%%%%%%%%%%%%%%%%%%%%%%%%%%%%%%%%%%%
%%%%%%%%%%%%%%%%%%%%%%%%%%%%%%%%%%%%%%%%%%%
%%%%%%%%%%%%%%%%%%%%%%%%%%%%%%%%%%%%%%%%%%%
\section*{ACKNOWLEDGEMENT}
%%%%%%%%%%%%%%%%%%%%%%%%%%%%%%%%%%%%%%%%%%%%%%%%%%%
%%%%%%%%%%%%%%%%%%%%%%%%%%%%%%%%%%%%%%%%%%%%%%%%%%%
%%%%%%%%%%%%%%%%%%%%%%%%%%%%%%%%%%%%%%%%%%%%%%%%%%%
YM would like to thank T. Wiseman for giving an opportunity 
to start this study and useful discussion 
when he was staying at Imperial College London as the
Erasmus Mundus PhD fellow.
He also thanks to T. Kitamura for useful comments.
KM acknowledges S. Mukohyama, N. Ohta, and S.M. Sibiryakov 
for discussions in the early stage of the present study.
This work is supported by a Waseda University Grant for 
Special Research Projects
(project number : 2015S-083)
and  by Grants-in-Aid from the 
Scientific Research Fund of the Japan Society for the Promotion of Science 
(No. 25400276). 

\newpage

%===============================%
%<<<<<<<<<<<< APPENDIX >>>>>>>>>>>>>>%
%======================================%
%%%%%%%%%%%%%%%%%%%%%%%%%%%%%%%%%%%%%%%%%%%
%%%%%%%%%%%%%%%%%%%%%%%%%%%%%%%%%%%%%%%%%%%
%%%%%%%%%%%%%%%%%%%%%%%%%%%%%%%%%%%%%%%%%%%
%%%%%%%%%%%%%%%%%%%%%%%%%%%%%%%%%%%%%%%%%%%
\appendix
%%%%%%%%%%%%%%%%%%%%%%%%%%%%%%%%%%%%%%%%%%%%%%%%%%%
%%%%%%%%%%%%%%%%%%%%%%%%%%%%%%%%%%%%%%%%%%%%%%%%%%%
%%%%%%%%%%%%%%%%%%%%%%%%%%%%%%%%%%%%%%%%%%%%%%%%%%%
\section{The disformal transformation in a static and spherically symmetric spacetime with asymptotic flatness.}
\label{trans_law_EF}
%%%%%%%%%%%%%%%%%%%%%%%%%%%%%%%%%%%%%%%%%%%%%%%%%%%
%%%%%%%%%%%%%%%%%%%%%%%%%%%%%%%%%%%%%%%%%%%%%%%%%%%
%%%%%%%%%%%%%%%%%%%%%%%%%%%%%%%%%%%%%%%%%%%%%%%%%%%
In this section,  we shall show the transformation law in 
Eddington-Finkelstein like ansatz (\ref{EF_ansatz}) under the disformal transformation.
We would also like to confirm that the spacetime properties, i.e., 
time independence , spherically symmetry and asymptotically flatness are held.

The disformal transformation we consider is 
 \begin{eqnarray}
 \hat{g}_{\mu \nu} = g_{\mu \nu} +(1-\sigma)u_\mu u_\nu \,,~
 \hat{u}^{\mu} = \sigma^{-1/2} u^\mu \,. \label{EF_graviton}
 \end{eqnarray}
 where the original metric $g_{\mu \nu}$ and aether $u^\mu$ are given by 
(\ref{EF_ansatz}).
Then, the each components of $\hat{g}_{\mu \nu}$ and $\hat{u}^\mu$ is given by
\begin{eqnarray}
&& d\hat{s}^2 = -\hat{T}(r)dv^2 + 2\hat{B}(r)dvdr +\hat{f}(r)dr^2 + r^2 d\Omega^2\,,
 \notag \\ \\
&& \hat{u}^\mu = \left( \hat{a}(r), \hat{b}(r),0,0 \right) \,,
\end{eqnarray}
where, 
\begin{eqnarray}
\hat{T}(r) &:=& T(r) - \left( 1-\sigma \right) \left[ { 1+a(r)^2 T(r) \over 
2a(r) } \right]^2 \,, \label{hatT} \\
\hat{B}(r) &:=& {1 \over 2} B(r) \left[ 1+ \sigma -(1-\sigma) a(r)^2 T(r) 
\right] \,, \\
\hat{f}(r) &:=& (1-\sigma) a(r)^2 B(r)^2 \,, \\
\hat{a}(r) &:=& \sigma^{-1/2} a(r) \,, \\
\hat{b}(r) &:=& {a(r)^2 T(r)-1 \over 2\sigma^{1/2}a(r)B(r) }\,.
\end{eqnarray}
From these form, we find that the time independence and spherical symmetry are
held after transformation.
Hence the horizon radius for the propagating degree of freedom 
whose sound speed is unity in this frame is given by a null surface $\hat{T}(r)=0$.
 
Note that the $(r,r)$ component of the metric $\hat{g}_{rr} = \hat{f}(r)$ is generated unlike original Eddington-Finkelstein type ansatz.
It is not quite unnatural considering the geometrical meaning of the disformal transformation.
More specifically, the transformation can be interpreted as 
a rescaling of timelike separation between 
two spacelike hypersurface with fixing three-dimensional space.
Then, the light cone whose opening 
angle is $90^\circ$ in original $(g,u)$ frame
is distorted after disformal transformation.
Hence, the null coordinate $v$ in original $(g, u)$ frame is 
no longer null in $(\hat{g}, \hat{u})$ frame.

 We then introduce 
new coordinate system $(v^*, r)$
in which $v^*$ becomes a null coordinate.
$v^*$ is defined by  
\begin{eqnarray}
dv^* &=& \sigma^{1/2} \left[ dv - {\hat{B}(r) - \sqrt{\hat{B}(r)^2 + \hat{T}(r)\hat{f}(r)} \over \hat{T}(r)} dr \right] \notag \\
&=& \sigma^{1/2} \left[ dv + {2(1-\sigma^{1/2}) a(r)^2 B(r) \over 1+\sigma^{1/2}-(1-\sigma^{1/2})a(r)^2T(r)}dr \right] \,. \notag \\
\end{eqnarray} 
Then, the metric and the aether field are transformed into 
\begin{eqnarray}
&& d\hat{s}^2 = -\hat{T}^*(r) dv^{*2} +2\hat{B}^*(r) dv^* dr + r^2 d\Omega^2\,, \notag \\
&& \hat{u}^\mu = \left( \hat{a}^*(r), \hat{b}^*(r),0,0 \right) \,,
\end{eqnarray} 
where, 
\begin{eqnarray}
\hat{T}^* (r) &=& \sigma^{-1}T(r) + \left( 1-\sigma ^{-1}\right) \left[ { 1+a(r)^2 T(r) \over 
2a(r) } \right]^2\,, \notag \\ \label{hatTs} \\
\hat{B}^* (r) &=& B(r)\,, \\
\hat{a}^*(r) &=&  {2 \sigma^{1/2} a(r) \over 1+ \sigma^{1/2} -(1-\sigma^{1/2})a(r)^2 T(r) }\,, \\
\hat{b}^*(r) &=& {a(r)^2 T(r)-1 \over 2 \sigma^{1/2} a(r) B(r)}\,.
\end{eqnarray}
Since Eddington-Finkelstein type metric is restored, i.e., the $(r,r)$ component of the metric is vanished, 
we can regard $v^*$ as a null coordinate in $(\hat{g},\hat{u})$ frame.

We now focus on the asymptotic property. 
Substituting (\ref{BC}), we find
\begin{eqnarray}
\hat{T}^*(r) &=& 1+{T_1 \over r} + \mathcal{O}(r^{-2})\,, \\
\hat{B}^*(r) &=& 1+ \mathcal{O}(r^{-2}) \,,\\
\hat{a}^*(r) &=& 1+{(1- \sigma^{1/2})T_1 \over 2 \sigma^{1/2} r} + \mathcal{O}(r^{-2}) \,, \\
\hat{b}^*(r) &=& \mathcal{O}(r^{-2}) \,.
\end{eqnarray}
Thus, it is found that the asymptotic flatness is held
even if the disformal transformation is performed.
Additionally, the mass of the spherical object, 
i.e., Noether charge with respect to time translational symmetry 
is also invariant under the disformal transformation.

%%%%%%%%%%%%%%%%%%%%%%%%%%%%%%%%%%%%%%%%%%%%%%%%%%%
%%%%%%%%%%%%%%%%%%%%%%%%%%%%%%%%%%%%%%%%%%%%%%%%%%%
%%%%%%%%%%%%%%%%%%%%%%%%%%%%%%%%%%%%%%%%%%%%%%%%%%%
\section{The invariance of spatial curvature under disformal transformation} 
\label{sec_field_redefinition}
%%%%%%%%%%%%%%%%%%%%%%%%%%%%%%%%%%%%%%%%%%%%%%%%%%%
%%%%%%%%%%%%%%%%%%%%%%%%%%%%%%%%%%%%%%%%%%%%%%%%%%%
%%%%%%%%%%%%%%%%%%%%%%%%%%%%%%%%%%%%%%%%%%%%%%%%%%%
 As we mentioned, the action of \ae-theory has an invariance except each 
coupling constants.  
In this section, we shall show the transformation law in detail.
For convenience, we define a new tensoral quantity 
$X^{\alpha}_{~\beta \gamma}$ as a change of Christoffel symbol,
\begin{eqnarray}
\hat{\Gamma}^{\alpha}_{~\beta \gamma} &=& \Gamma^{\alpha}_{~\beta \gamma} 
+ X^{\alpha}_{~\beta \gamma} \,, \notag \\
X^{\alpha}_{~\beta \gamma} &:=& (1-\sigma) \bigg[ \sigma^{-1} u^\alpha
 \mathcal{K}_{\beta \gamma}-\dot{u}^\alpha u_\beta u_\gamma  \bigg] \,.
\end{eqnarray}
where, $\mathcal{K}_{\mu \nu} := 
\gamma^{~~\alpha}_{(\mu} \nabla_\alpha u_{\nu)}$ is 
an extrinsic curvature associated with the aether.
Note that $\mathcal{K}_{\mu \nu}=\mathcal{K}_{\nu \mu}$ is imposed by 
the hypersurface orthogonality of the aether.
Then, covariant derivative of the aether is given by
\begin{eqnarray}
\hat{\nabla}_\alpha \hat{u}_\beta = \sigma^{1/2} 
\left[ \nabla_\alpha u_\beta -\left(1- \sigma^{-1} \right)
\mathcal{K}_{\alpha \beta} \right]\,,
\end{eqnarray}
and, we find the aether is invariant,
\begin{eqnarray}
\hat{\dot{u}}_\mu = \hat{u}^\alpha \hat{\nabla}_\alpha \hat{u}_\mu 
= \dot{u}_\mu \,.
\end{eqnarray}
thus, the $c_{13}$, $c_2$ and $c_{14}$ terms in the action (\ref{full_action})
 are transformed into 
\begin{eqnarray}
(\hat{\nabla}_\alpha \hat{u}^\beta)(\hat{\nabla}_\beta \hat{u}^\alpha) 
&=&\sigma^{-1} (\nabla_\alpha u^\beta)(\nabla_\beta u^\alpha)\,, \notag \\
(\hat{\nabla} \cdot \hat{u})^2 &=& \sigma^{-1}(\nabla \cdot u)^2 \,, \notag \\
\hat{\dot{u}}^2 &=& \dot{u}^2 \,.
\end{eqnarray}
The four dimensional Riemann tensor $R^{\alpha}_{~\mu \beta \nu} $ is 
transformed into
\begin{eqnarray}
\hat{R}^{\alpha}_{~\mu \beta \nu} &=& R^{\alpha}_{~\mu \beta \nu}  
+\nabla_\beta X^{\alpha}_{~\mu \nu} -\nabla_\nu X^{\alpha}_{~\mu \beta} \notag \\ 
&& + X^{\alpha}_{~\gamma \beta}X^{\gamma}_{~\mu \nu} 
 - X^{\alpha}_{~\gamma \mu}X^{\gamma}_{~\nu \beta} \,, \label{Riemann_trans}
\end{eqnarray}
and thus, we find the transformation law of Ricci scalar $R$ as follows : 
\begin{eqnarray}
\hat{R} &=& R - \left( 1- \sigma^{-1} \right) 
\left[ (\nabla_\alpha u^\beta)(\nabla_\beta u^\alpha) - (\nabla \cdot u)^2 
 \right] \notag \\
&&+(\mathrm{total~derivative~term})\,,
\end{eqnarray} 
Since the total derivative term in the action can be integrated out, 
we shall abbreviate it.

Turning our attention to the transformation law 
of the spatial three-curvature $\mathcal{R}_{\mu \nu}$.
From the Gauss-Codazzi relation, the spatial curvature can be expressed 
in terms of $g_{\mu \nu}$ and $u^\mu$ :
\begin{eqnarray}
\mathcal{R}_{\mu \nu} = \gamma^{\beta}_{~\mu} \gamma^{\delta}_{~\nu} 
\gamma^{\gamma}_{~\alpha} R^{\alpha}_{~\beta \gamma \delta} 
+\mathcal{K}_{\mu \alpha}\mathcal{K}^{\alpha}_{~\nu} - \mathcal{K} 
\mathcal{K}_{\mu \nu}  \,, \label{TR_trans}
\end{eqnarray}
Since the three metric $\gamma_{\mu \nu}$ is invariant, we have only to 
consider the terms associated with $X^{\alpha}_{~\beta \gamma}$ 
in (\ref{Riemann_trans}) to see the transformation of the first term.
Then,
\begin{eqnarray}
\gamma^{\beta}_{~\mu} \gamma^{\delta}_{~\nu} \gamma^{\gamma}_{~\alpha} 
\nabla_\gamma X^{\alpha}_{~\beta \delta} &=& -\left( 1- \sigma^{-1} \right)
\mathcal{K}\mathcal{K}_{\mu \nu} \,, \notag \\
\gamma^{\beta}_{~\mu} \gamma^{\delta}_{~\nu} \gamma^{\gamma}_{~\alpha} 
\nabla_\delta X^{\alpha}_{~\beta \gamma} &=& -\left( 1- \sigma^{-1} \right)
\mathcal{K}_{\mu \alpha} \mathcal{K}^{\alpha}_{~\nu} \,, \notag \\
\gamma^{\beta}_{~\mu} \gamma^{\delta}_{~\nu} \gamma^{\gamma}_{~\alpha} 
X^{\alpha}_{~\eta \gamma} X^{\eta}_{~\beta \delta} &=& 0  \,, \notag \\
\gamma^{\beta}_{~\mu} \gamma^{\delta}_{~\nu} \gamma^{\gamma}_{~\alpha}
 X^{\alpha}_{~\eta \delta} X^{\eta}_{~\beta \gamma} &=& 0 \,,
\end{eqnarray}
The last two terms of RHS in (\ref{TR_trans}) are transformed into
\begin{eqnarray}
\hat{\mathcal{K}}_{\mu \alpha} \hat{\mathcal{K}}^{\alpha}_{~\nu} &=& 
\sigma^{-1} \mathcal{K}_{\mu \alpha}\mathcal{K}^{\alpha}_{~\nu} \,, 
\notag \\
\hat{\mathcal{K}} \hat{\mathcal{K}}_{\mu \nu} &=& \sigma^{-1} \mathcal{K}
 \mathcal{K}_{\mu \nu} \,.
\end{eqnarray}
Thus, we can find $\mathcal{R}_{\mu \nu}$ is invariant under 
the transformation and obtain (\ref{cc_trans_law}).

Since $\mathcal{R}_{\mu \nu}$ and $\dot{u}^\mu$ are invariant under 
the disformal transformation which is a rescaling of timelike separation 
between two spacelike hypersurface, we expect that its spatial covariant 
derivatives are also invariant.
In order to confirm it, we examine the transformation law of 
a purely spatial tensoral quantity $\mathcal{Z}_{\mu \nu \cdots}$ so that 
$\hat{\mathcal{Z}}_{\mu \nu \cdots}=\mathcal{Z}_{\mu \nu \cdots}$ and 
$u^{\alpha} \mathcal{Z}_{\alpha \nu \cdots}
=u^{\alpha} \mathcal{Z}_{\mu \alpha}= \cdots =0$.
Then,
\begin{eqnarray}
\hat{\mathcal{D}}_{\mu} \hat{\mathcal{Z}}_{\nu \rho \cdots} &=& 
\hat{\gamma}_{\mu}^{~\alpha} \hat{\gamma}_{\nu}^{~\beta} \hat{\gamma}_{\rho}^{~\gamma} \cdots \hat{\nabla}_{\alpha} \hat{\mathcal{Z}}_{\beta \gamma \cdots} \notag \\
&=& 
\gamma_{\mu}^{~\alpha} \gamma_{\nu}^{~\beta} \gamma_{\rho}^{~\gamma} \cdots \Big[ \nabla_{\alpha} \mathcal{Z}_{\beta \gamma \cdots} 
+ X^{\delta}_{~\alpha \beta} \mathcal{Z}_{\delta \gamma \cdots} \notag \\
&&~~~~~~~~~~~~~~~~~~ + X^{\delta}_{~\alpha \gamma} \mathcal{Z}_{\beta \delta \cdots} + \cdots \Big] \,, \notag \\
&=& \mathcal{D}_{\mu}\mathcal{Z}_{\nu \rho}\,. 
\end{eqnarray}
The terms proportional to $X^\alpha_{~\beta \gamma}$ in the second line 
are vanished due to the spatial property of $\mathcal{Z}_{\mu \nu \cdots}$.
Thus, we conclude that all of the spatial higher derivative terms 
derived by non-projectable HL gravity are invariant under disformal transformation.

%%%%%%%%%%%%%%%%%%%%%%%%%%%%%%%%%%%%%%%%%%%%%%%%%%%
%%%%%%%%%%%%%%%%%%%%%%%%%%%%%%%%%%%%%%%%%%%%%%%%%%%
%%%%%%%%%%%%%%%%%%%%%%%%%%%%%%%%%%%%%%%%%%%%%%%%%%%
\section{The regularity conditions of horizon}
\label{horizon_regularity}
%%%%%%%%%%%%%%%%%%%%%%%%%%%%%%%%%%%%%%%%%%%%%%%%%%%
%%%%%%%%%%%%%%%%%%%%%%%%%%%%%%%%%%%%%%%%%%%%%%%%%%%
%%%%%%%%%%%%%%%%%%%%%%%%%%%%%%%%%%%%%%%%%%%%%%%%%%%
We  illustrate the detail of the regularity on the black hole horizons 
with/without $z \neq 1$ Lifshitz scaling terms.
In order to investigate the behavior near horizon, we  focus 
on the highest $r$-derivative terms in the basic equations.

\subsection{Einstein-aether case : $\beta_1=\beta_2=g_2=0$}
In this case, the set of the evolution equations, $(v, v)$ and 
$(\theta, \theta)$ components of (\ref{ae_eq_1}) and $s^\mu$ component 
of (\ref{ae_eq_2}), can be simplified into following form :
\begin{eqnarray}
T'' &=& T''[T,T',B,a,a']\,, \label{Teq} \\
B' &=& B'[T,T',B,a,a']\,, \label{Beq}\\
a'' &=& a''[T,T',B,a,a']\,, \label{aeq}
\end{eqnarray}
To see the cause of singular behavior on the scalar-graviton horizon, 
we expand (\ref{Teq})-(\ref{aeq}) around $r_{\rm{SH}}=0$. We then find
\begin{eqnarray}
T''(r) &=& {T_{[\ae]}(r_{\rm{SH}}) \over T_{\rm{S}}(r)} 
+\sum^{3}_{n=0} T_{[n]}(r_{\rm{SH}}) [T_{\rm{S}}(r)]^n\,, 
\\
B'(r) &=& {B_{[\ae]}(r_{\rm{SH}}) \over T_{\rm{S}}(r)} 
+\sum^{1}_{n=0} B_{[n]}(r_{\rm{SH}}) [T_{\rm{S}}(r)]^n\,, 
\\
a''(r) &=& {a_{[\ae]}(r_{\rm{SH}}) \over T_{\rm{S}}(r)} 
+\sum^{2}_{n=0} a_{[n]}(r_{\rm{SH}}) [T_{\rm{S}}(r)]^n\,.
\end{eqnarray}
where $T_{\rm{S}}(r)=T_{\rm{S}}'(r_{\rm{SH}}) (r-r_{\rm{SH}})$ and
 $T_{[\ae]}$, $B_{[\ae]}$, $a_{[\ae]}$, $T_{[n]}$, $B_{[n]}$ and 
$a_{[n]}$ are functionals with respect to
 $T'(r_{\rm{SH}})$, $B(r_{\rm{SH}})$, $a(r_{\rm{SH}})$ and $a'(r_{\rm{SH}})$.
Obviously, the irregularity on the scalar-graviton horizon is due to 
the coefficient of $T_{\rm{S}}(r)^{-1}$ terms.
As a result, for regularity,
 all of $T_{[\ae]}$,$B_{[\ae]}$ and $a_{[\ae]}$ must vanish,
 otherwise the evolution equations diverge at $r=r_{\rm{SH}}$.  
Although the explicit form of the coefficients of $T_{\rm{S}}(r)^{-1}$ terms are 
quite complicated for general setting of the coupling constants $c_{13}$, 
$c_{2}$ and $c_{14}$,  it can be reduced to slightly simplified form 
considering the disformal transformation (\ref{disformal}).
By the disformal transformation with $\sigma = c_{\rm{S}}^2$,  we can set 
the sound speed of the scalar graviton to be 
 unity without loss of generality. 
In this frame, the scalar-graviton horizon coincides with the metric horizon,
 i.e., $T_{\rm{S}}(r_{\rm{SH}}) = T(r_{\rm{SH}})=0$.

Transforming into the $c_{\rm{S}}^2 =1$ frame reduces 
the three-dimensional parameter 
space of coupling constants $(c_{13},c_{2}, c_{14})$ into two-dimensional 
one. In other words,
 one of these coupling constants is expressed by the other 
two. 
More explicitly, for example, 
we can eliminate $c_{2}$ from the evolution equations as follows:
Performing simple calculation, 
we find  $c_{\rm{S}}^2 =1$ for any $c_{13}$ and $c_{14}$, if we set 
\begin{eqnarray}
c_2 = {-2c_{13}+2c_{14}-c_{13}^2 c_{14} \over 2- 4c_{14}+3c_{13}c_{14}}\,. 
\label{c2_cs_unity}
\end{eqnarray} 
In this frame,  
we finally obtain the explicit forms of $T_{[\ae]}$,
$B_{[\ae]}$ and $a_{[\ae]}$ as
\begin{widetext}
\begin{eqnarray}
T_{[\ae]} &=& -\left[{ 2c_{13}-2c_{14} +c_{13}c_{14} -2(1-c_{13})c_{14} 
T'(r_{\rm{SH}})a^2(r_{\rm{SH}}) r_{\rm{SH}} \over 2c_{14} (1-c_{13}) B(r_{\rm{SH}})a^2(r_{\rm{SH}})r_{\rm{SH}}} \right] 
B_{[\ae]} \,, \\
B_{[\ae]} &=& \left[ { c_{14}B(r_{\rm{SH}}) \over 8r(c_{14}-2)(2-4c_{14}+3c_{13}c_{14})
 a^4(r_{\rm{SH}}) } \right] \sum_{n=0}^{6} B_{[\ae, n]} a^n(r_{\rm{SH}}) \,, \label{reg_con} \\
a_{[\ae]} &=&  \left[{ (2-c_{13})(2-c_{14})a(r_{\rm{SH}}) -2(1-c_{13})c_{14}a'(r_{\rm{SH}})r_{\rm{SH}}
 \over
 2c_{14} (1-c_{13}) B(r_{\rm{SH}})r_{\rm{SH}}} \right] B_{[\ae]} \,.
\end{eqnarray}
where  
\begin{eqnarray}
B_{[\ae,0]} &:=& \left[ 4c_{13} -4( 1 +c_{13} -c_{13}^2 )c_{14} 
+(4-3c_{13})c_{14}^2 \right]a'(r_{\rm{SH}}) r_{\rm{SH}}^2  \,, \notag \\
B_{[\ae,1]} &:=& 8(1-c_{13})^2 c_{14} a'(r_{\rm{SH}})r_{\rm{SH}} \,, \notag \\
B_{[\ae,2]} &:=& -4 (1-c_{13}) (2-2c_{14} +c_{13}c_{14}  ) \notag \,, \\
B_{[\ae,3]} &:=& 2\left[ -8+4c_{13}  +4(2-c_{13})^2 c_{14} 
-(4 - 3c_{13})c_{14}^2 \right]T'(r_{\rm{SH}}) a'(r_{\rm{SH}}) r_{\rm{SH}}^2 \notag \,, \\
B_{[\ae,4]} &:=& 8 (1-c_{13})^2 c_{14} T'(r_{\rm{SH}})r_{\rm{SH}} \,, \notag \\
B_{[\ae,5]} &:=& 0 \,, \notag \\
B_{[\ae,6]} &:=& \left[ 4c_{13} -4(1+c_{13}-c_{13}^2 )c_{14} 
+(4-3c_{13})c_{14}^2 \right]T'^2(r_{\rm{SH}}) r_{\rm{SH}}^2 \,,
\end{eqnarray}
\end{widetext}
Thus,  we conclude that $B_{[\ae]}=0$ should be imposed
for regularity of the scalar-graviton horizon.

As for the regularity on the universal horizon, we also expand 
the basic equations for $T$, $B$ and $a$  around the universal horizon
$r_{\rm UH}$, where $(u\cdot \xi)(r_{\rm UH}) =0$.
We then find there is no term with negative power of $U(r):=U'(r_{\rm UH})
(r-r_{\rm UH})$.
This means that the universal horizon is always regular, if it exists,
 unlike 
the scalar-graviton horizon.
\\

%%%%%%%%%%%%%%%%%%%%%%%%%%%%%%%%%%%%%%%%%%%%%%%%%%%%%%%%%%%%%%%
%%%%%%%%%%%%%%%%%%%%%%%%%%%%%%%%%%%%%%%%%%%%%%%%%%%%%%%%%%%%%%%
 \subsection{The case with $\dot{u}^4$ term 
: $\beta_1 \neq 0$ and $g_2=\beta_2=0$}
%%%%%%%%%%%%%%%%%%%%%%%%%%%%%%%%%%%%%%%%%%%%%%%%%%%%%%%%%%%%%%%
%%%%%%%%%%%%%%%%%%%%%%%%%%%%%%%%%%%%%%%%%%%%%%%%%%%%%%%%%%%%%%%
 The set of the evolution equations can be decomposed into $T(r)$, $B(r)$ 
and $a(r)$ equations similar to the Einstein-aether's case
 (\ref{Teq})-(\ref{aeq}),
 i.e.,  the linear-order differential equation for $B(r)$ and the second-order 
differential equations for $T(r)$ and $a(r)$.
The expanded equations around $r_{\rm SH}=0$ are given by 
\begin{eqnarray}
T''(r) &=& \sum_{n\geq 0} T_{[n]}^{(\beta_1)}(r_{\rm{SH}}) [T_{\rm{S}}(r)]^n\,, 
\\
B'(r) &=& \sum_{n \geq 0} B_{[n]}^{(\beta_1)}(r_{\rm{SH}}) [T_{\rm{S}}(r)]^n\,, 
\\
a''(r) &=& \sum_{n \geq 0} a_{[n]}^{(\beta_1)}(r_{\rm{SH}}) [T_{\rm{S}}(r)]^n\,.
\end{eqnarray}
where $T_{\rm{S}}(r)=T_{\rm{S}}'(r_{\rm{SH}}) (r-r_{\rm{SH}})$ and
 $T_{[n]}^{(\beta_1)}$, $B_{[n]}^{(\beta_1)}$ and 
$a_{[n]}^{(\beta_1)}$ are functionals with respect to
 $T'(r_{\rm{SH}})$, $B(r_{\rm{SH}})$, $a(r_{\rm{SH}})$ and $a'(r_{\rm{SH}})$.
 Since all of these equations have no terms with negative power of $T_{\rm{S}}(r)$ 
in the expansion near the scalar-graviton horizon, 
it is always regular without tuning $a_2$.
Similarly, it can be confirmed that the universal horizon is always regular in the same way.
Namely, the black hole solutions turn to depend on the mass parameter $T_1$
as well as the extra parameter $a_2$. 
Actually we find the black hole solutions in a certain range of $a_2$ 
in Section \ref{a4_BH}.
 \newpage

%%%%%%%%%%%%%%%%%%%%%%%%%%%%%%%%%%%%%%%%%%%%%%%%%%%%%%%%%%%%%%%
%%%%%%%%%%%%%%%%%%%%%%%%%%%%%%%%%%%%%%%%%%%%%%%%%%%%%%%%%%%%%%%
\subsection{The case with $\mathcal{R}^2$ and/or $\dot{u}^2 \mathcal{R}$ term 
:\\  $g_2 \neq 0$ and/or $\beta_2 \neq 0$}
%%%%%%%%%%%%%%%%%%%%%%%%%%%%%%%%%%%%%%%%%%%%%%%%%%%%%%%%%%%%%%%
%%%%%%%%%%%%%%%%%%%%%%%%%%%%%%%%%%%%%%%%%%%%%%%%%%%%%%%%%%%%%%%
In this case we find that $T'''(r)$, $B'''(r)$ and $a'''(r)$, which are 
the highest $r$-derivatives in the equations, appear  only in the
$(\theta,\theta)$ component of (\ref{ae_eq_1}).
This means that the evolution equations cannot be separated into 
$T(r)$, $B(r)$ and $a(r)$ equations unlike the Einstein-aether
only  with the 
 $\dot{u}^4$ term.
Therefore, we just focus on the coefficients of $T'''(r)$, $B'''(r)$ 
and $a'''(r)$ terms in the $(\theta,\theta)$ component of (\ref{ae_eq_1}).
To see the behavior near the universal horizon where $u \cdot \xi =0$, 
we shall express these terms  using  $U(r):=U'(r_{\rm UH})
(r-r_{\rm UH})$ instead of 
 $T(r):=T'(r_{\rm{SH}})(r-r_{\rm{SH}})$:
\begin{eqnarray}
0 \approx \Theta_{T} T'''(r) + \Theta_{B} B'''(r)+ \Theta_{a} a'''(r) \,,
\end{eqnarray}
where  $\Theta_{T}$, $\Theta_{B}$ and $\Theta_{a}$ are the functional 
which are given by
\begin{eqnarray}
\Theta_{T} &:=& {a(r_{\rm UH}) U^2(r) 
[ 2g_2 U(r) - \beta_2 r_{\rm UH}
 U'(r_{\rm UH}) ] \over r_{\rm UH}^3 B^4(r_{\rm UH})}
 \notag \,, \\
\Theta_{B} &:=& {4 g_2 U^4(r) \over r_{\rm UH}^3 B^5(r_{\rm UH})}
\notag \,, 
%\\
\end{eqnarray}
and 
\begin{widetext}
\begin{eqnarray}
\Theta_{a} &:=& -{2 U^2(r) [1+a(r_{\rm UH})U(r)] 
[ 2g_2 U(r) - \beta_2 r_{\rm UH} U'(r_{\rm UH}) ] 
\over r_{\rm UH}^3 a^2(r_{\rm UH}) B^4(r_{\rm UH})}  
\notag \,, 
%\\ 
\end{eqnarray}
\end{widetext}
 Clearly, it is impossible to avoid the divergence of this equation 
on the universal horizon because all of the coefficients
$\Theta_{T}, \Theta_{B}$ and $ \Theta_{a}$ must not vanish 
simultaneously at $r_{\rm UH}$  for regularity.
 Therefore, we conclude that the universal horizon is always singular.
 As a result, the thunderbolt singularity appears if $\mathcal{R}^2$
 and/or $\dot{u}^2 \mathcal{R}$ terms are joined.

%======================================%
%<<<<<<<<<<<< BIBLIOGRAPHY >>>>>>>>>>>>%
%======================================%
%%%%%%%%%%%%%%%%%%%%%%%%%%%%%%%%%%%%%%%%%%%%%%%%%%
%%%%%%%%%%%%%%%%%%%%%%%%%%%%%%%%%%%%%%%%%%%%%%%%%%


\begin{thebibliography}{99}
\bibitem{Hawking_Ellis}
R. Penrose, Phys. Rev. Lett. {\bf 14}, 57 (1965); 
S.W. Hawking, Proc. Roy. Soc. Lond., {\bf A300}, 187 (1967); 
S.W. Hawking and R. Penrose,Proc. Roy. Soc. Lond., {\bf A314}, 529 (1970); 
S.W. Hawking and G.F.R. Ellis, {\it The large scale structure of space-time}
 (Cambridge Univ., 1973).

\bibitem{loop}
See for example, C. Rovelli, {\it Quantum Gravity} (Cambridge Univ., 2004).

\bibitem{triangulation}
%Discrete Lorentzian Quantum Gravity
R. Loll, Nucl. Phys. Proc. Suppl. {\bf 94}, 96-107 (2001)[arXiv:hep-th/0011194].

\bibitem{string}
M.B. Green, J.H. Schwarz, and E. Witten, {\it Superstring Theory}, in 2 vols., (Cambridge Univ., 1987);
J. Polchinski, {\it String Theory}, in 2 vols., (Cambridge Univ., 1998).

\bibitem{HL_original_paper}
P. Ho\v{r}ava,  
%"Quantum Gravity at a Lifshitz Point", 
Phys. Rev. D {\bf 79}, 084008 (2009)[arXiv:0901.3775[hep-th]].

\bibitem{renormalizability}
D. Anselmi and M. Halat,
%"Renormalization of Lorentz violating theories”,
Phys. Rev. D {\bf 76}, 125011 (2007)[arXiv:0707.2480 [hep-th]];
M. Visser, 
%"Lorentz symmetry breaking as a quantum field theory regulator”,
Phys. Rev. D {\bf 80}, 025011 (2009)[arXiv:0902.0590[hep-th]];
D. Orlando and S. Reffert,
%”On the Renormalizability of Horava-Lifshitz-type Gravities”,
Class. Quant. Grav. {\bf 26}, 155021 (2009)[arXiv:0905.0301[hep-th]]
R. Iengo, J. G. Russo and M. Serone, 
%”Renormalization group in Lifshitz-type theories”,
JHEP {\bf 0911}, 020 (2009) [arXiv:0906.3477 [hep-th]];
M. Visser,
%”Power-counting renormalizability of generalized Horava gravity”
arXiv:0912.4757[hep-th];
M. Eune, W. Kim and E. J. Son, 
%”Effective potentials in the Lifshitz solar field theory”,
Phys. Lett. B {\bf 703}, 100 (2011)[arXiv:1105.5194 [hep-th]];
D. L. Lopez Nacir, F. D. Mazzitelli and L. G. Trombetta, 
%"Lifshitz scalar fields: one loop renormalization in curved background",
Phys. Rev. D {\bf 85}, 024051 (2012) [arXiv:1111.1662 [hep-th]];
M. Colombo, A. E. Gumrukcuoglu  and T. P. Sotiriou,
%”Horava gravity with mixed derivative terms”,
Phys. Rev. D {\bf 91}, 044021 (2015)[arXiv:1410.6360[hep-th]];
T. Fujimori, T. Inami, K. Izumi and T. Kitamura,
%Power-counting and Renormalizability in Lifshitz Scalar Theory
Phys. Rev. D {\bf 91}, 125007 (2015)[arXiv:1502.01820[hep-th]].

\bibitem{HL_cosmological_sin_avoicance}
R. H. Brandenberger,
%"Matter bounce in Ho\UTF{0159}ava-Lifshitz cosmology"
Phys. Rev. D {\bf 80}, 043516 (2009)[arXiv:0904.2835[hep-th]];
K. Maeda, Y. Misonoh, T. Kobayashi,
%"Oscillating Universe in Horava-Lifshitz Gravity"
Phys. Rev. D {\bf 82}, 064024 (2010)[arXiv:1006.2739[hep-th]];
Y. Misonoh, K.Maeda, T. Kobayashi,
%"Oscillating Bianchi IX Universe in Horava-Lifshitz Gravity"
Phys. Rev. D {\bf 84}, 064030 (2011)[arXiv:1104.3978[hep-th]].

\bibitem{HL_BH_ref}
E. Barausse and T. P. Sotiriou, 
%A no-go theorem for slowly rotating black holes in Horava-Lifshitz gravity
Phys. Rev. Lett. {\bf 109} 181101 (2012)
[Phys. Rev. Lett. {\bf 110}, no. 3, 039902 (2013)]
[arXiv:1207.6370[gr-qc]];
E. Barausse and T. P. Sotiriou, 
%Slowly rotating black holes in Horava-Lifshitz gravity
Phys. Rev. D {\bf 87}, 087504 (2013)[arXiv:1212.1334[gr-qc]];
A. Wang,
%"Stationary and slowly rotating spacetimes in Ho\UTF{0159}ava-Lifshitz gravity",
Phys. Rev. Lett. {\bf 110}, 091101 (2013)[arXiv:1212.1876[hep-th]];
J. Greenwald, J. Lenells, V. H. Satheeshkumar and A.Wang,
%"Gravitational collapse in Ho\UTF{0159}ava-Lifshitz theory",
Phys. Rev. D88, 024044 (2013)[arXiv:1304.1167[hep-th]];
T.P. Sotiriou, I.Vega and D.Vernieri,
%"Rotating black holes in three-dimensional Ho\UTF{0159}ava gravity"
Phys. Rev. D {\bf 90}, 044046 (2014)[arXiv:1405.3715[gr-qc]];

\bibitem{EA_original_paper}
T. Jacobson and D. Mattingly,  
%"Gravity with a dynamical preferred frame", 
Phys. Rev. D {\bf 64}, 024028 (2001)[arXiv:gr-qc/0007031].

\bibitem{khronon_theory}
T. Jacobson, 
%"Extended Horava gravity and Einstein-aether theory"
Phys. Rev. D {\bf 81}, 101502 (2010)
[Phys. Rev. D. {\bf 82}, 129901 (2010)]
[arXiv: 1001.4823[hep-th]].

\bibitem{EAHLBH}
E. Barausse, T. Jacobson and T.P. Sotiriou, 
%"Black holes in Einstein-aether and Ho\v{r}ava-Lifshitz gravity"
Phys. Rev. D{\bf 83}, 124043 (2011)[arXiv:1104.2889[gr-qc]].

\bibitem{UH_BH}
D. Blas and S. Sibiryakov,
%"Ho\v{r}ava gravity versus thermodynamics: The black hole case", 
Phys. Rev. D {\bf 84}, 124043 (2011)[arXiv:1110.2195[hep-th]].

\bibitem{mec_UH}
P. Berglund, J. Bhattacharyya and D. Mattingly,
%"Mechanics of universal horizons"
Phys. Rev. D {\bf 85}, 124019 (2012)[arXiv:1202.4497[hep-th]].

\bibitem{maximally_EA}
J. Bhattacharyya and D. Mattingly,
%"Universal horizons in maximally Symmetric space"
Int. J. Mod. Phys. D{\bf23} (2014) 1443005[arXiv:1408.6479 [hep-th]].

\bibitem{high_dim_UH}
K. Lin, F. Shu, A. Wang, Q. Wu,
%"High-dimensional Lifshitz-type spacetimes, 
%universal horizons and black holes in Ho\v{r}ava-Lifshitz gravity"
Phys. Rev. D {\bf91}, 044003 (2015)[arXiv:1404.3413[gr-qc]]

\bibitem{UH_LV}
K. Lin, E. Abdalla, R. Cai, A. Wang,
%"Universal horizons and black holes in gravitational theories 
%with broken Lorentz symmetry"
Inter. J. Mod. Phys. D{\bf23}, (2014) 1443004[arXiv:1408.5976[gr-qc]]

\bibitem{newlook}
K. Lin, O. Goldoni, M.F. da Silva, A. Wang,
%"New look at black holes: Existence of universal horizons"
Phys. Rev. D {\bf91}, 024047 (2015)[arXiv:1410.6678[gr-qc]]

\bibitem{charged_BH}
S. Janiszewski, A. Karch, B. Robinson and D. Sommer,
%"Charged black holes in Horava gravity"
JHEP {\bf1404}, 163 (2014)[arXiv:1401.6479[hep-th]].

 \bibitem{C_BH}
 C. Ding, A. Wang and X. Wang,
 %"Charged Einstein-aether black holes and Smarr formula"
 arXiv:1507.06618[gr-qc].

\bibitem{UH_tunneling}
P. Berglund, J. Bhattacharyya and D. Mattingly,
%“Towards Thermodynamics of Universal Horison in Einstein-aether Theory”
Phys. Rev. Lett. {\bf 110}, 071301 (2013)[arXiv:1210.4940[hep-th]].

\bibitem{UH_Wald}
A. Mohd, 
%“On the thermodynamics of universal horizons in Einstein-\AE ther theory”, 
arXiv:1309.0907[gr-qc].

\bibitem{ray_tracing}
B. Cropp, S. Liberati, A. Mohd, M. Visser,
%“Ray tracing Einstein-\AE ther black holes: Universal 
%versus Killing horizons”
Phys. Rev. D {\bf 89}, 064061 (2014)[arXiv:1312.0405[gr-qc]].

\bibitem{dynamical_UH}
M. Saravani, N. Afshordi, and R.B. Mann,
%“Dynamical emergence of universal horizons during the formation
% of black holes”
Phys. Rev. D {\bf 89}, 084029 (2014)[arXiv:1310.4143[gr-qc]].

\bibitem{formation_UH}
M. Tian, X. Wang, M.F. da Silva, A. Wang,
%“Gravitational collapse and formation of universal horizons”
arXiv:1501.04134[gr-qc].

\bibitem{NPHL_action}
S. Carloni, E. Elizalde and P.J. Silva,  
%"Matter couplings in Ho\v{r}ava-Lifshitz and their cosmological application", 
Class. Quant. Grav. {\bf 28}: 195002 (2011)[arXiv:1009.5319[hep-th]].

\bibitem{ae_wave}
T. Jacobson and D. Mattingly,  
%"Einstein-Aether Waves", 
Phys. Rev. D {\bf 70}, 024003 (2004)[arXiv:gr-qc/0402005].

\bibitem{ae_res}
B.Z. Foster, 
%"Metric Redefinitions in Einstein-{\AE}there Theory", 
Phys. Rev. D {\bf 72}, 044017 (2005)[arXiv:gr-qc/0502066].

\bibitem{ae-HL_equality}
E. Barausse, T.P. Sotiriou,
%"Black holes in Lorentz-violating gravity theories"
Class. Quant. Grav. {\bf 30} 244010 (2013)[arXiv:1307.3359[gr-qc]].

\bibitem{initial_const}
T. Jacobson,
%“Initial value constraints with tensor matter”
Class. Quant. Grav. {\bf 28} 245011 (2011) [arXiv:1108.1496[gr-qc]].

\bibitem{EABH}
C. Elling and T. Jacobson, 
%"Black holes in Einstein-aether theory"
Class. Quant. Grav. {\bf 23}, 5643-5660 (2006) [arXiv:gr-qc/0604088].

\bibitem{noether_charge_ae}
B.Z. Foster, 
%“Noether charges and black hole mechanics in Einstein-aether theory”, 
Phys Rev. D {\bf 73}, 024005 (2006) [arXiv:gr-qc/0509121].

\bibitem{Hamiltonian_HL}
W. Donnelly and T. Jacobson, 
%"Hamiltonian structure of Ho\v{r}ava Gravity"
Phys. Rev. D { \bf 84}, 104019 (2011) [arXiv: 1106.2131[hep-th]].

\bibitem{GW_HL}
D. Blas and H. Sanctuary,  
%"Gravitational Radiation in  Ho\v{r}ava Gravity"
Phys. Rev. D { \bf 84}, 064004 (2011) [arXiv: 1105.5149[gr-qc]].

\bibitem{g_b_healthy}
%"Models of non-relativistic quantum gravity: the good, the bad and the healthy"
D.Blas, O.Pujolas and S.Sibiryakov, 
JHEP {\bf 1104}, 018 (2011) [arXiv:1007.3503[hep-th]].

\bibitem{Hawking_Stewart}
S.W. Hawking and J.M.Stewart, 
%“Naked and thunderbolt singularities in black hole evaporation”
Nucl. Phys. B{\bf 400} 393-415 (1993) [arXiv:hep-th/9207105].

\bibitem{CGHS}
C.G.Callan, S.B.Giddings, J.A.Harvey and A.Strominger,
%“Evanescent black holes”
Phys. Rev. D {\bf 45} 1005 (1992) [arXiv:hep-th/9111056].

\bibitem{NS_TB}
A.Ishibashi and A.Hosoya,
%“Naked singularity and a thunderbolt”
Phys. Rev. D {\bf 66}  104016 (2002) [arXiv:gr-qc/0207054].

\bibitem{Wald_entropy}
V. Iyer and R.M. Wald, 
%“Some Properties of Noether Charge and a Proposal for 
%Dynamical Black Hole Entropy”, 
Phys. Rev. D {\bf 50}, 846 (1994)[arXiv:gr-qc/9403028]

\bibitem{NC_is_entropy}
R.M. Wald, 
%“Black Hole Entropy is Noether Charge”
Phys. Rev. D {\bf 48}, 3427 (1993)[arXiv:gr-qc/9307038]

\end{thebibliography}
\end{document}